\documentclass[12pt,english]{article}
\usepackage{natbib}
\usepackage{verbatim}
\usepackage[T1]{fontenc}
\usepackage{lmodern}
\usepackage{pifont}
\usepackage[T1]{fontenc}
\usepackage{makecell}
\usepackage[latin9]{inputenc}
\usepackage[letterpaper]{geometry}
\usepackage{amsmath}
\usepackage{color}
\usepackage{float}
\usepackage{graphicx}
\usepackage{rotating}
\usepackage{adjustbox}
\usepackage{flafter}
\usepackage{setspace}
\usepackage{etoolbox}
\usepackage{amssymb}
\usepackage{mathrsfs}
\usepackage{multirow}
\usepackage{enumerate}
\usepackage{hyphenat}
\usepackage{threeparttable}
\usepackage{footmisc}
\usepackage[colorlinks,linkcolor=blue,anchorcolor=blue,citecolor=blue,urlcolor=blue]{hyperref}
\usepackage{color}
\usepackage{babel}
\usepackage{multirow,booktabs}
\usepackage{tabularx}
\usepackage{array}
\usepackage{dcolumn}
\usepackage{ulem}
\usepackage{lipsum} % For blind text
\usepackage{chngcntr} % Required for counter manipulation

\AtBeginEnvironment{thebibliography}{%
  \setstretch{1.1}% was 1; a little more breathing room
  \setlength{\parsep}{0pt}%
  \setlength{\parskip}{0pt}%
}

\allowdisplaybreaks[4]

\counterwithin{table}{section}

\bibpunct{(}{)}{;}{a}{,}{,}
\normalfont
\DeclareFontShape{T1}{lmr}{bx}{sc} { <-> ssub * cmr/bx/sc }{}

\RequirePackage{lscape}
\DeclareMathAlphabet{\mathscrbf}{OMS}{mdugm}{b}{n}
\newtheorem{theorem}{Theorem}

\newtheorem{assumption}{Assumption}

\newtheorem{proposition}{Proposition}
\newtheorem{remark}{Remark}

\DeclareMathOperator*{\plim}{plim}

\numberwithin{equation}{section}
\numberwithin{theorem}{section}
\numberwithin{corollary}{section}
\numberwithin{proposition}{section}
\numberwithin{remark}{section}
\makeatletter

\allowdisplaybreaks[4]
\makeatother
\newcolumntype{Y}{>{\centering\arraybackslash}X}
\makeatother
\usepackage{fancyhdr}
\newcommand{\footscriptsize}{\fontsize{8pt}{9pt}\selectfont}

\newcommand{\blind}{0}

\newcommand{\xmark}{$\times$}

\begin{document}

\def\spacingset#1{\renewcommand{\baselinestretch}%
{#1}\small\normalsize} \spacingset{1}

%%%%%%%%%%%%%%%%%%%%%%%%%%%%%%%%%%%%%%%%%%%%%%%%%%%%%%%%%%%%%%%%%%%%%%%%%%%%%%

\if0\blind
{
  \title{\bf IV Estimation of Heterogeneous Spatial Dynamic Panel Models with Interactive Effects\thanks{
    We would like to thank \'Aureo de Paula, Paul Elhorst, Ayden Higgins, Arturas Juodis and seminar participants at Birmingham, BI Norwegian Business School, Exeter, Kent, King's College London, Michigan State  University, University of Southampton, as well as attendees of the 2023 AMEF, 2024 IAAE, and 2025 CFE conferences for their valuable comments and discussions. Jia Chen's work was supported by Grant SRG2024-00022-FSS and APAEM Seed Grant
from the University of Macau. Takashi Yamagata's work was partially supported by JSPS KAKENHI (grant numbers 25K00625, 25K05036 and 25H00544).}}
  \author{Jia Chen\hspace{.2cm}\\
   Department of Economics \& APAEM, University of Macau\\
     \\
    Guowei Cui \\
    Department of Economics, Huazhong University of Science and Technology\\
     \\
    Vasilis Sarafidis \\
    Department of Economics, Finance and Accounting, Brunel University London \\
    \\
    Takashi Yamagata \\
    Department of Economics and Related Studies, University of York \& \\
        Graduate School of Economics and Management, Tohoku University
    }
  \maketitle
} \fi

\if1\blind
{
  \bigskip
  \bigskip
  \bigskip
  \begin{center}
    {\LARGE\bf IV Estimation of Heterogeneous Spatial Dynamic Panel Models with Interactive Effects}
\end{center}
  \medskip
} \fi

\bigskip
\begin{abstract}
This paper develops a Mean Group Instrumental Variables (MGIV) approach for spatial dynamic panel data models with interactive effects, under large $N$ and $T$ asymptotics. Unlike existing approaches that typically impose slope-parameter homogeneity, MGIV accommodates cross-sectional heterogeneity in slope coefficients. The proposed estimator is linear, making it computationally efficient and robust. Furthermore, it enables asymptotically valid inferences without requiring bias correction. The Monte Carlo experiments indicate strong finite-sample performance of the MGIV approach across various sample sizes and parameter configurations. The practical utility of the methodology is illustrated through an application to regional economic growth in Europe. Our results provide evidence of conditional convergence in regional growth dynamics. Spillovers are shown to play a dominant role, particularly for investment rates where nearly four-fifths of the total effect on GDP per capita growth originates from neighboring regions.
\end{abstract}

\noindent%
\vfill

\newpage
\spacingset{1.8} % DON'T change the spacing!

\section{Introduction}\label{sec-intro}

Economic outcomes are often shaped by three distinct but interconnected mechanisms: temporal dynamics, spatial interactions, and exposure to aggregate shocks. Temporal dynamics arise from persistence in behaviour, such as adjustment costs or habit formation, whereby past outcomes influence current decisions (e.g., \citealt{Hamermesh1995}, \citealt{JappelliPistaferri2017}). Spatial interactions reflect the influence of peers, networks, and spillover mechanisms (e.g., \citealt{Case1991}, \citealt{Manski1993}, \citealt{BramulleEtal2009}), where the behavior of one entity is affected by others. These effects are further complicated by aggregate, pervasive shocks, such as technological change, global market fluctuations, or policy interventions, whose impact may differ across individual entities and are commonly captured through interactive fixed effects (e.g., \citealt{SarafidisWansbeek2021}).

In many empirical applications, these three mechanisms operate simultaneously, and their effects may differ substantially across cross-sectional entities. For instance, regional economic growth depends on its own past performance, is influenced by neighbouring regions through investment and labour mobility, and is affected by aggregate shocks such as business cycles or EU-wide policy changes. Importantly, the strength of these effects is unlikely to be homogeneous across regions, reflecting differences in economic structure, geography, or institutional environments. These considerations motivate the need for econometric models that jointly accommodate temporal dynamics, spatial dependence, interactive effects, and heterogeneous responses.

 In the context of large-$T$ panels, spatial dynamic models with additive fixed effects were developed by \citet{YuEtAl2008}, \citet{Korniotis2010}, \citet{LeeYu2014}, among others, where spatial interdependence is captured through a pre-specified $N \times N$ adjacency matrix $\mathbf{W}$. Such additive error-components specifications are nested within models with interactive effects (e.g., \citet{AnselinGalloJayet2008}), and therefore impose a more restrictive form of unobserved heterogeneity.\footnote{Comprehensive overviews of spatial panel data models with additive effects can be found in \citet{Elhorst2014} and \citet{LeeYu2015}}

Recent advances that incorporate interactive effects into spatial panel models include \citet{ShiLee2017}, \citet{BaiLi2021}, \citet{CuiSarafidisYamagata2023}, \citet{AquaroEtal2021}, \citet{ChenEtAl2022}, \citet{HigginsMartellosio2023}, and \citet{BaltagiShu2025}.\footnote{In \citet{HigginsMartellosio2023}, the adjacency matrix $\mathbf{W}$ is only partially observed, while \citet{BaltagiShu2025} allow for endogenous network formation. These settings address a different source of complexity compared to the present framework, which maintains $\mathbf{W}$ as observed and focuses on slope heterogeneity.} To clarify how the present paper relates to this literature, Table~\ref{tab:literature_comparison} summarises key contributions along three key modelling dimensions: slope heterogeneity, temporal dynamics, and interactive effects.
The table reveals a clear pattern. Most existing contributions impose slope homogeneity across cross-sectional units. This restriction can be unduly strong in many applications. When slope coefficients are heterogeneous, dynamic pooled estimators generally fail to consistently estimate population-average effects, even for large $T$; see \citet{RobertsonSymons1992} and \citet{PesaranSmith1995}. At the same time, the few contributions that allow for heterogeneity do so under important limitations. \citet{AquaroEtal2021} assume a purely idiosyncratic error structure, abstracting from additive or interactive effects. On the other hand, \citet{ChenEtAl2022} focus on static panels without temporal dynamics and rely on a CCE-type approach (\citealt{Pesaran2006}) that imposes a rank condition on factor loadings.\footnote{Extending the approach of \citet{ChenEtAl2022} to dynamic models is non-trivial, as it requires the construction of factor proxies from suitable cross-sectional averages of the dependent variable, the number of which grows with $T$. As noted on page 56 of their paper, the finite-sample bias arising from the correlation between cross-sectional averages and idiosyncratic errors is likely to be exacerbated in the presence of spatial lags. This issue becomes especially pronounced when the spatial weighting matrix remains relatively dense as the sample size increases, which often occurs when $\mathbf{W}$ is specified based on geographic proximity.}

\begin{table}[htbp]
\centering
\caption{Comparison with spatial literature for large-$T$ panels}
\label{tab:literature_comparison}
\begin{threeparttable}
\small
\begin{tabular}{p{3.6cm}ccccc}
\hline
Paper
& \makecell{Slope\\heterogeneity}
& Dynamics
& \makecell{Interactive\\effects}
& Method
& \makecell{Treatment of unobserved\\heterogeneity} \\
\hline
\citet{AquaroEtal2021}
  & \checkmark & \xmark & \xmark & QMLE & \makecell{Idiosyncratic\\errors only}\\

\citet{BaiLi2021}
  & \xmark & \checkmark & \checkmark & QMLE & \makecell{Joint QMLE of\\interactive effects}\\

\citet{BaltagiShu2025}
  & \xmark & \checkmark & \checkmark & QMLE & \makecell{Joint QMLE of\\interactive effects}\\

\citet{ChenEtAl2022}
  & \checkmark & \xmark & \checkmark & IV & CCE \\

\citet{CuiSarafidisYamagata2023}
  & \xmark & \checkmark & \checkmark & IV & PCA \\

\citet{HigginsMartellosio2023}
  & \xmark & \checkmark & \checkmark & QMLE & \makecell{PCA within\\penalised QMLE}\\

\citet{Korniotis2010}
  & \xmark & \checkmark & \xmark & GMM & \makecell{Additive FE}\\

\citet{LeeYu2014}
  & \xmark & \checkmark & \xmark & GMM & \makecell{Additive FE}\\

\citet{ShiLee2017}
  & \xmark & \checkmark & \checkmark & QMLE & \makecell{Joint QMLE of\\interactive effects}\\

\citet{YuEtAl2008}
  & \xmark & \checkmark & \xmark & QMLE & \makecell{Additive FE}\\

\textbf{This paper}
  & \checkmark & \checkmark & \checkmark & IV & PCA \\
\hline
\end{tabular}

\begin{tablenotes}
\footnotesize
\item \textit{Notes:} The table focuses on large-$T$ panel data
frameworks and highlights the main dimensions along which the present
approach differs from the existing literature. Contributions developed
under fixed-$T$ asymptotics are not reported.
\end{tablenotes}
\end{threeparttable}
\end{table}

The present paper contributes to this literature by developing an IV framework under large $N,T$ asymptotics that accommodates heterogeneous slopes, temporal dynamics, and interactive effects. Specifically, we propose a Mean Group Instrumental Variables (MGIV) estimator, extending the framework of \citet{NorkuteEtal2021} to allow for spatial interactions and the identification of heterogeneous spillover effects. The presence of both spatial and temporal lags introduces additional sources of endogeneity relative to \citet{NorkuteEtal2021} and \citet{ChenEtAl2022}, rendering the asymptotic analysis considerably more involved.

Our approach constructs valid instruments by projecting out common factors from exogenous covariates using Principal Components Analysis (PCA), following \citet{Bai2003}. This yields consistent individual-specific IV estimators, which are then averaged to obtain the MGIV estimator. The resulting estimator is $\sqrt{N}$-consistent and asymptotically normally distributed for the population mean effects as $N,T \to \infty$, and is asymptotically unbiased, so that standard inference procedures apply without the need for bias correction. From a practical perspective, MGIV is linear and computationally efficient, making it well suited for empirical applications with large datasets.

Allowing for individual-specific parameters in a spatial dynamic setting represents an important departure from conventional homogeneous specifications. As noted by \citet{LesageChih2016}: ``\textit{space-time panel data samples covering longer time spans allow us to produce parameter estimates for all $N$ spatial units, an exciting point of departure for future work. Allowing for heterogeneous coefficients for each spatial unit holds a natural appeal when contrasted with conventional static spatial panel models.}''

We illustrate the practical relevance of our method by estimating a regional spatial growth model, focusing on the magnitude of regional growth spillovers in Europe. Differences in industrial structure, urbanization, and geography give rise to heterogeneous responses across regions. For instance, growth drivers in industrial Bavaria (Germany) likely differ from agricultural Thessaly (Greece), as do spillover effects in urban Greater London and rural Lapland (Finland). Our results provide evidence of conditional convergence in regional growth dynamics. Spillovers play a dominant role, particularly for investment rates, where roughly four-fifths of the total impact on GDP per capita growth is attributable to neighboring regions, emphasizing the importance of inter-regional investment linkages. Human capital and R\&D spillovers further reinforce the role of knowledge diffusion and innovation in fostering growth across regions.

Throughout, for an $m\times m$ matrix $\mathbf{A}=(a_{ij})_{1\leq i,j\leq m}$, we denote its trace by $\mathrm{tr} (\mathbf{A})=\sum_{i=1}^m a_{ii}$. For an $m\times n$ matrix $\mathbf{B}=(b_{ij})_{1\leq i\leq m,1\leq j\leq n}$, denote its column sum norm by $\|\mathbf{B}\|_1=\max_{1\le j\le n}\sum_{i=1}^m |b_{ij}|$, its row sum norm by $\|\mathbf{B}\|_\infty=\max_{1\le i\le m}\sum_{j=1}^n|b_{ij}|$, and its Frobenius norm by $\|\mathbf{B}\|=\sqrt{\mathrm{tr} (\mathbf{B}^{\prime }\mathbf{B})}$. Also define $\mathbf{P}_{\mathbf{B}}=\mathbf{%
B}(\mathbf{B}^{\prime }\mathbf{B})^{-1}\mathbf{B}^\prime$ and $\mathbf{M}_{\mathbf{B}}=\mathbf{I}_m-\mathbf{P}_{\mathbf{B}}$, where $\mathbf{I}_m$ is the $m\times m$ identity matrix. Denote by $C$ a generic positive constant whose value may not be the same at each appearance, and $\delta_{N\!T}^2=\min\{N,T\}$. We use $N,T \rightarrow \infty$ to denote that $N$ and $T$ pass to infinity jointly, and $\plim$ to denote the probability limit.

\section{Model and Estimation Approach}\label{sec-method}

We study the following spatial dynamic panel data model:
\begin{equation}  \label{spm_scalar}
\begin{split}
y_{it}&=\psi_i \sum_{j=1}^Nw_{ij}y_{jt}+\rho_i y_{i,t-1}+\mathbf{x}%
_{it}^{\prime }\boldsymbol{\beta}_i+u_{it};
 \\
u_{it}&=\boldsymbol{\lambda}_i^{\prime }\mathbf{f}_t+\boldsymbol{\phi}_i^{\prime }\boldsymbol{g}_t+\varepsilon_{it}, \quad i=1,2,\ldots, N,\  t=1,2,\ldots, T.
\end{split}%
\end{equation}
In the above, $y_{it}$ denotes the outcome of interest for individual unit $i$ at time $t$, while the $k\times 1$ vector $\mathbf{x}_{it}=\left(x_{1it},\dots, x_{kit}\right)^{\prime}$ contains observed idiosyncratic characteristics. $\sum_{j=1}^N w_{ij}y_{jt}$ is the so-called ``spatial-lag'', a weighted sum of neighbours' outcomes, where $w_{ij}$ denotes the weight assigned to neighbour $j$ in relation to individual $i$. These weights capture the connectedness structure among individuals and are specified within the $N \times N$ adjacency matrix $\mathbf{W}=[w_{ij}]$, where $w_{ii}=0$ for all $i$, such that no individual is treated as its own neighbour; further regularity conditions on $\mathbf{W}$ are provided in Section~\ref{subsec-assump}.
In line with standard practice in the spatial econometrics literature, the adjacency matrix $\mathbf{W}$ is treated as fixed and known. This corresponds to a setting in which the network structure captures predetermined linkages, such as geographic proximity or institutional connections, and is treated as exogenous to the outcome equation. Following \citet{BramulleEtal2009}, this amounts to maintaining that the network is formed independently of the unobserved components affecting outcomes, and is not jointly determined with $y_{it}$.

The associated slope coefficients of the model ($\rho_{i}$, $\psi_{i}$, and $\boldsymbol{\beta}_{i}$) are allowed to vary across cross-sectional entities, capturing heterogeneous responses in persistence, spillovers and direct effects (e.g., \citealt{KelejianPiras2017}, \citealt{JingEtAl2018}, \citealt{ArbiaEtal2010} and \citealt{ArbiaEtal2021}).\footnote{The main results of this paper naturally extend to models incorporating ``contextual effects'' (e.g., \citealt{Manski1993}), also referred to as the spatial Durbin model (\citealt{Elhorst2014}). This is further discussed in Remark \ref{remark:Durbin_model}.} Slope heterogeneity is modelled through a random-coefficient framework, whereby these coefficients are viewed as draws from a common population distribution. Conditional on their realisations, however, they are treated as fixed (albeit unknown) for each unit. In this sense, the randomness pertains to the cross-sectional distribution of heterogeneity, rather than to the interpretation of individual-specific parameters.

The remaining unobserved components are treated as stochastic. Specifically, $\mathbf{f}_t$ and $\boldsymbol{g}_t$ denote $r_1\times 1$ and $r_2\times1$ vectors of unobserved common factors that vary over time and capture aggregate shocks affecting all entities, with individual-specific loadings $\boldsymbol{\lambda}_i$ and $\boldsymbol{\phi}_i$ reflecting heterogeneous sensitivities. Lastly, $\varepsilon_{it}$ is a purely idiosyncratic error.

The model in \eqref{spm_scalar} is relevant in a range of empirical settings where temporal persistence, spatial interactions, and common shocks are all expected to matter simultaneously. For example, in regional economic growth, current output may depend on past performance, be influenced by neighbouring regions through trade, investment, or labour mobility, and respond to economy-wide shocks such as technological change, policy interventions, or global market fluctuations (e.g., \citealt{Bai2009}, \citealt{SarafidisWansbeek2012}). Similar considerations arise in housing markets and financial networks, where outcomes may display persistence over time, spillovers across interconnected entities, and heterogeneous responses to common shocks. In such settings, it is natural to allow the strength of persistence, direct effects, and spillovers to vary across entities, reflecting differences in labour market conditions, housing supply constraints, demographic composition, and exposure to migration and trade networks.

There are three primary sources of endogeneity in model \eqref{spm_scalar}. First, $\sum_{j=1}^N w_{ij}y_{jt}$ is endogenous by construction. This term essentially represents the formal specification of an equilibrium outcome of a spatial interaction process, wherein the value of the dependent variable for one individual unit is simultaneously determined alongside that of its neighbours. This reciprocal interdependence underscores the networked nature of these relationships, as highlighted by \citet{Elhorst2021}.
Second, the lagged dependent variable, $y_{i,t-1}$ is also endogenous due to the presence of incidental parameters (e.g., \citealt{Nickell1981} and \citealt{PhillipsSull2007}).
Third, endogeneity also arises from the potential dependence of the covariates on the latent factors. To account for this, and consistent with the frameworks of \citet{Pesaran2006}, \citet{BaiNg2006}, and \citet{NorkuteEtal2021} among many others, we assume
\begin{equation}  \label{spm_X_scalar}
\mathbf{x}_{it}=\boldsymbol{\Gamma}_{i}^{\prime }\mathbf{f}_t+\mathbf{v}%
_{it}.
\end{equation}
This specification reflects the fact that many observed covariates are themselves shaped by common aggregate forces, such as macroeconomic conditions, technological change, or policy shifts, and therefore inherit a factor structure analogous to that of the dependent variable. The associated loadings capture heterogeneous sensitivities of individual units to these common influences. That is, $\mathbf{x}_{it}$ is influenced by $\mathbf{f}_{t}$, with the associated loadings represented by $\boldsymbol{\Gamma}_{i}$, an $r_1 \times k$ matrix. Importantly, for the sake of generality, we allow some of the latent factors governing $y_{it}$ and $\mathbf{x}_{it}$ to differ. This distinction justifies the inclusion of the additional term $\boldsymbol{\phi}_i^{\prime }\boldsymbol{g}_t$ in model \eqref{spm_scalar}.\footnote{Alternatively, one may assume that the additional term $\boldsymbol{\phi}_i^{\prime }\boldsymbol{g}_t$ enters the process for $\mathbf{x}_{it}$ rather than $u_{it}$. This case is conceptually simpler, since all factors directly affecting $y_{it}$ would then be spanned by $\mathbf{x}_{it}$.} The loadings $\boldsymbol{\Gamma}_{i}$ are permitted to exhibit correlation with both $\boldsymbol{\lambda}_i$ and $\boldsymbol{\phi}_i$, further accommodating potential interdependencies that may arise in the components of the model. Finally, $\mathbf{v}_{it}$ denotes the idiosyncratic error component for $\mathbf{x}_{it}$.

Stacking the equations in (\ref{spm_scalar}) over $t$ yields
\begin{equation}  \label{pm_vectori}
\begin{split}
\mathbf{y}_{i}&=\psi_i \sum_{j=1}^{N} w_{ij} \mathbf{y}_{j}+\rho_i \mathbf{y}_{i,-1} +%
\mathbf{X}_{i}\boldsymbol{\beta}_i+\mathbf{F}\boldsymbol{\lambda}_i+\mathbf{G%
}\boldsymbol{\phi}_i+\boldsymbol{\varepsilon}_{i}, \\
\mathbf{X}_{i}&=\mathbf{F}\boldsymbol{\Gamma}_{i}+\mathbf{V}_{i},\ \ \ i=1,\ldots,N,
\end{split}%
\end{equation}
where  $\mathbf{y}_{i}=(y_{i1},\ldots,y_{iT})^{\prime }$, $\mathbf{y}_{j}=(y_{j1},\ldots,y_{jT})^{\prime}$, $\mathbf{y}%
_{i,-1}=(y_{i0},\ldots,y_{i,T-1})^{\prime }$, $\mathbf{X}_{i}=(\mathbf{x}_{i1},\cdots,%
\mathbf{x}_{iT})^{\prime }$, $\mathbf{F}=(\mathbf{f}_1,\cdots,\mathbf{f}%
_T)^{\prime }$, $\mathbf{G}=(\boldsymbol{g}_1,\cdots,\boldsymbol{g}%
_T)^{\prime }$, $\boldsymbol{\varepsilon}_{i}=(\varepsilon_{i1},\cdots,%
\varepsilon_{iT})^{\prime }$ and $\mathbf{V}_{i}=(\mathbf{v}_{i1},\ldots,%
\mathbf{v}_{iT})^{\prime }$.

Define $\boldsymbol{\theta}
_i=(\psi_i,\rho_i,\boldsymbol{\beta}_i^{\prime })^{\prime }$, $\mathbf{C}
_{i}=(\sum_{j=1}^{N} w_{ij} \mathbf{y}_{j},\mathbf{y}_{i,-1},\mathbf{X}_{i})$, and $\mathbf{u}_{i}=\mathbf{F}\boldsymbol{\lambda}_i+\mathbf{G}\boldsymbol{%
\phi}_i+\boldsymbol{\varepsilon}_{i}$.  Then, the first equation in (\ref{pm_vectori}) can be reformulated as
\begin{equation}
\mathbf{y}_{i}=\mathbf{C}_{i}\boldsymbol{\theta}_i+\mathbf{u}_{i}.
\end{equation}

We use the method of Instrumental Variables to estimate $\boldsymbol{\theta}_{i}$. To this end, define the ``defactoring'' matrices that project out $\mathbf{F}$ and  $\mathbf{F}_{-1}=(\mathbf{f}_0,\ldots,\mathbf{f}_{T-1})^\prime$ as $\mathbf{M}_{{\mathbf{F}}}=\mathbf{I}_T - \mathbf{%
F}(\mathbf{F}^{\prime }\mathbf{F})^{-1}\mathbf{F}^\prime$ and $\mathbf{M}_{{\mathbf{F}_{-1}}}=\mathbf{I}_T - \mathbf{%
F}_{-1}(\mathbf{F}_{-1}^{\prime }\mathbf{F}_{-1})^{-1}\mathbf{F}_{-1}^\prime$. Letting $\mathbf{X}_{i,-1}=(\mathbf{x}_{i0},\ldots,\mathbf{x}_{i,T-1})^\prime$, further define
\begin{equation}\label{eq:Z_matrix}
{\mathbf{Z}}_{i}=\left(\sum_{j=1}^Nw_{ij}\mathbf{M}_{{\mathbf{F}}}\mathbf{X%
}_{j},\ \mathbf{M}_{{\mathbf{F}}}\mathbf{M}_{{\mathbf{F}}_{-1}}\mathbf{X}_{i,-1},\ \mathbf{M}_{%
{\mathbf{F}}}\mathbf{X}_{i}\right),
\end{equation}
whose elements are instruments for the elements of $\mathbf{C}_i$.\footnote{Loosely speaking, the term $\sum_{j=1}^Nw_{ij}\mathbf{M}_{\mathbf{F}}\mathbf{X}_{j}$ instruments $\sum_{j=1}^{N} w_{ij} \mathbf{y}_{j}$, the term $\mathbf{M}_{\mathbf{F}} \mathbf{M}_{\mathbf{{F}}_{-1}}\mathbf{X}_{i,-1}$ instruments $\mathbf{y}_{i,-1}$, and ${\mathbf{M}_{\mathbf{{F}}}}\mathbf{X}_{i}$ instruments $\mathbf{X}_{i}$.} It is straightforward to see that, due to the de-factorisation, we have
\begin{equation}\label{eq-IV}
\mathbf{Z}_{i}=\left(\sum_{j=1}^Nw_{ij}\mathbf{M}_{{\mathbf{F}}}\mathbf{V
}_{j},\ \mathbf{M}_{{\mathbf{F}}}\mathbf{M}_{{\mathbf{F}}_{-1}}\mathbf{V}_{i,-1},\ \mathbf{M}_{%
{\mathbf{F}}}\mathbf{V}_{i}\right) \ \ \ {\rm and}\ \ \ \mathbf{Z}_{i}^\prime \mathbf{u}_i=\mathbf{Z}_{i}^\prime \big(\mathbf{G}\boldsymbol{%
\phi}_i+\boldsymbol{\varepsilon}_{i}\big),
\end{equation}
where $\mathbf{V}_{i,-1}=(\mathbf{v}_{i0},\ldots,\mathbf{v}_{i,T-1})^\prime$. Under the assumption that $\{\mathbf{v}_{is}, 1\leq i\leq N, 0\leq s\leq T\}$ is uncorrelated with $\{\mathbf{f}_s, 0\leq s\leq T\}$, $\{\mathbf{g}_s, 1\leq s\leq T\}$, $\{\boldsymbol{\phi}_i,1\leq i\leq N\}$, and $\{\varepsilon_{is},1\leq i\leq N, 1\leq s\leq T\}$,  %$\{\mathbf{f}_s, 0\leq s\leq T\}$ and $\{\mathbf{v}_{is}, 1\leq i\leq N, 0\leq s\leq T\}$ are uncorrelated with $\{\mathbf{g}_s, 1\leq s\leq T\}$ and $\{\varepsilon_{is},1\leq i\leq N, 1\leq s\leq T\}$,
$\mathbf{Z}_{i}$ is exogenous and therefore provides valid instruments for $\mathbf{C}_i$. Instrument relevance follows from the presence of the idiosyncratic component $\mathbf{v}_{it}$ in $\mathbf{x}_{it}$, which ensures that the defactored instruments remain informative about the endogenous regressors after projection onto the orthogonal complement of the factor space. This mechanism is explicitly explored in the Monte Carlo analysis, where we consider different configurations governing the relative importance of idiosyncratic and factor-driven variation (e.g., $\pi_u \in \{1/4, 3/4\}$).

\begin{remark}\label{remark:z_choice}
An alternative IV strategy would be to use $\mathbf{M}_{\mathbf{F}_{-1}}\mathbf{X}_{i,-1}$ as instruments, which are valid under the maintained assumptions. However, the formulation in \eqref{eq:Z_matrix}--\eqref{eq-IV} effectively premultiplies the model by $\mathbf{M}_{\mathbf{F}}$, removing the factor component $\mathbf{F}$ from the regression error. Consequently, the moment conditions depend only on $\mathbf{G}\boldsymbol{\phi}_i+\boldsymbol{\varepsilon}_i$, leading to more efficient estimation.
\end{remark}

Since $\mathbf{F}$ and $\mathbf{F}_{-1}$ are not observed, we estimate them by applying PCA on the covariance matrices of $\mathbf{X}_i$ and $\mathbf{X}_{i,-1}$, $1\leq i\leq N$. For identification purposes, we assume $T^{-1}\mathbf{F}^\prime\mathbf{F}= \mathbf{I}_{r_{1}}$ and $T^{-1}\mathbf{F}_{-1}^\prime\mathbf{F}_{-1}= \mathbf{I}_{r_{1}}$. Then, the estimates $\widehat{\mathbf{F}}$ and $\widehat{\mathbf{F}}_{-1}$ are obtained as $\sqrt{T}$ times the eigenvectors corresponding to the $r_1$ largest eigenvalues of the $T \times T$ matrices $(NT)^{-1}\sum_{i=1}^N \mathbf{X}_{i}\mathbf{X}%
_{i}^{\prime }$ and $(NT)^{-1}\sum_{i=1}^N \mathbf{X}_{i,-1}\mathbf{X}_{i,-1}^{\prime }$, respectively.\footnote{For simplicity and without loss of generality, $r_1$ is treated as known. However, in practice it can be consistently estimated using established methods in the literature, such as the information criterion approach of \citet{BaiNg2002} or the eigenvalue methods of \citet{AhnHorenstein2013}. In the Monte Carlo analysis, we adopt the latter and estimate the number of factors in each replication.} Note that, since the factors are extracted from observed covariates, no estimation error arises in $\widehat{\mathbf{F}}$ or $\widehat{\mathbf{F}}_{-1}$ that is associated with the estimation of any of the slope coefficients in (\ref{spm_scalar}). Furthermore, the factor loadings
$\boldsymbol{\Gamma}_i$ can be estimated as $\widehat{\boldsymbol{\Gamma}}_i=T^{-1}\widehat{%
\mathbf{F}}^{\prime }\mathbf{X}_{i}$.

Subsequently, we construct feasible instruments for $\mathbf{C}_{i}$ as
\begin{equation}\label{eq:Z_hat_matrix}
\widehat{\mathbf{Z}}_{i}=\left(\sum_{j=1}^Nw_{ij}\mathbf{M}_{\widehat{\mathbf{F}}}\mathbf{X%
}_{j},\ \mathbf{M}_{\widehat{\mathbf{F}}}\mathbf{M}_{\widehat{\mathbf{F}}_{-1}}\mathbf{X}_{i,-1},\ \mathbf{M}_{\widehat{\mathbf{F}}}\mathbf{X}_{i}\right),
\end{equation}
and the resulting individual-specific IV estimator of $\boldsymbol{\theta}_i$ is given by
\begin{equation}\label{individual-est}
\widehat{\boldsymbol{\theta}}_i=\left(\widehat{\mathbf{A}}_i^{\prime }%
\widehat{\mathbf{B}}_i^{-1}\widehat{\mathbf{A}}_i\right)^{-1}\widehat{%
\mathbf{A}}_i^{\prime }\widehat{\mathbf{B}}_i^{-1}\widehat{\mathbf{c}}_{y,i},
\end{equation}
where
\begin{equation}\label{notation1}
\widehat{\mathbf{A}}_i=T^{-1}\widehat{\mathbf{Z}}_{i}^{\prime }
\mathbf{C}_{i},\ \ \ \ \widehat{\mathbf{B}}_i=T^{-1} \widehat{\mathbf{Z}}%
_{i}^{\prime }\widehat{\mathbf{Z}}_{i},\ \ \ \ \widehat{\mathbf{c}}_{y,i}=T^{-1}
\widehat{\mathbf{Z}}_{i}^{\prime }\mathbf{y}_{i}.
\end{equation}

\begin{remark}\label{remark:defactoring_u}
Alternatively, a two-stage individual-specific IV procedure can be considered. After obtaining the individual estimates $\widehat{\boldsymbol{\theta}}_{i}$, $1\leq i\leq N$, the residuals $\widehat{\mathbf{u}}_{i}=\mathbf{y}_{i}-\mathbf{C}_{i}%
\widehat{\boldsymbol{\theta}}_{i}$ are used to estimate the latent factors in $u_{it}$. This is achieved by extracting the eigenvectors associated with the largest $r_1+r_2$ eigenvalues of the $T \times T$ matrix $(NT)^{-1}\sum_{i=1}^N \widehat{\mathbf{u}}_{i}\widehat{\mathbf{u}}_{i}^{\prime }$. However, as each $\widehat{\boldsymbol{\theta}}_{i}$ is obtained from an individual-specific time series IV regression, $\widehat{\boldsymbol{\theta}}_{i}$ is $\sqrt{T}$-consistent rather than $\sqrt{NT}$-consistent. This implies that the estimation of the full set of factors, including $\mathbf{g}_t$, may become highly inefficient. Consequently, this estimator is not pursued further here. Note that the estimator of $\mathbf{F}$, described earlier, does not face this issue, as it can be directly obtained from the raw data $\left\{\mathbf{X}_{i}\right\}_{i=1}^{N}$.
\end{remark}

\begin{remark}\label{remark:Durbin_model}
Under a similar set of assumptions as in Section \ref{subsec-assump}, the model in Eq. \eqref{pm_vectori} can be extended to include additional spatio-temporal lags, such as $\sum_{j=1}^{N} w_{ij} \mathbf{y}_{j,-1}$, or spatial lags of the covariates, $\sum_{j=1}^{N} w_{ij} \mathbf{X}_{j}$. The spatial-time lag $\sum_{j=1}^{N} w_{ij} \mathbf{y}_{j,-1}$ can be instrumented by $\sum_{j=1}^Nw_{ij}\mathbf{M}_{{\mathbf{F}}} \mathbf{M}_{{\mathbf{F}_{-1}}} \mathbf{X}_{j,-1}$. For spatially lagged covariates, valid instruments can be constructed using higher orders of spatial and/or time lags of defactored $\mathbf{X}_i$, such as  $\sum_{j=1}^Nw_{ij}\mathbf{M}_{{\mathbf{F}}}\mathbf{M}_{{\mathbf{F}_{-2}}} \mathbf{X}_{j,-2}$ or $\sum_{j=1}^N \big(\sum_{k=1}^N w_{ik}w_{kj}\big)\mathbf{M}_{{\mathbf{F}}} \mathbf{M}_{{\mathbf{F}_{-1}}} \mathbf{X}_{j,-1}$\footnote{Note that $\big(\sum_{k=1}^N w_{ik}w_{kj}\big)$ is the $(i,j)$th element of $\mathbf{W}^2$, where $\mathbf{W}$ is the spatial weighting matrix.}. Given the flexibility in the choice of instruments, the Monte Carlo analysis in Section \ref{sec-MC} evaluates an estimator that constructs optimal instruments using the Lasso-based procedure of \citet{BelloniChenChernozhukovHansen2012}.
\end{remark}

The individual-specific IV estimator $\widehat{\boldsymbol{\theta}}_i$ in Eq. \eqref{individual-est} provides estimates of individual-level effects. In some applications, we may also be interested in the mean effects, which can be estimated by the Mean Group IV (MGIV) estimator, defined as
\begin{equation}\label{eq:MGest}
\widehat{\boldsymbol\theta}_{MG}=\frac{1}{N}\sum\limits_{i=1}^N\widehat{\boldsymbol\theta}_i.
\end{equation}
Under the random-coefficient model, $\boldsymbol{\theta}_i=\boldsymbol{\theta}+%
\mathbf{e}_i$, where $\boldsymbol{\theta}=\mathbb{E}(\boldsymbol{\theta}_i)$ and $\mathbf{e}_i$ is a random error, we establish the consistency of $\widehat{\boldsymbol\theta}_{MG}$ for $\boldsymbol{\theta}$ in Section \ref{subsec-theory}. Alternatively, we may consider the pooled estimator, defined as
\begin{equation}\label{pool-est}
\widehat{\boldsymbol{\theta}}_P=\left(\widehat{\mathbf{A}}^{\prime }%
\widehat{\mathbf{B}}^{-1}\widehat{\mathbf{A}}\right)^{-1}\widehat{%
\mathbf{A}}^{\prime }\widehat{\mathbf{B}}^{-1}\widehat{\mathbf{c}}_{y},
\end{equation}
where $\widehat{\mathbf{A}}=\frac{1}{N}\sum_{i=1}^N\widehat{\mathbf{A}}_i$, $\widehat{\mathbf{B}}=\frac{1}{N}\sum_{i=1}^N\widehat{\mathbf{B}}_i$, and $\widehat{\mathbf{c}}_{y}=\frac{1}{N}\sum_{i=1}^N\widehat{\mathbf{c}}_{y,i}$. However, under the random-coefficient model for heterogeneity, the pooled estimator is inconsistent for $\boldsymbol{\theta}$ (see \citealt{RobertsonSymons1992}, \citealt{PesaranSmith1995}). We provide a brief proof of the inconsistency in Section~S3 of the Online Supplement. On the other hand, when the coefficients are homogeneous, i.e., $\boldsymbol{\theta}_i\equiv\boldsymbol{\theta}$, the pooled estimator $\widehat{\boldsymbol{\theta}}_P$ is consistent (see \citealt{CuiSarafidisYamagata2023}). We refer the reader to \citet{CuiSarafidisYamagata2023} for details on estimation in the homogeneous case of our model and the properties of the corresponding IV estimators.

\section{Assumptions and Asymptotic Theory}\label{sec-theory}

In this section, we examine the limiting properties of the individual-specific IV estimator, $\hat{\boldsymbol{\theta}}_i$, and the Mean Group IV (MGIV) estimator, $\widehat{\boldsymbol\theta}_{MG}$. To facilitate the asymptotic analysis, we first introduce
some assumptions.

\subsection{Assumptions}\label{subsec-assump}
\begin{assumption}[idiosyncratic error in $\mathbf{y}$]
\label{assumption-idioiny} The idiosyncratic error $\varepsilon_{it}$ has zero mean and uniformly bounded fourth-order moments. Furthermore, for each $i$, $\{\varepsilon_{it}, t=1,2,\ldots\}$ is a stationary process with absolutely summable serial covariances. %linear process, $\varepsilon_{it}=\sum_{\ell=0}^\infty p_{i\ell}\zeta_{i,t-\ell}$, where $\zeta_{it}$ is i.i.d. over both $i$ and $t$ and has zero mean and finite fourth-order moments, and $p_{i\ell}$ are coefficients satisfying $\sup_{i}\sum_{\ell=0}^\infty \vert p_{i\ell}\vert\leq C$.

\end{assumption}

\begin{assumption}[idiosyncratic error in $\mathbf{x}$]
\label{assumption-idioinx}  The idiosyncratic error $\mathbf{v}_{it}$ is independent from $\varepsilon_{js}$ for all $i,j,s,t$, and has zero mean and uniformly bounded fourth-order moments. %$\mathbb{E}(\mathbf{v}_{it})=\mathbf{0}$ and $\sup_{i,t}
%\mathbb{E}\|\mathbf{v}_{it}\|^{8} \le C$
%{\color{red}[Is 8th order moment needed?]}.
For each $i$, $\{\mathbf{v}_{it}, t=1,2,\ldots\}$ is a stationary process with absolutely summable serial covariances. Denoting $\boldsymbol{\Sigma}_{ij,st}= \mathbb{E}%
\left(\mathbf{v}_{is}\mathbf{v}_{jt}^{\prime }\right)$, it holds that
\begin{enumerate}
%\item \label{assumption-idioinx1}

\item \label{assumption-idioinx2}  $N^{-1}T^{-1}\sum_{i=1}^N
\sum_{j=1}^N\sum_{s=1}^T\sum_{t=1}^T\|\boldsymbol{\Sigma}_{ij,st}\|\le C$; furthermore, for all $i,j,s,t$, there exist $\bar{\sigma}_{ij}$ and $%
\tilde{\sigma}_{st}$ such that $\|\boldsymbol{\Sigma}_{ij,st}\| \le \bar{\sigma}_{ij}$, $\left\|\boldsymbol{\Sigma}_{ij,st}\right\| \le \tilde{%
\sigma}_{st}$, $T^{-1}\sum_{s=1}^T\sum_{t=1}^T
\tilde{\sigma}_{st} \le C$, and
 $\sum_{j=1}^N\bar{\sigma}_{ij} \le C$;

%\item \label{assumption-idioinx3}  for all $s,t$, $\mathbb{E}%
%\left\|N^{-1/2}\sum_{i=1}^N (\mathbf{v}_{is}\mathbf{v}_{it}^{\prime }-\boldsymbol{%
%\Sigma}_{ii,st})\right\|^4 \le C$, and for all $i,j$, \\$\mathbb{E}%
%\left\|T^{-1/2}\sum_{t=1}^T (\mathbf{v}_{it}\mathbf{v}_{jt}^{\prime }-\boldsymbol{%
%\Sigma}_{ij,tt})\right\|^4 \le C$; {\color{red}[Is this needed?]}

\item \label{assumption-idioinx4}  %for all $j$ and $s$, $\mathbb{E}%
%\left\|N^{-1/2}T^{-1}\sum_{s_1=1}^T\sum_{i=1}^N \sum_{t=1}^T\mathbf{f}_{s_1}\mathbf{v}_{is_1}^\prime \big[\mathbf{v}_{it}%
%\mathbf{v}_{js}^{\prime }-\boldsymbol{\Sigma}_{ij,ts}\big]\right\|^2\le C$ and

%{\color{red}$\sum_{k_1=1}^N\sum_{k_2=1}^N\sum_{t_1=1}^T\sum_{t_2=1}^T\left\Vert\mathrm{Cov}\big(\mathbf{v}_{i,t_1-s_1}\mathbf{v}_{k_1,t_1}^\prime, \mathbf{v}_{j,t_2-s_2}\mathbf{v}_{k_2,t_2}^\prime\big)\right\Vert\leq CNT$ for all $i,j=1,\ldots,N$ and $s_1,s_2=0,1,\ldots$,}
%$\mathbb{E}%
%\left\|N^{-1/2}T^{-1/2}\sum_{i=1}^N \sum_{t=1}^T\mathbf{\Gamma}_i \big[\mathbf{v}_{it}%
%\mathbf{v}_{js}^{\prime }-\boldsymbol{\Sigma}_{ij,ts}\big]\right\|^2\le C$, and for all $s$,
%$$\mathbb{E}\left\|
%N^{-1/2}T^{-1/2}\sum_{i=1}^N\sum_{t=1}^T\mathbf{h}_t[\mathbf{v}_{is}^{\prime
%}\mathbf{v}_{it}-\mathbb{E}(\mathbf{v}_{is}^{\prime }\mathbf{v}_{it})]\right\|^2
%\le C,\ \ \ \mbox{where }\mathbf{h}_t=(\mathbf{f}_t^\prime, \mathbf{g}_t^\prime)^\prime;$$
for all $i,j_1,j_2$, $T^{-1}\left\Vert\sum_{t=1}^T\sum_{s=1}^T \mathbb{E}\left[\left(\mathbf{v}_{i t}\mathbf{v}_{j_1 t}^\prime-\boldsymbol{\Sigma}_{ij_1,tt}\right)\left(\mathbf{v}_{i s}\mathbf{v}_{j_2 s}^\prime-\boldsymbol{\Sigma}_{i j_2,ss}\right)^\prime\right]\right\Vert\leq C$;

\item \label{assumption-idioinx5}  $N^{-1}T^{-2}\sum_{i=1}^N
\sum_{j=1}^N\sum_{s_1=1}^T\sum_{s_2=1}^T\sum_{t_1=1}^T\sum_{t_2=1}^T|\mathrm{%
cov}(\mathbf{v}_{is_1}^{\prime }\mathbf{v}_{is_2} ,\mathbf{v}_{jt_1}^{\prime
}\mathbf{v}_{jt_2} )| \le C$.
\end{enumerate}

\end{assumption}

\begin{assumption}[factors]
\label{assumption-fac}  The factors satisfy that
\begin{enumerate}
\item \label{assumption-factor1}
$\mathbf{f}_t$ and $\mathbf{g}_t$ are independent from $\mathbf{v}_{is}$ and $\varepsilon_{is}$ for all $i,s,t$;

\item\label{assumption-factor2}  Both $\{\mathbf{f}_t,t=1,2,\ldots\}$ and $\{\mathbf{g}_t,t=1,2,\ldots\}$ are stationary processes with  $\mathbb{E}\|\mathbf{f}_t\|^{8} \le C$ and $\mathbb{E%
}\|\mathbf{g}_t\|^{8} \le C$. %There exist positive definite matrices %$\boldsymbol{\Sigma}
%_{F}$ and $\boldsymbol{\Sigma}_{G}$ such that  %$\plim\limits_{T\rightarrow\infty}T^{-1}\mathbf{F}^{\prime }\mathbf{F} =\boldsymbol{\Sigma}_{F}$ and $\plim\limits_{T\rightarrow\infty}T^{-1}\mathbf{G}^{\prime }%
%\mathbf{G} = \boldsymbol{\Sigma}_{G}$.
\end{enumerate}
\end{assumption}

\begin{assumption}[loadings]
\label{assumption-loadings}
The factor loadings satisfy
\begin{enumerate}
\item\label{assumption-loadings1} $\boldsymbol{\Gamma}_{i}$, $%
\boldsymbol{\lambda}_{i}$ and  $\boldsymbol{\phi}${$_{i}$} are independent
 from $\varepsilon_{jt}$,  $\mathbf{v}_{jt}$, $\mathbf{f}_t$, and $%
\mathbf{g}_t$ for all $i,j,t$; %{\color{red}[is independence from $\mathbf{f}_t$ and $\mathbf{g}_t$ necessary?]}

\item\label{assumption-loadings2}$\boldsymbol{\Gamma}_{i}\sim\mathrm{i.i.d}(\boldsymbol{\Gamma
},\boldsymbol{\Sigma}_{\Gamma})$,  $\boldsymbol{\lambda}_{i}\sim \mathrm{%
i.i.d}(\mathbf{0},\boldsymbol{\Sigma}_{\lambda})$,  $\boldsymbol{\phi}${$%
_{i}\sim \mathrm{i.i.d}(\mathbf{0},\boldsymbol{\Sigma}_{\phi})$}, where $\boldsymbol{\Gamma}$ is an $r_1\times k$ matrix, $
\boldsymbol{\Sigma}_{\Gamma}$ is positive definite and  $\boldsymbol{\Sigma}%
_{\lambda}$ and {$\boldsymbol{\Sigma}_{\phi}$} are  positive semi-definite; furthermore, $%
\mathbb{E}\left\|\boldsymbol{\Gamma}_{i}\right\|^4 \le C$,  $\mathbb{E}\left\|%
\boldsymbol{\lambda}_{i}\right\|^4 \le C$ and $\mathbb{E}\left\|\boldsymbol{%
\phi}_{i}\right\|^4 \le C$. %{\color{red}[is zero mean assumption, especially for $\boldsymbol{\Gamma}_i$, necessary?]}
\end{enumerate}
\end{assumption}

\begin{assumption}[weighting matrix]
\label{assumption-weight}  The weight matrix, $\mathbf{W}=(w_{ij})_{N\times N}=(\mathbf{w}_1,\ldots,\mathbf{w}_N)^\prime$, satisfies that

\begin{enumerate}
\item \label{assumption-weight-diag}  all diagonal elements of $\mathbf{W}$
are zero;

\item \label{assumption-weight-bound} $\mathbf{W}$ has bounded row and column sum norms, i.e., $\|\mathbf{W}\|_\infty<C$ and $\|\mathbf{W}\|_1<C$;

\item \label{assumption-weight-inv}   $\mathbf{I}%
_{N}-\boldsymbol{\Psi} \mathbf{W}$ is invertible, where $\boldsymbol{\Psi}=\mathrm{diag}(\psi_1,\ldots,\psi_N)$;

\item \label{assumption-weight-correlation} $\sup\limits_{i}|\psi_i|<\min\{1/\|\mathbf{W}\|_\infty,\ 1/\|\mathbf{W}\|_1\}$; %$\sup\limits_{1\le i\le N}|\rho_{i}|+\sup\limits_{1\le i\le
%N}|\psi_{i}|\|\mathbf{W}\|_1< 1$.% $\sum_{\ell=0}^{\infty}\big\|%
%\lbrack \boldsymbol{\rho}(\mathbf{I}_{N}-\boldsymbol{\Psi} \mathbf{W}%
%)^{-1}]^{\ell} \big\|_{\infty} \le C$, and  $\sum_{\ell=0}^{\infty}\big\|%
%\lbrack \boldsymbol{\rho}(\mathbf{I}_{N}-\boldsymbol{\Psi} \mathbf{W}%
%)^{-1}]^{\ell} \big\|_{1} \le C$.

\item \label{assumption-weight-rho} denoting $\rho_{w}= \frac{\sup_i|{\rho}_i|}{1-\sup_i|{\psi}_i|\cdot\|\mathbf{W}^\prime\|_{1}}$, then there exists a constant $0<c_w<1$ such that $0<\rho_w\leq c_w$.
\end{enumerate}
\end{assumption}

Before presenting the next assumption, we introduce some notation. First, notice that under Assumptions \ref{assumption-idioiny}--\ref{assumption-loadings}, we have $\mathbb{E}\big(\mathbf{Z}_i^\prime\mathbf{G}\boldsymbol{%
\phi}_i\boldsymbol{\varepsilon}_{i}^\prime\mathbf{Z}_i\big)=\mathbb{E}\big[\mathbf{Z}_i^\prime\mathbf{G}\mathbb{E}\big(\boldsymbol{%
\phi}_i\boldsymbol{\varepsilon}_{i}^\prime\big)\mathbf{Z}_i\big]=\mathbb{E}\big[\mathbf{Z}_i^\prime\mathbf{G}\mathbb{E}\big(\boldsymbol{%
\phi}_i\big)\mathbb{E}\big(\boldsymbol{\varepsilon}_{i}^\prime\big)\mathbf{Z}_i\big]=\mathbf{0}$ and hence,
\begin{align}
\mathrm{Var}\left(\frac{1}{\sqrt{T}}{\mathbf{Z}}_{i}^{\prime }\mathbf{u}_i\right)=\frac{1}{T}\mathbb{E}\big[{\mathbf{Z}}_{i}^{\prime }\big(\mathbf{G}\boldsymbol{%
\phi}_i+\boldsymbol{\varepsilon}_{i}\big)\big(\mathbf{G}\boldsymbol{%
\phi}_i+\boldsymbol{\varepsilon}_{i}\big)^\prime{\mathbf{Z}}_{i}\big]=\frac{1}{T}\mathbb{E}\big[{\mathbf{Z}}_{i}^{\prime }\left(\mathbf{G}\boldsymbol{\Sigma}_{\boldsymbol{\phi}}\mathbf{G}^\prime+\boldsymbol{\Sigma}_{\varepsilon,i}\right)\mathbf{Z}_i\big],\notag
\end{align}
where $\boldsymbol{\Sigma}_{\boldsymbol{\phi}}$ is the variance of $\boldsymbol{\phi}_i$ defined in Assumption \ref{assumption-loadings}\ref{assumption-loadings2} and $\boldsymbol{\Sigma}_{\varepsilon,i}=\mathbb{E}\big(\boldsymbol{\varepsilon}_i\boldsymbol{\varepsilon}_i^\prime\big)$. Then, define
\begin{align}
{\mathbf{A}}_{i}=T^{-1}{\mathbf{Z}}_{i}^{\prime }
\mathbf{C}_{i},\quad\quad {\mathbf{B}}_{i}=T^{-1} {\mathbf{Z}}%
_{i}^{\prime }{\mathbf{Z}}_{i},\quad\quad \boldsymbol{\Phi}_{i}=\frac{1}{T}\mathbb{E}\big[{\mathbf{Z}}_{i}^{\prime }\left(\mathbf{G}\boldsymbol{\Sigma}_{\boldsymbol{\phi}}\mathbf{G}^\prime+\boldsymbol{\Sigma}_{\varepsilon,i}\right)\mathbf{Z}_i\big].\notag
\end{align}

The next assumption imposes some regularity conditions on $\mathbf{A}_i$ and $\mathbf{B}_i$ to ensure the identification of each individual coefficient vector $\boldsymbol{\theta}_i$. Following \citet{Bai2009}, it also imposes a high-level condition on the sampling distribution of the score $T^{-1/2}\mathbf Z_i^\prime\mathbf u_i$.

\begin{assumption}[identification and score regularity]
\label{assumption-ident}  It holds that
\begin{enumerate}
\item \label{assumption-ident-fullrank}  the matrices ${\mathbf{A}}_{i}$ and ${\mathbf{B}}_{i}$ have full column ranks for each $i$ and $T$;
\item \label{assumption-ident-limit} for each $i$, there exist a $3k\times (k+2)$ full-column-rank matrix ${\mathbf{A}}_{i,0}$ and two $3k\times 3k$ positive definite matrices ${\mathbf{B}}_{i,0}$ and $\boldsymbol{\Phi}_{i,0}$, such that $\plim\limits_{N,T\rightarrow\infty}{\mathbf{A}}_{i}={\mathbf{A}}_{i,0}$, $\plim\limits_{N,T\rightarrow\infty}{\mathbf{B}}_{i}={\mathbf{B}}_{i,0}$, and $\lim\limits_{N,T\rightarrow\infty} \boldsymbol{\Phi}_{i}=\boldsymbol{\Phi}_{i,0}$.
\item \label{assumption-ident-clt} for each $i$,
$T^{-1/2}\mathbf{Z}_i^\prime\mathbf{u}_i
\stackrel{d}{\longrightarrow}
N\!\left(\mathbf{0},\boldsymbol{\Phi}_{i,0}\right)$,
as $N,T\rightarrow\infty$.
\end{enumerate}
\end{assumption}

% ORIGINAL ASSUMPTION WITHOUT CLT

% \begin{assumption}[identification]
% \label{assumption-ident}  It holds that
% \begin{enumerate}
% \item \label{assumption-ident-fullrank}  the matrices ${\mathbf{A}}_{i}$ and ${\mathbf{B}}_{i}$ have full column ranks for each $i$ and $T$;

% %\item \label{assumption-ident-moment} \sout{for each $i$, {\color{red}${\mathbf{A}}_{i}$ and ${\mathbf{B}}_{i}$ are bounded in probability}; }

% \item \label{assumption-ident-limit} for each $i$, there exist a $3k\times (k+2)$ full-column-rank matrix ${\mathbf{A}}_{i,0}$ and two $3k\times 3k$ positive definite matrices ${\mathbf{B}}_{i,0}$ and $\boldsymbol{\Phi}_{i,0}$, such that $\plim\limits_{T\rightarrow\infty}{\mathbf{A}}_{i}={\mathbf{A}}_{i,0}$, $\plim\limits_{T\rightarrow\infty}{\mathbf{B}}_{i}={\mathbf{B}}_{i,0}$, and $\lim\limits_{T\rightarrow\infty} \boldsymbol{\Phi}_{i}=\boldsymbol{\Phi}_{i,0}$.
% \end{enumerate}
% \end{assumption}

\begin{assumption}[random coefficients]
\label{assumption-random}  The heterogeneous coefficient vector $\boldsymbol{\theta}_i$ is random and satisfies that
\begin{enumerate}
\item \label{assumption-random-dist}
$\boldsymbol{\theta}_i$ follows the random-coefficient model $\boldsymbol{\theta}_i=\boldsymbol{\theta}+%
\mathbf{e}_i$, where $\boldsymbol{\theta}=(\psi,\rho,\boldsymbol{\beta}%
^{\prime })^{\prime }$ and $\mathbf{e}_i$ is a random error that is independently and identically
distributed with zero mean and variance $\boldsymbol{\Sigma}_{%
\boldsymbol{\theta}}$; furthermore, $\mathbf{e}_i$ is uniformly bounded in probability and independent of $%
\boldsymbol{\Gamma}_j$, $\boldsymbol{\lambda}_j$, $\boldsymbol{\phi}_j$, $%
\varepsilon_{jt}$, $\mathbf{v}_{jt}$, $\mathbf{f}_t$, and $\mathbf{g}_t$ for
all $i, j, t$; %{\color{red}[also need $\mathbf{e}_i$ uniformly bounded in probability or $\max_{1\leq i\leq N} \Vert\mathbf{e}_i\Vert=O_p(\log N)$]}

\item \label{assumption-random-inv}   $\mathbf{I}%
_{N}-\boldsymbol{\Psi} \mathbf{W}$ is invertible a.s.;

\item \label{assumption-random-correlation} $\sup\limits_{i}|\psi_i|<\min\{1/\|\mathbf{W}\|_\infty,\ 1/\|\mathbf{W}\|_1\}$ a.s.;

\item \label{assumption-random-rho} $0<\rho_w<1$ a.s. and $\mathbb{E}\left[\frac{1}{\left(1-\rho_w\right)^2}\right]\leq C$, where $\rho_w$ was defined in Assumption \ref{assumption-weight}\ref{assumption-weight-rho}.
\end{enumerate}
\end{assumption}

Assumptions \ref{assumption-idioiny}--\ref{assumption-ident} are standard in the panel with interactive fixed effects and spatial econometrics literature. Here, we highlight their main implications. Assumption \ref{assumption-idioiny} imposes some moment conditions as well as restrictions on the cross-sectional heteroskedasticity and time-series dependence on the idiosyncratic error $\varepsilon_{it}$.
%To simplify the asymptotic analysis, we assume $\varepsilon_{it}$ is cross-sectionally uncorrelated. This assumption is not unreasonable given that our model explicitly accounts for both local spatial dependence (through the spatial lag term) and global cross-sectional dependence (through the interactive fixed effects term) in the dependent variable. Hence, the cross-sectional dependence that is remaining in the idiosyncratic error $\varepsilon_{it}$ would be negligible. On the other hand, we allow for heteroskedasticity across $i$ but constrain the variances with an upper bound. %\footnote{Note that under Assumption \ref{assumption-idioiny}, $\mathrm{Var}(\varepsilon_{it})=\sigma_\zeta^2\sum_{\ell=0}^\infty p_{i\ell}^2\leq C$ for all $i$, where $\sigma_\zeta^2=\mathrm{Var}(\zeta_{it})$. }.
We also allow for weak time-series dependence in $\varepsilon_{it}$ through stationarity with absolutely summable serial covariances. Such conditions are commonly used in the literature (e.g., \citealt{ChenEtAl2022} and \citealt{Pesaran2006}). Since our identification strategy relies on exogenous covariates rather than lagged dependent variables as instruments, weak serial correlation in $\varepsilon_{it}$ does not compromise instrument validity under the maintained conditions.
In addition, richer temporal dynamics can be accommodated by including further lags of the dependent variable where appropriate.
% This assumption is standard in the dynamic panel data literature (e.g., \citet{AlvarezArellano2003}, \citet{ShiLee2017}), where temporal dependence is captured through lagged dependent variables.
%In addition, serial dependence of an autoregressive type can be accommodated by including further lags of the dependent variable in the model.
Assumption \ref{assumption-idioinx} allows for weak cross-sectional and time-series dependence in $\mathbf{v}_{it}$, subject to restrictions that govern the strength of such dependence. For example, Assumption \ref{assumption-idioinx}\ref{assumption-idioinx2} implies that $\mathbf{v}_{it}$ has absolutely summable serial covariances as well as spatial covariances with other cross-sectional units. Heteroskedasticity in $\mathbf{v}_{it}$ is also allowed. %This distinction in the dependence structures of $\varepsilon_{it}$ and $\mathbf{v}_{it}$ is important, as after defactoring, the dynamics of the regressors are driven by $\mathbf{v}_{it}$.
Assumptions \ref{assumption-fac} and \ref{assumption-loadings} are common in the factor model literature (e.g., \citealt{Bai2003}) and ensure that the latent factors and loadings are well behaved. In particular, the factors can admit serial dependence. The loadings, $\boldsymbol{\lambda}_i$ and $\boldsymbol{\phi}_i$ are assumed to have zero mean, reflecting that they are components of the composite error term in the $y$ equation, while the loadings $\boldsymbol{\Gamma}_i$ are allowed to have a non-zero mean. The requirement that $\boldsymbol{\Sigma}_\Gamma$ is positive definite ensures that the factor space is sufficiently spanned by the regressors for identification.

Assumption \ref{assumption-weight} places standard restrictions on the adjacency matrix $\mathbf{W}$ (see, e.g., \citealt{KelejianPrucha2001}). Assumption \ref{assumption-weight}\ref{assumption-weight-diag} imposes a normalization so that no individual is treated as its own neighbour. Assumption \ref{assumption-weight}\ref{assumption-weight-bound} requires the row and column sums of $\mathbf{W}$ to be uniformly bounded, a condition satisfied by commonly used weighting schemes such as row-normalized contiguity matrices or distance-based matrices with sufficiently fast decay. This restriction rules out the presence of units that exert pervasive spatial effects and ensures that strong spatial interactions remain mostly local. Situations involving strong pervasive influences are instead accommodated through the interactive effects structure. Assumption \ref{assumption-weight}\ref{assumption-weight-inv} ensures that $(\mathbf{I}_N - \boldsymbol{\Psi}\mathbf{W})$ is invertible, so that the spatial structure admits a unique reduced-form representation. Assumption \ref{assumption-weight}\ref{assumption-weight-correlation} further restricts the magnitude of the spatial coefficients, ensuring that spatial feedback effects remain bounded. Assumption \ref{assumption-weight}\ref{assumption-weight-rho} imposes the condition $0<\rho_w<1$, which ensures that the spatial dynamic system is uniformly stable. Together, these conditions guarantee that the resulting process for $y_{it}$ is both cross-sectionally and temporally stable and possesses well-defined moments.

Assumption \ref{assumption-ident}\ref{assumption-ident-fullrank} ensures identification of the individual-specific parameters. Formally, the full-rank conditions imposed on the sample moment matrices guarantee that the instruments are sufficiently informative about the endogenous regressors. Instrument relevance arises from the presence of exogenous covariates with nonzero effects. In particular, lagged covariates are informative about $y_{i,t-1}$ even when $\rho_i = 0$, since $y_{i,t-1}$ depends on past covariates through the structural equation at time $t-1$. Similarly, spatially weighted covariates are informative about $\sum_{j=1}^N w_{ij} y_{jt}$ even when $\psi_i = 0$, since neighbouring outcomes depend on their own covariates. Importantly, this requires only that for each $i$, not all elements of $\boldsymbol{\beta}_i$ are zero.
Assumptions \ref{assumption-ident}\ref{assumption-ident-limit}--\ref{assumption-ident-clt} are used to establish the asymptotic distribution of the individual-specific estimator. The high-level condition in \ref{assumption-ident}\ref{assumption-ident-clt} can be replaced by more primitive assumptions controlling the serial dependence of $\mathbf{z}_{it}(\mathbf{g}_t^\prime\boldsymbol{\phi}_i+\varepsilon_{it})$, where $\mathbf{z}_{it}$ is the $t$-th column of $\mathbf{Z}_i^\prime$, so that the central limit theorem for $T^{-1/2}\mathbf{Z}_i^\prime\mathbf{u}_i=T^{-1/2}\sum_{t=1}^T\mathbf{z}_{it}(\mathbf{g}_t^\prime\boldsymbol{\phi}_i+\varepsilon_{it})$ holds. But to simplify the exposition and the proof of Theorem \ref{individual-asydist}, we follow \citet{Bai2009} and impose the limiting normal distribution for $T^{-1/2}\mathbf{Z}_i^\prime\mathbf{u}_i$ directly.

Finally, Assumption \ref{assumption-random} specifies a random-coefficient framework for slope heterogeneity, as in, e.g., \citet{Pesaran2006}. This is required when we study the asymptotic properties of the MGIV estimator. It is worth noting that Assumption \ref{assumption-random}\ref{assumption-random-inv}--\ref{assumption-random-rho} are the random-coefficient version of Assumption \ref{assumption-weight}\ref{assumption-weight-inv}--\ref{assumption-weight-rho}. They ensure that, for each realisation of the random coefficients, the spatial dynamic system is stable almost surely.

\subsection{Asymptotic Theory}\label{subsec-theory}
The following theorem establishes the asymptotic properties of the individual-specific estimator $\widehat{\boldsymbol{\theta}}_i$.

\begin{theorem}\label{individual-asydist}
If Assumptions \ref{assumption-idioiny}--\ref{assumption-weight} and \ref{assumption-ident}\ref{assumption-ident-fullrank} hold, then the individual-specific IV estimator, $\widehat{\boldsymbol{\theta}}_i$, defined in (\ref{individual-est}) is consistent as $N,T\rightarrow\infty$.
If, in addition, Assumptions
\ref{assumption-ident}\ref{assumption-ident-limit}--\ref{assumption-ident-clt}
hold and $T/N^2\rightarrow0$ as $N,T\rightarrow\infty$,
%If, in addition, Assumption \ref{assumption-ident}\ref{assumption-ident-limit} holds and $T/N^2\rightarrow0$ as $N,T\rightarrow\infty$,
then $\widehat{\boldsymbol{\theta}}_i$ has the following asymptotic distribution
\begin{equation*}
\sqrt{T}(\widehat{\boldsymbol{\theta}}_i-\boldsymbol{\theta}_i)\stackrel{d}{\longrightarrow} N\left(\mathbf{0},\ \boldsymbol{\Sigma}_{i,0}\right), \ \ {\rm as\ }N,T\rightarrow\infty,
\end{equation*}
where $\boldsymbol{\Sigma}_{i,0}=\big(\mathbf{A}_{i,0}^\prime\mathbf{B}_{i,0}^{-1}\mathbf{A}_{i,0}\big)^{-1}\mathbf{A}_{i,0}^\prime\mathbf{B}_{i,0}^{-1}\boldsymbol{\Phi}_{i,0}\mathbf{B}_{i,0}^{-1}\mathbf{A}_{i,0}\big(\mathbf{A}_{i,0}^\prime\mathbf{B}_{i,0}^{-1}\mathbf{A}_{i,0}\big)^{-1}$.
\end{theorem}

\begin{remark}
The large $N$ requirement is indispensable because the validity of the instruments used in the IV estimator $\widehat{\boldsymbol{\theta}}_{i}$ requires consistent estimation (up to a rotation) of the factors $\mathbf{f}_t$. For the consistency of $\widehat{\boldsymbol{\theta}}_{i}$, $N$ and $T$ can grow to infinity at any rate. However, for the asymptotic centred normal distribution to hold, we require $T/N^2\rightarrow 0$. This is because the bias in each individual estimator that arises from factor estimation is of order $O\big(T^{-1/2}\delta_{NT}^{-1}+\delta_{NT}^{-2}\big)$ (see equation (\ref{eq-theorem1proof1}) in Appendix A). The condition $T/N^2\rightarrow 0$ ensures that this bias is asymptotically negligible when scaled by $\sqrt{T}$. It is commonly used in the derivation of asymptotic distributions for individual estimators of heterogeneous panel data models with interactive fixed effects (see, e.g., \citealt{Pesaran2006}, \citealt{ChenEtAl2022} and \citealt{CaoEtAl2024}) %It is more general than the condition in \citet{NorkuteEtal2021} and \cite{CuiSarafidisYamagata2023}, which requires $T/N\rightarrow c$ for some constant $0 < c < \infty$.
and allows $T$ to grow faster than $N$ (but slower than $N^{2}$), proportionally to $N$, or slower than $N$. We also note that the condition $T/N^2\rightarrow 0$ is weaker than $T/N\rightarrow c$ for some $0<c<\infty$, which is often imposed for the asymptotic normality of the pooled estimator of homogeneous dynamic panel data models with interactive fixed effects (e.g., \citealt{MoonWeidner2017}, \citealt{NorkuteEtal2021}, and \citealt{CuiSarafidisYamagata2023}). The main reason for the weaker relative-growth-rate condition in Theorem \ref{individual-asydist} is the slower convergence rate of the individual estimator (i.e., $\sqrt{T}$). The pooled estimator of homogeneous coefficients in \citet{MoonWeidner2017}, \citet{NorkuteEtal2021}, and \citet{CuiSarafidisYamagata2023} has a convergence rate of $\sqrt{NT}$ and therefore requires stronger conditions to make the factor estimation bias negligible in the asymptotic distribution.
\end{remark}

The asymptotic covariance matrix, $\boldsymbol{\Sigma}_{i,0}$ can be estimated by
\begin{align}
\widehat{\boldsymbol{\Sigma}}_{i}=\big(\widehat{\mathbf{A}}_{i}^\prime \widehat{\mathbf{B}}_{i}^{-1}\widehat{\mathbf{A}}_{i}\big)^{-1}\widehat{\mathbf{A}}_{i}^\prime \widehat{\mathbf{B}}_{i}^{-1}\widehat{\boldsymbol{\Phi}}_{i}\widehat{\mathbf{B}}_{i}^{-1}\widehat{\mathbf{A}}_{i}\big(\widehat{\mathbf{A}}_{i}^\prime\widehat{\mathbf{B}}_{i}^{-1}\widehat{\mathbf{A}}_{i}\big)^{-1},\label{Sec3insert1}
\end{align}
where $\widehat{\boldsymbol{\Phi}}_{i}$ is an estimator of $\boldsymbol{\Phi}_{i,0}$, which can be constructed using the Newey-West type procedure (see \citealt{NeweyWest1987})
\begin{align}
\widehat{\boldsymbol{\Phi}}_{i}=\widehat{\boldsymbol{\Phi}}_{i}(0)+\sum_{h=1}^{L_T}K\left(\frac{h}{L_T+1}\right)\left[\widehat{\boldsymbol{\Phi}}_{i}(h)+\widehat{\boldsymbol{\Phi}}_{i}(h)^\prime\right],\label{Sec3insert2}
\end{align}
where $L_T$ is the bandwidth, $K(\cdot)$ is a kernel and $\widehat{\boldsymbol{\Phi}}_{i}(h)=\frac{1}{T}\sum_{t=h+1}^T\hat{\tilde{u}}_{it}\hat{\tilde{u}}_{i,t-h}\hat{\mathbf{z}}_{it}\hat{\mathbf{z}}_{i,t-h}^\prime$, in which $\hat{\tilde{u}}_{it}$ is the $t$-th element of the vector $\widehat{\widetilde{\mathbf{u}}}_i:=\mathbf{M}_{\widehat{\mathbf{F}}}\widehat{\mathbf{u}}_{i}$ with $\widehat{\mathbf{u}}_{i}:=\mathbf{y}_{i}-\mathbf{C}_{i}\widehat{\boldsymbol{\theta}}_i$, and $\hat{\mathbf{z}}_{it}$ is the $t$-th column of $\widehat{\mathbf{Z}}_{i}^{\prime }$.

In the following proposition, we establish the consistency of the  covariance estimator $\widehat{\boldsymbol{\Sigma}}_{i}$.

\smallskip
\begin{proposition}\label{individual-Cov-Consistency}
Suppose that the following conditions hold
\begin{description}
\item (i) the kernel $K(\cdot)$ is bounded on its support $[0,1]$ and is right-continuous at zero with $K(0)=1$;

\item (ii) the bandwidth $L_T$ satisfies $L_T\rightarrow\infty$ and $L_T/\delta_{NT}\rightarrow0$ as $N,T\rightarrow\infty$;

\item (iii) denoting $\mathbf{s}_{it}=\mathbf{z}_{it}\tilde{u}_{it}$, where $\mathbf{z}_{it}$ is the $t$-th column of $\mathbf{Z}_i^\prime$ and $\tilde{u}_{it}$ is the $t$-th element of $\widetilde{\mathbf{u}}_{i}:=\mathbf{M}_{\mathbf{F}}\mathbf{u}_i=\mathbf{M}_{\mathbf{F}}\big(\mathbf{G}\boldsymbol{%
\phi}_i+\boldsymbol{\varepsilon}_{i}\big)$, for each $i$, we have the following
\begin{align}
%\sup_{0\leq h\leq L_T}\sum_{r=-\infty}^\infty \left\Vert\mathrm{Cov}\big(\mathrm{vec}(\mathbf{s}_{it}\mathbf{s}_{i,t-h}^\prime), \mathrm{vec}(\mathbf{s}_{i,t-r}\mathbf{s}_{i,t-h-r}^\prime)\big)\right\Vert \leq C,\label{Sec3insert3}
\sup_{0\leq h\leq L_T}\,\sum_{t=h+1}^T\sum_{s=h+1}^T \left\Vert\mathrm{Cov}\big(\mathrm{vec}(\mathbf{s}_{it}\mathbf{s}_{i,t-h}^\prime),\ \mathrm{vec}(\mathbf{s}_{i,s}\mathbf{s}_{i,s-h}^\prime)\big)\right\Vert \leq CT,\label{Sec3insert3}
\end{align}
where $\mathrm{vec}(\cdot)$ is the vectorisation operator.
\end{description}
Then, under Assumptions \ref{assumption-idioiny}--\ref{assumption-weight}
and \ref{assumption-ident}\ref{assumption-ident-fullrank}--\ref{assumption-ident-limit},
the covariance matrix estimator $\widehat{\boldsymbol{\Sigma}}_{i}$ is consistent
for $\boldsymbol{\Sigma}_{i,0}$ as $N,T\rightarrow\infty$.

%Then, under Assumptions \ref{assumption-idioiny}--\ref{assumption-ident}, the covariance matrix estimator $\widehat{\boldsymbol{\Sigma}}_{i}$ is consistent for $\boldsymbol{\Sigma}_{i,0}$ as $N,T\rightarrow\infty$.
\end{proposition}

Assumptions (i)--(iii) in the above proposition are used in establishing the consistency of $\widehat{\boldsymbol{\Phi}}_{i}$ to ${\boldsymbol{\Phi}}_{i,0}$. The conditions on $K(\cdot)$ are standard for Newey-West type variance estimators. The conditions on the bandwidth $L_T$ ensure that the estimation error in $\widehat{\boldsymbol{\Phi}}_{i}$ that is due to the truncation at $L_T$ and the estimation of factors and individual coefficients $\boldsymbol{\theta}_i$ is asymptotically negligible. Condition (iii) on the other hand constrains the sampling error of the Newey-West estimator so that it is also negligible.

Proposition \ref{individual-Cov-Consistency} indicates that we can use the asymptotic distribution in Theorem \ref{individual-asydist} and the covariance estimate $\widehat{\boldsymbol{\Sigma}}_{i}$ to carry out inference about coefficients of interest for a specific cross-sectional unit.
 \bigskip

Next, we consider the asymptotic properties of the MGIV estimator $\widehat{\boldsymbol\theta}_{MG}$ defined in (\ref{eq:MGest}). To this end, we strengthen Assumption \ref{assumption-ident} to the following assumption.
\begin{assumption}
\label{assumption-uniident}
\ It holds that
\begin{enumerate}
\item \label{assumption-uniident-unifullrank} there exists a positive constant $c>0$ such that
$$
\inf\limits_i\, \lambda_{\min}(\mathbf{B}_i)\geq c\quad\rm{and}\quad \inf\limits_i\, \lambda_{\min}(\mathbf{A}_i^\prime\mathbf{B}_i^{-1}\mathbf{A}_i)\geq c,
$$
where $\lambda_{\min}(\cdot)$ denotes the smallest eigenvalue of a square matrix;

%\item \label{assumption-uniident-unibounded} {\color{red}${\mathbf{A}}_{i}$ and ${\mathbf{B}}_{i}$ are uniformly bounded in probability};

\item \label{assumption-uniident-uniorder} $\mathbb{E}\left\Vert\frac{1}{\sqrt{NT}}\sum\limits_{i=1}^N(\mathbf{A}_i^\prime\mathbf{B}_i^{-1}\mathbf{A}_i)^{-1}\mathbf{A}_i^\prime\mathbf{B}_i^{-1}\mathbf{Z}_i^\prime\big(\mathbf{G}\boldsymbol{%
\phi}_i+\boldsymbol{\varepsilon}_{i}\big)\right\Vert^2\leq C$.

\end{enumerate}
\end{assumption}

\begin{theorem}\label{MG-asydist}
If Assumptions \ref{assumption-idioiny}--\ref{assumption-loadings}, \ref{assumption-weight}\ref{assumption-weight-diag}--\ref{assumption-weight-bound}, \ref{assumption-random} and \ref{assumption-uniident}\ref{assumption-uniident-unifullrank} hold, then the MGIV estimator, $\widehat{\boldsymbol\theta}_{MG}$, is consistent for the population mean $\boldsymbol{\theta}$. If, in addition, Assumption \ref{assumption-uniident}\ref{assumption-uniident-uniorder} holds and $N/T^2\rightarrow0$, then the MGIV estimator has the following asymptotic distribution
\begin{equation*}
\sqrt{N}(\widehat{\boldsymbol\theta}_{MG}-\boldsymbol{\theta})\stackrel{d}{\longrightarrow} N\left(\mathbf{0},\ \boldsymbol{\Sigma}_{\boldsymbol\theta}\right),\ \ {\rm as\ }N,T\rightarrow\infty,
\end{equation*}
where $\boldsymbol{\Sigma}_{\boldsymbol\theta}$ was defined in Assumption \ref{assumption-random}\ref{assumption-random-dist}. Furthermore,
\begin{align}
\widehat{\boldsymbol{\Sigma}}_{\boldsymbol\theta}=\frac{1}{N-1}\sum\limits_{i=1}^N \left(\widehat{\boldsymbol{\theta}}_i-\widehat{\boldsymbol\theta}_{MG}\right)\left(\widehat{\boldsymbol{\theta}}_i-\widehat{\boldsymbol\theta}_{MG}\right)^\prime \notag
\end{align}
is a consistent estimator of $\boldsymbol{\Sigma}_{\boldsymbol\theta}$.
\end{theorem}

%\begin{remark}
%We can estimate $\boldsymbol{\Sigma}_{\boldsymbol\theta}$ by
%\begin{align}
%\widehat{\boldsymbol{\Sigma}}_{\boldsymbol\theta}=\frac{1}{N-1}\sum\limits_{i=1}^N \left(\widehat{\boldsymbol{\theta}}_i-\widehat{\boldsymbol\theta}_{MG}\right)\left(\widehat{\boldsymbol{\theta}}_i-\widehat{\boldsymbol\theta}_{MG}\right)^\prime. \notag
%\end{align}
%Following the proofs of Theorem 5 (ii) in \citet{NorkuteEtal2021}, we can show that $\widehat{\boldsymbol{\Sigma}}_{\boldsymbol\theta}$ is consistent for $\boldsymbol{\Sigma}_{\boldsymbol\theta}$ under the assumptions of Theorem \ref{MG-asydist}.
%\end{remark}

\begin{remark}

Assumption \ref{assumption-uniident}\ref{assumption-uniident-unifullrank} is a uniform version of Assumption \ref{assumption-ident}\ref{assumption-ident-fullrank}. It ensures that $\mathbf{B}_i$ and $(\mathbf{A}_i^\prime\mathbf{B}_i^{-1}\mathbf{A}_i)$ are uniformly invertible. Assumption \ref{assumption-uniident}\ref{assumption-uniident-uniorder} is a high-level condition used in the derivation of the asymptotic distribution in Theorem \ref{MG-asydist}. Together with the condition $N/T^2\rightarrow0$, it guarantees that the average convergence rate of the individual estimators, i.e., $\frac{1}{N}\sum_{i=1}^N\big(\widehat{\boldsymbol{\theta}}_i-\boldsymbol{\theta}_i\big)$, is $o_p(N^{-1/2})$ so that the asymptotic distribution of $\big(\widehat{\boldsymbol\theta}_{MG}-\boldsymbol{\theta}\big)$ is determined by the partial sum of the idiosyncratic errors $\mathbf{e}_i$ in the random-coefficient model (see equation (\ref{eq-theorem2proof12}) in Appendix A). More specifically, the relative-growth-rate condition of $N/T^2\rightarrow0$ makes the bias arising from factor estimation asymptotically negligible when scaled by $N^{1/2}$. %It is weaker than $T/N\rightarrow c$ for some $0<c<\infty$, which is often imposed for the asymptotic normality of the pooled estimator of homogeneous dynamic panel data models with interactive fixed effects (e.g., \citealt{MoonWeidner2017}, \citealt{NorkuteEtal2021}, and \citealt{CuiSarafidisYamagata2023}) or the asymptotic normality of the mean-group estimator of heterogeneous dynamic panel data models with interactive fixed effects (e.g., \citealt{ChudikPesaran2015}). %We further note that alternatively, we can replace the pair of conditions, i.e., Assumption \ref{assumption-uniident}\ref{assumption-uniident-uniorder} and $N/T^2\rightarrow0$, with a stronger relative-growth-rate condition of $N/T\rightarrow0$, which makes the convergence rate $O_p(T^{-1/2})$ of each individual estimator $\widehat{\boldsymbol{\theta}}_i$ decay faster than $N^{-1/2}$.  %It can be weakened to $\mathbb{E}\left\Vert\frac{1}{\sqrt{NT}}\sum\limits_{i=1}^N(\mathbf{A}_i^\prime\mathbf{B}_i^{-1}\mathbf{A}_i)^{-1}\mathbf{A}_i^\prime\mathbf{B}_i^{-1}\mathbf{Z}_i^\prime\big(\mathbf{G}\boldsymbol{%
%\phi}_i+\boldsymbol{\varepsilon}_{i}\big)\right\Vert^2=o(T)$ without requiring changes to the proof of Theorem \ref{MG-asydist}.

Theorems \ref{individual-asydist} and \ref{MG-asydist} indicate that under the required assumptions, both the individual-specific and MGIV estimators are correctly centered around the true parameters without necessitating any bias correction. Intuitively, this property holds because the idiosyncratic errors $\mathbf{v}_{it}$ in the $\mathbf{x}$ equation are assumed to be independent of the composite error term $u_{it}$ in the $y$ equation.  Consequently, the defactored $\mathbf{x}_{it}$ are uncorrelated with $u_{it}$ and therefore are used to construct the IV estimators.
\end{remark}

\begin{remark}
Thus far, we have assumed that $\mathbf{X}_{i}$ is exogenous with respect to the purely idiosyncratic error, $\boldsymbol{\varepsilon}_{i}$. This assumption will be violated if, for example, $\mathbf{X}_{i}$ is subject to reverse causality or measurement error. However, our estimation method can accommodate such sources of endogeneity, provided that valid external instruments are available.\footnote{For a discussion of this issue in the context of CCE estimation, see \citet{HardingLamarche2011}.} To formalise this, let $\mathbf{X}_{i}=\left(\mathbf{X}_{i}^{\left(exog\right)}, \mathbf{X}_{i}^{\left(endog\right)}\right)$, where $\mathbf{X}_{i}^{\left(exog\right)}$ and $\mathbf{X}_{i}^{\left(endog\right)}$ refer to the sets of strictly exogenous and endogenous covariates, respectively, with respect to $\boldsymbol{\varepsilon}_{i}$. Suppose $\mathbf{X}_{i}^{\left(exog\right)}$ and $\mathbf{X}_{i}^{\left(endog\right)}$ have dimensions $T \times k^{\left(exog\right)}$ and $T \times k^{\left(endog\right)}$, respectively, with $k^{\left(exog\right)}+k^{\left(endog\right)}=k$, and there exists $\mathbf{X}_{i}^{\left(ext\right)}$ that contains $k^{\left(ext\right)}$  external instruments for $\mathbf{X}_{i}^{\left(endog\right)}$, $k^{\left(ext\right)}\geq k^{\left(endog\right)}$. Note that $\mathbf{X}_{i}^{\left(ext\right)}$ can still be correlated with the factors in the $y$ equation, i.e., it may be subject to a similar data generating process as in Eq. \eqref{spm_X_scalar}, so long as it remains strictly exogenous with respect to $\boldsymbol{\varepsilon}_{i}$. Define $\mathbf{X}^{+}_{i}=(\mathbf{X}_{i}^{\left(exog\right)}, \mathbf{X}_{i}^{\left(ext\right)})$, a $T \times k^{+}$ matrix with $k^{+}= k^{\left(exog\right)}+k^{\left(ext\right)}$, and $\widehat{\mathbf{F}}^{+}_{x}$, a $T\times {r}^{+}_{x}$ matrix that contains $\sqrt{T}$ times the eigenvectors corresponding to the ${r}^{+}_{x}$ largest eigenvalues of the $T\times T$ matrix $\sum_{i=1}^{N}\mathbf{X}^{+}_{i}\left(\mathbf{X}^{+}_{i}\right)^{\top}/NT$. The associated defactoring matrices can be defined analogously to those in Section \ref{sec-method}, with $\widehat{\mathbf{F}}$ and $\widehat{\mathbf{F}}_{-1}$ replaced by $\widehat{\mathbf{F}}^{+}_{x}$ and $\widehat{\mathbf{F}}^{+}_{x,-1}$, respectively. Under this framework, the matrix of instruments retains the same structure as in \eqref{eq:Z_matrix}, with $\mathbf{X}_{i}$ and $\mathbf{X}_{i,-1}$ replaced by $\mathbf{X}^{+}_{i}$ and $\mathbf{X}^{+}_{i,-1}$.
\end{remark}

\section{Monte Carlo Experiments\label{sec-MC}}

We investigate the finite sample behaviour of the proposed approach by means of Monte Carlo experiments. We shall focus on the mean, bias, RMSE, empirical size and power of the t-test.

\subsection{Design}\label{subsec:Design}
\setlength{\belowdisplayskip}{0pt} \setlength{\belowdisplayshortskip}{0pt}
\setlength{\abovedisplayskip}{0pt} \setlength{\abovedisplayshortskip}{0pt}
We consider the following heterogeneous, spatial dynamic panel data model:
\begin{align}\label{DGP_sim_1}
y_{it}  & =\alpha_{i}+\rho_{i} y_{i,t-1}+\psi_{i} \sum_{j=1}^Nw_{ij}y_{jt}+\sum_{\ell=1}^{k}\beta_{\ell,i}x_{\ell it}+u_{it}\text{; } \quad
u_{it}=\sum_{s=1}^{r_{y}}\varphi_{si} f_{st}+\varepsilon_{it},
\end{align}
$i=1,...,N$, $t=-49,...,T$, where
\begin{equation}
f_{st}=\varrho_{fs}f_{s,t-1}+(1-\varrho_{fs}^{2})^{1/2}\zeta_{st}\label{mcfac}\text{,}\ \ \ \  s=1,\ldots,r_y,
\end{equation}
with ${\zeta_{st}\sim i.i.d.N(0,1)}$. We set $k=2$, $r_{y}=3$ and $\varrho_{fs}=0.5$.

The spatial weighting matrix, $\mathbf{W}=[w_{ij}]$ is a row-normalised rook matrix of circular form (\citealt{KapoorKelejianPrucha2007}), such that its $i$th row, $1<i<N$, has non-zero entries in positions $i-1$ and $i+1$, whereas the non-zero entries in rows $1$ and $N$ are in positions $(1,2)$, $(1,N)$, and $(N,1)$, $(N,N-1)$, respectively. This matrix is row normalized so that all of its nonzero elements are equal to $1/2$.

The idiosyncratic error, $\varepsilon_{it}$, is non-normal and heteroskedastic across both $i$ and $t$, such that $\varepsilon_{it}=\varsigma_{\varepsilon}\sigma_{it}(\epsilon_{it}-1)/\sqrt{2}$ with $\epsilon_{it}\sim i.i.d.\chi_{1}%
^{2}$ and $\sigma_{it}^{2}=\eta_{i}\phi_{t}$, where $\eta_{i}\sim i.i.d.\chi
_{2}^{2}/2$ and $\phi_{t}=t/T$ for $t=0,1,...,T$ and unity for $t<0$. The value of $\varsigma_{\varepsilon}$ will be specified later.

\setlength{\belowdisplayskip}{0pt} \setlength{\belowdisplayshortskip}{0pt}
\setlength{\abovedisplayskip}{0pt} \setlength{\abovedisplayshortskip}{0pt}
The stochastic process for the covariates is given by %
\begin{equation}
x_{\ell it}=\mu_{\ell i}+\sum_{s=1}^{r_{x}}\gamma_{\ell si} f_{st}+v_{\ell
it};\label{mcx} \text{ }i=1,2,...,N; \quad t=-49,-48,...,T,
\end{equation}
for $\ell=1,2$. We set $r_{x}=2$. Thus, the first two factors in $u_{it}$, i.e., $f_{1t}$ and $f_{2t}$, also drive the data generating process (DGP) for $x_{\ell it}$, $\ell=1,2$. However, $f_{3t}$ does not enter the DGP of $x_{\ell it}$, $\ell=1,2$.\footnote{Observe that, using notation of earlier sections, $f_{3t}=g_{t}$ in Eq. \eqref{spm_scalar}.}

\setlength{\belowdisplayskip}{0pt} \setlength{\belowdisplayshortskip}{0pt}
\setlength{\abovedisplayskip}{0pt} \setlength{\abovedisplayshortskip}{0pt}
The idiosyncratic errors, $v_{\ell
it}$, in the DGP of $x_{\ell it}$ are serially correlated, such that
\begin{equation}
v_{\ell it}=\varrho_{\upsilon,\ell}v_{\ell i,t-1}+(1-\varrho_{\upsilon,\ell}%
^{2})^{1/2}\varpi_{\ell it}\text{; }\ \ \ \varpi_{\ell it}\sim i.i.d.N(0,\varsigma
_{\upsilon}^{2})\text{,}\label{mcvit}
\end{equation}
with $\varrho_{\upsilon,\ell}=\varrho_{\upsilon}=0.5$ for $\ell=1,2$.

\setlength{\belowdisplayskip}{0pt} \setlength{\belowdisplayshortskip}{0pt}
\setlength{\abovedisplayskip}{0pt} \setlength{\abovedisplayshortskip}{0pt}
The individual-specific fixed effects in the $y$ and $\mathbf{x}$ equations are generated as correlated and zero-mean random variables. Specifically,
\begin{equation}\label{individual_specific_effects}
\alpha_{i}\sim i.i.d.N(0,\left(  1-\varrho \right)  ^{2}); \quad
\mu_{\ell i}=\varrho_{\mu,\ell}\alpha_{i}+(1-\varrho_{\mu,\ell}^{2})^{1/2}\omega_{\ell i}\text{, }%
\end{equation}
where $\omega_{\ell i}\sim i.i.d.N(0,\left(  1-\varrho \right)^{2})$, for $\ell=1,2$. We set $\varrho=\varrho_{\mu,\ell}=0.5$ for $\ell=1,2$.

\noindent
\setlength{\belowdisplayskip}{0pt} \setlength{\belowdisplayshortskip}{0pt}
\setlength{\abovedisplayskip}{0pt} \setlength{\abovedisplayshortskip}{0pt}
The factor loadings in the $y$ equation are generated as $\varphi_{si}\sim i.i.d.N(0,1)$ for $s=1,...,r_{y}(=3)$, and the factor loadings in the $\mathbf{x}$ equation are
drawn as
\begin{equation}
\gamma_{{1}si}=\varrho_{\gamma,{1}s}\varphi_{3i}+(1-\varrho_{\gamma,{1}s}^{2})^{1/2}%
\xi_{{1}si}\text{; }\ \ \ \xi_{{1}si}\sim i.i.d.N(0,1);
\label{corx1}
\end{equation}
\begin{equation}
\gamma_{{2}si}=\varrho_{\gamma,{2}s}\varphi_{si}+(1-\varrho_{\gamma,{2}s}^{2})^{1/2}%
\xi_{{2}si}\text{; }\ \ \ \xi_{{2}si}\sim i.i.d.N(0,1);
\label{corx2}
\end{equation}
for $s=1,...,r_{x}(=2)$. The process in Eq. (\ref{corx1}) allows the factor loadings to $f_{1t}$ and $f_{2t}$ in the DGP of $x_{1it}$ to be correlated with the factor loadings to $f_{3t}$, which does not enter the DGP of $\mathbf{x}_{it}$. On the other hand, Eq. (\ref{corx2}) ensures that the factor loadings to $f_{1t}$ and $f_{2t}$ in $x_{2it}$ are correlated with the factor loadings to the same factors in the DGP of $y_{it}$. We consider $\varrho_{\gamma,11}=\varrho_{\gamma,12} \in \left \{0\text{, }0.5\right \}$, whilst $\varrho_{\gamma,21}=\varrho_{\gamma,22}=0.5$.

It is straightforward to see that the average variance of $\varepsilon_{it}$ depends only on $\varsigma_{\varepsilon}^{2}$.
Let $\pi_{u}$ denote the proportion of the average variance of $u_{it}$ that is due
to $\varepsilon_{it}$. That is, we define $\pi_{u}:=\varsigma_{\varepsilon}^{2}/\left(  r_{y} +\varsigma_{\varepsilon}^{2}\right)$. Thus, for example, $\pi_{u}=3/4$ means that the
variance of the idiosyncratic error accounts for 75\% of the total variance in
$u_{it}$. In this case, most of the variation in the composite error term $u_{it}$ is due to the
idiosyncratic component and the factor structure has relatively minor
significance.
%\setlength{\belowdisplayskip}{0pt} \setlength{\belowdisplayshortskip}{0pt}
%\setlength{\abovedisplayskip}{0pt} \setlength{\abovedisplayshortskip}{0pt}
%Solving in terms of $\varsigma_{\varepsilon}^{2}$ yields %
%\begin{equation}
%\varsigma_{\varepsilon}^{2}=\frac{\pi_{u}}{\left(  1-\pi_{u}\right)  } r_{y}\label{mcpiu}  \text{.}%
%\end{equation}
We set $\varsigma_{\varepsilon}^{2}$ values such that $\pi_{u}\in \left \{1/4\text{, }3/4\right \}$.\footnote{These values of $\pi_u$ are motivated by the results in \citet{Sargent_Sims_1977}, in which they find that two common factors explain $86\%$ of the variation in unemployment rate and $26\%$ of the variation in residential construction.}

The slope coefficients are generated as $\rho _{i}=\rho +\eta _{\rho, i}$, $\psi_{i}=\psi+\eta _{\psi, i}$, $\beta _{1i}=\beta _{1}+\eta _{\beta _{1},i}$ and $\beta _{2i}=\beta _{2}+\eta _{\beta _{2},i}$. Following \citet{Bai2009}, we set $\rho=0.4$, $\psi=0.25$, and $\beta_{1}=3$ and $\beta_{2}=1$. In addition, we set $\eta _{\rho, i}\sim i.i.d.~U\left[ -c_{\rho},+c_{\rho}\right] $, $\eta _{\psi, i}\sim i.i.d.~U\left[ -c_{\psi},+c_{\psi}\right] $ and
\begin{equation*}
\eta _{\beta_{\ell},i}=\left[ {(2c_{\rho})^2}/12\right] ^{1/2} \rho_{\beta}\xi _{\beta _{\ell},i} + \left(1-\rho_{\beta}^{2}\right)^{1/2} \eta _{\rho, i}  ,
\end{equation*}%
where $\xi _{\beta_{\ell},i}$ is the standardised squared idiosyncratic error in $x_{\ell it}$, computed as%
\begin{equation*}
\xi _{\beta _{\ell },i}=\frac{\overline{v_{\ell i}^{2}}-\overline{v_{\ell }^{2}}}{\left[
N^{-1}\sum_{i=1}^{N}\left( \overline{v_{\ell i}^{2}}-\overline{v_{\ell }^{2}}\right)
^{2}\right] ^{1/2}},
\end{equation*}%
with $\overline{v_{\ell i}^{2}}=T^{-1}\sum_{t=1}^{T}v_{\ell it}^{2}$ and $\overline{%
v_{\ell }^{2}}=N^{-1}\sum_{i=1}^{N}\overline{v_{\ell i}^{2}}$, for $\ell =1,2$.
We set $c_{\rho}=0.2$, $c_{\psi}=0.15$, $\rho_{\beta}=0.4$.

\setlength{\belowdisplayskip}{0pt} \setlength{\belowdisplayshortskip}{0pt}
\setlength{\abovedisplayskip}{0pt} \setlength{\abovedisplayshortskip}{0pt}
We define the signal-to-noise ratio (SNR) conditional on the factor structure, the individual-specific effects and the spatial lag, as follows:
\begin{equation}
SNR:=\frac{\overline{\text{var}}\left[  \left(  y_{it}-\varepsilon_{it}\right)
|\mathcal{L}\right]  }{\overline{\text{var}}\left(  \varepsilon_{it}\right)
}=\frac{\left(  \frac{\beta_{1}^{2}+\beta_{2}^{2}}{1-\rho^{2}%
}\right)  \varsigma_{\upsilon}^{2}+\frac{\varsigma_{\varepsilon}^{2}}%
{1-\rho^{2}}-\varsigma_{\varepsilon}^{2}}{\varsigma_{\varepsilon
}^{2}}\text{,}%
\end{equation}
where $\mathcal{L}$ is the information set that contains the factor structure,
the individual-specific effects and the spatial lag\footnote{The reason for conditioning on these
variables is that they influence both the composite error of $y_{it}$, as well as the covariates.}, and $\overline{\text{var}}\left(
\varepsilon_{it}\right)  $ is the overall average of $E\left(  \varepsilon
_{it}^{2}\right)  $ over $i$ and $t$. Solving for $\varsigma_{\upsilon}^{2}$ yields
\begin{equation}
\varsigma_{\upsilon}^{2}= \varsigma_{\varepsilon}^{2} \left[SNR- \frac{\rho^{2}}{1-\rho^{2}}\right]%
\left( \frac{\beta_{1}^{2}+\beta_{2}^{2}}{1-\rho^{2}}\right)^{-1}\text{.}%
\end{equation}
Following \citet{JuodisSarafidis2018ER} and \citet{CuiSarafidisYamagata2023}, we set $\varsigma_{\upsilon}^{2}$ value so that $SNR=4$.

We evaluate the performance of the MGIV estimator of the population mean parameters $\boldsymbol{\theta}$, as defined in Eq. \eqref{eq:MGest}. We distinguish among the following sets of instruments:
\begin{equation}
\mathbf{\widehat{Z}}^{[1]}_{i}=\left(  {\mathbf{M}_{\widehat{\mathbf{F}}}}\mathbf{X}_{i},\quad \mathbf{M}_{\widehat{\mathbf{F}}} \mathbf{M}_{\widehat{\mathbf{F}}_{-1}}\mathbf{X}_{i,-1},\quad \sum_{j=1}^Nw_{ij}\mathbf{M}_{\widehat{\mathbf{F}}}\mathbf{X}_{j}, \quad \sum_{j=1}^Nw_{ij}\mathbf{M}_{\widehat{\mathbf{F}}} \mathbf{M}_{\widehat{\mathbf{F}}_{-1}}\mathbf{X}_{j,-1} \right)  \text{,}
\label{Z_matrix_1}%
\end{equation}
\begin{equation}
\mathbf{\widehat{Z}}^{[2]}_{i}=\left(  {\mathbf{M}_{\widehat{\mathbf{F}}}}\mathbf{X}_{i},\quad \mathbf{M}_{\widehat{\mathbf{F}}} \mathbf{M}_{\widehat{\mathbf{F}}_{-1}}\mathbf{X}_{i,-1},\quad \mathbf{M}_{\widehat{\mathbf{F}}} \mathbf{M}_{\widehat{\mathbf{F}}_{-2}}\mathbf{X}_{i,-2},\quad  \sum_{j=1}^Nw_{ij}\mathbf{M}_{\widehat{\mathbf{F}}}\mathbf{X}_{j} \right)
\label{Z_matrix_2}%
\end{equation}
and
\begin{equation}
\mathbf{\widehat{Z}}^{[3]}_{i}=\left(  {\mathbf{M}_{\widehat{\mathbf{F}}}}\mathbf{X}_{i},\quad \mathbf{M}_{\widehat{\mathbf{F}}} \mathbf{M}_{\widehat{\mathbf{F}}_{-1}}\mathbf{X}_{i,-1},\quad \sum_{j=1}^Nw_{ij}\mathbf{M}_{\widehat{\mathbf{F}}}\mathbf{X}_{j}, \quad \sum_{j=1}^N{w}_{ij}^{[2]}\mathbf{M}_{\widehat{\mathbf{F}}}\mathbf{X}_{j} \right)  \text{,}
\label{Z_matrix_3}%
\end{equation}
where ${w}_{ij}^{[2]}$ denotes the $(i,j)$th entry of $\mathbf{W}^{2}$. Therefore, $\mathbf{\widehat{Z}}^{[1]}_{i}$ is constructed from  $\mathbf{M}_{\widehat{\mathbf{F}}}\mathbf{X}_{i}$ and
$\sum_{j=1}^Nw_{ij}\mathbf{M}_{\widehat{\mathbf{F}}}\mathbf{X}_{j}$, as well as their first lags. In $\mathbf{\widehat{Z}}^{[2]}_{i}$, the lagged spatial instrument is replaced by the second lag of $\mathbf{M}_{\widehat{\mathbf{F}}}\mathbf{X}_{i}$, and thus it uses more information from lagged own variation. Finally,  $\mathbf{\widehat{Z}}^{[3]}_{i}$ replaces the lagged spatial instrument with a ``second-order'' spatial instrument, which corresponds to contemporaneous values of the neighbors of neighbors, i.e.\ the entries of $\mathbf{W}^{2}$. All these matrices are of dimension $T\times 4k$.
Given that the choice of instruments involves a degree of flexibility, we also consider constructing optimal instruments using the Lasso-based method proposed by \citet{BelloniChenChernozhukovHansen2012}. In particular, let $\mathbf{M}_{\widehat{\mathbf{F}}} \mathbf{M}_{\widehat{\mathbf{F}}_{-\kappa}}\mathbf{X}_{i,-\kappa}$ denote the $\kappa$th lag of $\mathbf{M}_{\widehat{\mathbf{F}}} \mathbf{X}_{i}$, for $\kappa=1,\dots,K_{1}$. Analogously, we include the spatially-weighted lags $\sum_{j=1}^Nw_{ij}\mathbf{M}_{\widehat{\mathbf{F}}} \mathbf{M}_{\widehat{\mathbf{F}}_{-\kappa}}\mathbf{X}_{j,-\kappa}$, for $\kappa=1,\dots,K_{2}$, as well as second- and higher-order lags $\sum_{j=1}^N{w}_{ij}^{[\kappa+1]}\mathbf{M}_{\widehat{\mathbf{F}}} \mathbf{X}_{j}$, for $\kappa=1,\dots,K_{3}$. We set $K_{1}=5$, $K_{2}=5$ and $K_{3}=3$, and formulate a matrix $\mathbf{\widehat{Z}}^{+}_{i}$, which collects all these instruments. The dimension of $\mathbf{\widehat{Z}}^{+}_{i}$ is $T \times K_{+}$, where $K_{+}=13k=26$.
Starting from this enlarged candidate set, we apply the Lasso procedure separately to each endogenous regressor, $\mathbf{y}_{i,-1}$ and $\sum_{j=1}^{N} w_{ij} \mathbf{y}_{j}$. Six instruments are treated as fixed and always included: $\mathbf{M}_{\widehat{\mathbf{F}}}\mathbf{X}_{i}$, $\mathbf{M}_{\widehat{\mathbf{F}}}\mathbf{M}_{\widehat{\mathbf{F}}_{-1}}\mathbf{X}_{i,-1}$, and $\sum_{j=1}^N w_{ij}\mathbf{M}_{\widehat{\mathbf{F}}}\mathbf{X}_{j}$. The remaining instruments, contained in $\mathbf{\widehat{Z}}^{+}_{i}$, are subject to a data-driven penalization that shrinks coefficients toward zero. Instruments with predictive content are thus retained, and the others are discarded. The final instrument matrix, denoted $\mathbf{\widehat{Z}}^{[L]}_{i}$, consists of the fixed instruments together with the union of the sets selected by the Lasso for the two endogenous regressors.

In addition to MGIV, we also consider the pooled two-stage IV (2SIV) estimator developed by \citet{CuiSarafidisYamagata2023}, the QMLE estimator developed by \citet{ShiLee2017}, and the CCEX-IV estimator of \citet{ChenEtAl2022}. 2SIV and QMLE are designed for spatial dynamic panel data models with \textit{homogeneous} parameters, whereas CCEX-IV is designed for spatial \textit{static} panel data models with heterogeneous parameters. Note that although 2SIV is inconsistent under coefficient heterogeneity, as is QMLE, it remains useful because the associated $J$-test has power against violations of slope-parameter homogeneity.

In terms of the sample size, we consider three cases. Case I sets $N=25\tau$ and $T=100\tau$ for $\tau=1,2,4$. This implies that while $N$ and $T$ increase by multiples of 2, the ratio of $N$ to $T$ remains equal to $1/4$ in all circumstances. Case II sets $N=100\tau$ and $T=25\tau$ for $\tau=1,2,4$. Therefore, $N/T=4$, as both $N$ and $T$ grow. Finally, Case III sets $N=T=50\tau $ for $\tau=1,2,4$. These choices allow us to consider different combinations of $(N,T)$ in relatively small and large sample sizes. Studying the performance of the estimators under these circumstances is important, as in many applications $N$ can be significantly larger than $T$ or vice versa. This is exemplified by our empirical application discussed in the following section. %Studying the performance of the estimator under these circumstances is valuable, as in many applications $N$ is often comparable to, if not significantly larger than, $T$. This is exemplified by our empirical application discussed in the following section.

All results are obtained based on 2,000 replications, and all tests are conducted at the 5\% significance level. For the power of the ``t-test'' on a population mean parameter, we specify $H_0:\rho=\rho^{0}+0.1$ (or $H_0:\psi=\psi^{0}+0.1$, or $H_0:\beta_{\ell}=\beta_{\ell}^{0}+0.1$ for $\ell=1,2$) against the two-sided alternative, where ${\rho^0,\psi^0,\beta_1^0,\beta_2^0}$ denote the true values of the population mean parameters.

\subsection{Results}

%``ARB'' denotes absolute relative bias, defined as $ARB \equiv \left(|\widehat{\theta}_{\ell}-\theta_{\ell}|/\theta_{\ell}\right)100$, where $\theta_{\ell}$ denotes the $\ell$th entry of $\boldsymbol{\theta}$.

Table \ref{tab:sim_results_rho_psi_pi3/4_MGIV} reports MGIV results for $\rho=0.4$ and $\psi=0.25$ when $\pi_u=3/4$ in terms of the estimation Mean, RMSE, ARB (Absolute Relative Bias), as well as empirical size (nominal level is $5\%$) and size-corrected power, which is computed based on the 2.5\% and 97.5\% quantiles of the empirical distribution of the t-ratio under the null hypothesis.\footnote{In the Appendix, Tables~\ref{tab:sim_results_rho_psi_pi1/4_MGIV}--\ref{tab:sim_results_beta2_MGIV} report results for $\rho = 0.4$ and $\psi = 0.25$ when $\pi_u = 1/4$, together with the complete set of results for $\beta_{1} = 3$ and $\beta_{2} = 1$. Tables~\ref{tab:sim_results_rho_IVi}--\ref{tab:sim_results_beta2_IVi} present results averaged across the individual-specific IV estimates, while Tables~\ref{tab:sim_results_rho_2SIV}--\ref{tab:sim_results_beta2_2SIV} report results for 2SIV. Results for QMLE and CCEX-IV are provided in Tables~\ref{tab:sim_results_QMLE}--\ref{tab:sim_results_CCEXIV}. For MGIV and 2SIV, the number of factors is estimated in each replication using the eigenvalue ratio criterion of \citet{AhnHorenstein2013}, applied to $(NT)^{-1}\sum_{i=1}^N \mathbf{X}_i \mathbf{X}_i'$. For computational efficiency, QMLE is implemented using the true number of factors, while CCEX-IV does not require explicit factor estimation.} We also report results on the power of the over-identifying restrictions test statistic, J-test, based on the corresponding (pooled) 2SIV estimator in \citet{CuiSarafidisYamagata2023}. This serves as a diagnostic for slope homogeneity, since rejection of the null indicates that the homogeneous specification is not supported by the data. ARB is defined as $ARB:= \left(|\widehat{\theta}_{\ell}-\theta_{\ell}|/\theta_{\ell}\right)\times100$, where $\theta_{\ell}$ denotes the $\ell$th entry of $\boldsymbol{\theta}=\left(\psi, \rho, \boldsymbol{\beta}' \right)'$. In the tables, ``MGIV$^{[\kappa]}$'' denotes the Mean Group IV estimator of $\boldsymbol{\theta}$ as defined in Eq. \eqref{eq:MGest} with the corresponding matrix of instruments given by $\mathbf{\widehat{Z}}^{[\kappa]}_{i}$ for $\kappa \in \{ 1,2,3,L\}$.

All MGIV estimators considered exhibit small bias across designs, with RMSE declining steadily as the sample grows. Empirical size is close to the nominal $5\%$ even when $T$ is modest, and quickly approaches $5\%$ as both $N$ and $T$ increase. The only exception is MGIV$^{[3]}$ in case II (with $N/T=4$), which displays substantial size distortions and relatively large ARB values.
Among the MGIV estimators with fixed instrument sets, MGIV$^{[2]}$, which relies more on own time-lag instruments than on spatial ones, consistently outperforms MGIV$^{[1]}$ and MGIV$^{[3]}$ in terms of ARB, size, and RMSE. A key distinction lies in the J-test's ability to detect slope-parameter heterogeneity: estimators that depend heavily on spatial instruments exhibit limited power, whereas MGIV$^{[2]}$ achieves satisfactory power that increases sharply with sample size. Hence, deviations from slope homogeneity are more easily detected when the instrument set expands along the time dimension.
The Lasso-based variant MGIV$^{[L]}$ performs similarly to MGIV$^{[2]}$ across all metrics. When selecting from the enlarged instrument set $\mathbf{\widehat{Z}}^{+}_{i}$, MGIV$^{[L]}$ tends to include additional own-lag instruments instead of spatial ones, yielding slightly stronger J-test power while maintaining comparable size and RMSE.

When $\pi_u=1/4$ (see Table~\ref{tab:sim_results_rho_psi_pi1/4_MGIV}, Appendix), results are broadly similar to those for $\pi_u=3/4$. The main difference is that the J-test has higher power, as a larger share of the error variation is driven by common factors rather than idiosyncratic noise. Similar patterns hold for the parameters $\beta_{1}$ and $\beta_{2}$ (Tables \ref{tab:sim_results_beta1_MGIV}-\ref{tab:sim_results_beta2_MGIV} in the Appendix). In a nutshell, MGIV performs well across all designs, exhibiting negligible bias, steadily declining RMSE with larger samples, correct empirical size, and robustness to variations in $\pi_{u}$. Among the fixed-instrument estimators, MGIV$^{[2]}$ dominates, while the Lasso-based MGIV$^{[L]}$ offers marginal yet consistent gains in J-test power.

Results for pooled estimators (reported in the Appendix) indicate that 2SIV, QMLE, and CCEX-IV exhibit substantial relative bias and pronounced size distortions, with a tendency to over-reject the null hypothesis in the associated $t$-tests.

%\begin{landscape}
%\begin{table}[!htbp]
\begin{sidewaystable}
\centering
\begin{threeparttable}
\footnotesize
\footscriptsize
\setlength{\tabcolsep}{2.7pt} % tighten if needed
\renewcommand{\arraystretch}{1.15}
\caption{Mean Group IV results for $\pi_{u}=3/4$}
\label{tab:sim_results_rho_psi_pi3/4_MGIV}
\begin{tabular}{l*{27}{c}}
%\toprule
\addlinespace[2pt]
\multicolumn{28}{c}{\small \textbf{Panel A ($\rho=0.4$)}}\\

& \multicolumn{6}{c}{\small \textbf{MGIV$^{[1]}$}} &  & \multicolumn{6}{c}{\small \textbf{MGIV$^{[2]}$}} &  & \multicolumn{6}{c}{\small \textbf{MGIV$^{[3]}$}} &  & \multicolumn{6}{c}{\small \textbf{MGIV$^{[L]}$}} \\
%\addlinespace[2pt]
\cmidrule(lr){2-7} \cmidrule(lr){9-14} \cmidrule(lr){16-21} \cmidrule(lr){23-28}

\multicolumn{28}{c}{\small \textbf{Case I:} $N=25\tau$, $T=100\tau$}\\

$\tau$ & Mean & RMSE & ARB & Size & Power & J-test &  & Mean & RMSE & ARB & Size & Power & J-test &  & Mean & RMSE & ARB & Size & Power & J-test &  & Mean & RMSE & ARB & Size & Power & J-test \\
\cmidrule(lr){2-7} \cmidrule(lr){9-14} \cmidrule(lr){16-21} \cmidrule(lr){23-28}

1 & 0.401 & 0.027 & 0.219 & 0.061 & 0.940 & 0.091 &  & 0.400 & 0.027 & 0.090 & 0.058 & 0.937 & 0.121 &  & 0.407 & 0.028 & 1.785 & 0.072 & 0.967 & 0.098 &  & 0.400 & 0.026 & 0.049 & 0.059 & 0.938 & 0.217 \\
2 & 0.402 & 0.018 & 0.363 & 0.063 & 1.000 & 0.068 &  & 0.401 & 0.018 & 0.130 & 0.058 & 1.000 & 0.235 &  & 0.405 & 0.019 & 1.241 & 0.077 & 1.000 & 0.075 &  & 0.401 & 0.018 & 0.118 & 0.058 & 1.000 & 0.321 \\
4 & 0.401 & 0.012 & 0.268 & 0.054 & 1.000 & 0.091 &  & 0.401 & 0.012 & 0.153 & 0.050 & 1.000 & 0.753 &  & 0.403 & 0.012 & 0.760 & 0.059 & 1.000 & 0.091 &  & 0.400 & 0.012 & 0.075 & 0.050 & 0.999 & 0.756 \\

\multicolumn{28}{c}{\small \textbf{Case II:} $N=100\tau$, $T=25\tau$}\\

$\tau$ & Mean & RMSE & ARB & Size & Power & J-test &  & Mean & RMSE & ARB & Size & Power & J-test &  & Mean & RMSE & ARB & Size & Power & J-test &  & Mean & RMSE & ARB & Size & Power & J-test \\
\cmidrule(lr){2-7} \cmidrule(lr){9-14} \cmidrule(lr){16-21} \cmidrule(lr){23-28}

1 & 0.397 & 0.020 & 0.685 & 0.064 & 0.997 & 0.063 &  & 0.397 & 0.020 & 0.823 & 0.060 & 0.997 & 0.110 &  & 0.403 & 0.020 & 0.837 & 0.072 & 0.998 & 0.068 &  & 0.401 & 0.022 & 0.167 & 0.049 & 0.981 & 0.058 \\
2 & 0.404 & 0.012 & 0.910 & 0.077 & 1.000 & 0.061 &  & 0.402 & 0.011 & 0.470 & 0.069 & 1.000 & 0.231 &  & 0.411 & 0.016 & 2.737 & 0.203 & 1.000 & 0.059 &  & 0.402 & 0.011 & 0.481 & 0.067 & 1.000 & 0.262 \\
4 & 0.403 & 0.007 & 0.725 & 0.076 & 1.000 & 0.074 &  & 0.402 & 0.007 & 0.377 & 0.058 & 1.000 & 0.781 &  & 0.409 & 0.011 & 2.153 & 0.249 & 1.000 & 0.077 &  & 0.403 & 0.007 & 0.625 & 0.057 & 0.999 & 0.790 \\

\multicolumn{28}{c}{\small \textbf{Case III:} $N=50\tau$, $T=50\tau$}\\

$\tau$ & Mean & RMSE & ARB & Size & Power & J-test &  & Mean & RMSE & ARB & Size & Power & J-test &  & Mean & RMSE & ARB & Size & Power & J-test &  & Mean & RMSE & ARB & Size & Power & J-test \\
\cmidrule(lr){2-7} \cmidrule(lr){9-14} \cmidrule(lr){16-21} \cmidrule(lr){23-28}

1 & 0.404 & 0.022 & 0.915 & 0.062 & 0.993 & 0.061 &  & 0.402 & 0.022 & 0.414 & 0.058 & 0.993 & 0.113 &  & 0.411 & 0.024 & 2.707 & 0.094 & 0.995 & 0.077 &  & 0.402 & 0.022 & 0.494 & 0.055 & 0.993 & 0.139 \\
2 & 0.403 & 0.014 & 0.680 & 0.058 & 1.000 & 0.055 &  & 0.401 & 0.013 & 0.293 & 0.054 & 1.000 & 0.233 &  & 0.408 & 0.016 & 2.087 & 0.099 & 1.000 & 0.063 &  & 0.401 & 0.013 & 0.285 & 0.055 & 1.000 & 0.282 \\
4 & 0.402 & 0.009 & 0.385 & 0.062 & 1.000 & 0.088 &  & 0.401 & 0.009 & 0.140 & 0.053 & 1.000 & 0.769 &  & 0.405 & 0.010 & 1.258 & 0.091 & 1.000 & 0.086 &  & 0.402 & 0.008 & 0.400 & 0.052 & 0.999 & 0.775 \\
\midrule
% ===== Panel B =====
\multicolumn{28}{c}{\small \textbf{Panel B ($\psi=0.25$)}}\\

& \multicolumn{6}{c}{\small \textbf{MGIV$^{[1]}$}} &  & \multicolumn{6}{c}{\small \textbf{MGIV$^{[2]}$}} &  & \multicolumn{6}{c}{\small \textbf{MGIV$^{[3]}$}} &  & \multicolumn{6}{c}{\small \textbf{MGIV$^{[L]}$}} \\
%\addlinespace[2pt]
\cmidrule(lr){2-7} \cmidrule(lr){9-14} \cmidrule(lr){16-21} \cmidrule(lr){23-28}

\multicolumn{28}{c}{\small \textbf{Case I:} $N=25\tau$, $T=100\tau$}\\

$\tau$ & Mean & RMSE & ARB & Size & Power & J-test &  & Mean & RMSE & ARB & Size & Power & J-test &  & Mean & RMSE & ARB & Size & Power & J-test &  & Mean & RMSE & ARB & Size & Power & J-test \\
\cmidrule(lr){2-7} \cmidrule(lr){9-14} \cmidrule(lr){16-21} \cmidrule(lr){23-28}

1 & 0.251 & 0.028 & 0.532 & 0.067 & 0.945 & 0.091 &  & 0.252 & 0.028 & 0.784 & 0.071 & 0.935 & 0.121 &  & 0.250 & 0.027 & 0.101 & 0.071 & 0.941 & 0.098 &  & 0.252 & 0.028 & 0.876 & 0.069 & 0.935 & 0.217 \\
2 & 0.250 & 0.016 & 0.038 & 0.061 & 1.000 & 0.068 &  & 0.250 & 0.016 & 0.092 & 0.057 & 1.000 & 0.235 &  & 0.250 & 0.015 & 0.197 & 0.056 & 1.000 & 0.075 &  & 0.250 & 0.016 & 0.092 & 0.057 & 1.000 & 0.321 \\
4 & 0.250 & 0.010 & 0.141 & 0.052 & 1.000 & 0.091 &  & 0.250 & 0.010 & 0.106 & 0.053 & 1.000 & 0.753 &  & 0.249 & 0.010 & 0.240 & 0.052 & 1.000 & 0.091 &  & 0.250 & 0.010 & 0.080 & 0.052 & 1.000 & 0.756 \\

\multicolumn{28}{c}{\small \textbf{Case II:} $N=100\tau$, $T=25\tau$}\\

$\tau$ & Mean & RMSE & ARB & Size & Power & J-test &  & Mean & RMSE & ARB & Size & Power & J-test &  & Mean & RMSE & ARB & Size & Power & J-test &  & Mean & RMSE & ARB & Size & Power & J-test \\
\cmidrule(lr){2-7} \cmidrule(lr){9-14} \cmidrule(lr){16-21} \cmidrule(lr){23-28}

1 & 0.254 & 0.030 & 1.680 & 0.064 & 0.926 & 0.063 &  & 0.253 & 0.029 & 1.236 & 0.053 & 0.931 & 0.110 &  & 0.252 & 0.028 & 0.893 & 0.053 & 0.939 & 0.068 &  & 0.252 & 0.035 & 0.758 & 0.053 & 0.857 & 0.058 \\
2 & 0.250 & 0.013 & 0.012 & 0.057 & 1.000 & 0.061 &  & 0.250 & 0.013 & 0.062 & 0.056 & 1.000 & 0.231 &  & 0.249 & 0.013 & 0.583 & 0.053 & 1.000 & 0.059 &  & 0.250 & 0.013 & 0.070 & 0.056 & 1.000 & 0.262 \\
4 & 0.250 & 0.007 & 0.023 & 0.053 & 1.000 & 0.074 &  & 0.250 & 0.007 & 0.087 & 0.055 & 1.000 & 0.781 &  & 0.249 & 0.007 & 0.387 & 0.057 & 1.000 & 0.077 &  & 0.250 & 0.007 & 0.120 & 0.055 & 0.999 & 0.790 \\

\multicolumn{28}{c}{\small \textbf{Case III:} $N=50\tau$, $T=50\tau$}\\

$\tau$ & Mean & RMSE & ARB & Size & Power & J-test &  & Mean & RMSE & ARB & Size & Power & J-test &  & Mean & RMSE & ARB & Size & Power & J-test &  & Mean & RMSE & ARB & Size & Power & J-test \\
\cmidrule(lr){2-7} \cmidrule(lr){9-14} \cmidrule(lr){16-21} \cmidrule(lr){23-28}

1 & 0.253 & 0.027 & 1.052 & 0.053 & 0.958 & 0.061 &  & 0.253 & 0.026 & 1.143 & 0.051 & 0.961 & 0.113 &  & 0.251 & 0.026 & 0.510 & 0.056 & 0.961 & 0.077 &  & 0.253 & 0.027 & 1.017 & 0.051 & 0.958 & 0.139 \\
2 & 0.250 & 0.013 & 0.068 & 0.052 & 1.000 & 0.055 &  & 0.250 & 0.014 & 0.125 & 0.044 & 1.000 & 0.233 &  & 0.249 & 0.013 & 0.269 & 0.051 & 1.000 & 0.063 &  & 0.250 & 0.014 & 0.118 & 0.045 & 1.000 & 0.282 \\
4 & 0.250 & 0.008 & 0.031 & 0.045 & 1.000 & 0.088 &  & 0.250 & 0.008 & 0.001 & 0.054 & 1.000 & 0.769 &  & 0.249 & 0.008 & 0.268 & 0.051 & 1.000 & 0.086 &  & 0.250 & 0.008 & 0.040 & 0.054 & 0.999 & 0.775 \\
% Row 36: (optional) final spacer or notes row
\multicolumn{28}{c}{}\\
\bottomrule
\end{tabular}
\begin{tablenotes}[para,flushleft]
\footnotesize
\textbf{Notes:} MGIV$^{[\kappa]}$ refers to the Mean Group IV estimator in Eq. \eqref{eq:MGest} based on the instrument matrix $\mathbf{\widehat{Z}}^{[\kappa]}_{i}$, for $\kappa \in
\{1,2,3,L \}$, as explained in Section \ref{subsec:Design}.
\end{tablenotes}
\end{threeparttable}
\end{sidewaystable}
%\end{table}
%\end{landscape}

\begin{figure}[h!]
    \centering
    \includegraphics[scale=0.75]{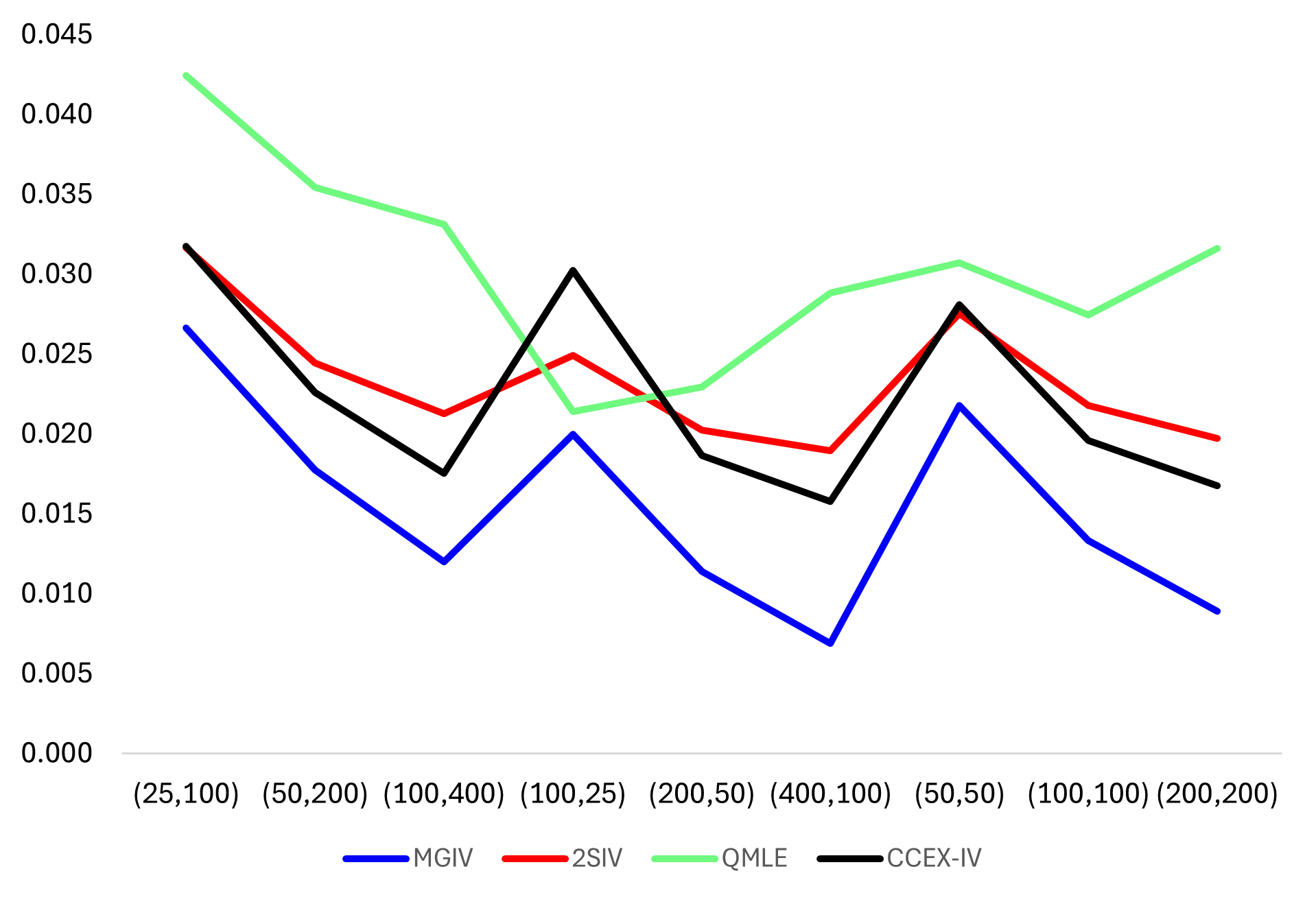}
    \caption{RMSE performance for $\rho=0.4$ across different (N,T) combinations ($\pi_u=3/4$)}
    \label{fig:rmse}
\end{figure}

Figure \ref{fig:rmse} illustrates the performance of MGIV, 2SIV, QMLE, and CCEX-IV in terms of RMSE across different $(N,T)$ combinations when $\rho = 0.4$ and $\pi_u = 3/4$. The MGIV estimator consistently achieves the lowest RMSE across all designs, with accuracy improving as the sample size increases. In contrast, 2SIV, CCEX-IV, and especially QMLE exhibit substantially higher RMSE, reflecting their sensitivity to slope heterogeneity (for 2SIV and QMLE) and dynamic structure (CCEX-IV).

\begin{figure}[h!]
    \centering
    \includegraphics[scale=0.75]{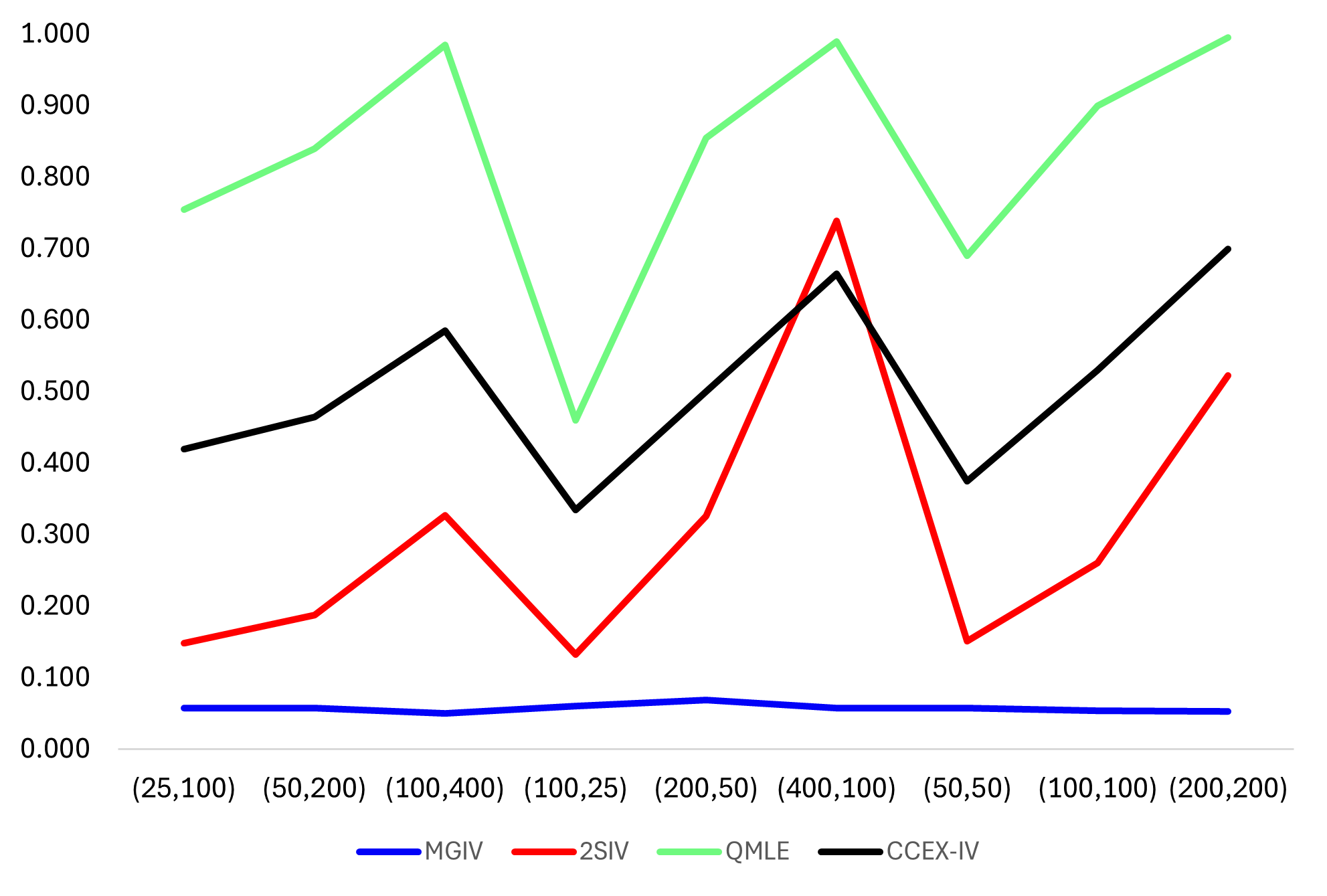}
    \caption{Size performance for $\rho=0.4$ across different (N,T) combinations ($\pi_u=3/4$)}
    \label{fig:size}
\end{figure}

Figure \ref{fig:size} reports the empirical size of the corresponding $t$-tests. MGIV delivers size close to the nominal level across all $(N,T)$ configurations, even in small samples. By contrast, 2SIV, QMLE and CCEX-IV exhibit severe size distortions, with rejection frequencies far exceeding the nominal level. These findings highlight the importance of accounting for slope heterogeneity and dynamic structure in achieving reliable inference, particularly when both features are present.

\section{Illustration: Empirical Spatial Growth Model}\label{sec-empirical}

This section applies the proposed methodology to estimate a heterogeneous spatial growth model for 266 EU regions over 2001-2018, focusing on the magnitude and channels of regional growth spillovers, a central theme in regional economic research.

\subsection{Model Specification and Estimation}

We study the following spatial Durbin model of regional growth:
\begin{equation}
\Delta y_{it} = \delta_{i} y_{i,t-1} + \psi_{0,i} \sum_{j=1}^{N} w_{ij} \Delta y_{jt} + \psi_{i} \sum_{j=1}^{N} w_{ij} y_{j,t-1} + \sum_{\ell=1}^{k} \beta_{\ell,i} x_{\ell it} + \sum_{\ell=1}^{k} \gamma_{\ell,i} \sum_{j=1}^{N} w_{ij} x_{\ell jt} + u_{it}\text{,}
\label{eq:empirical}
\end{equation}
where $i$ denotes an EU NUTS-2 region ($i=1,\ldots,266$) and $t$ a year ($t=1,\ldots,18$).\footnote{NUTS-2 regions provide harmonized data on GDP and other macroeconomic indicators, ensuring comparability across space and time. More granular NUTS-3 data are often incomplete or inconsistent, making NUTS-2 the preferred level of aggregation.}
The composite error term is specified as
\begin{equation}
u_{it} = \alpha_{i} + \boldsymbol{\lambda}_{i}^{\prime}\mathbf{f}_{t} + \boldsymbol{\varphi}_{i}^{\prime}\mathbf{g}_{t} + \varepsilon_{it}\text{.}
\label{eq:empirical_u}
\end{equation}

Conditional convergence arises when $\theta_{i}(=\delta_{i}+\psi_{i})<0$, whereas $\theta_{i}>0$ indicates conditional divergence. The case $\theta_{i}=0$ corresponds to the absence of long-run adjustment, so that the model is expressed entirely in first differences, with no spatial cointegration relationship binding regional growth paths \citep{YuEtAl2012}.

The dependent variable, $y_{it}=\ln(gdp_{it})$, denotes the natural logarithm of real GDP per capita in PPP-adjusted constant prices. Its lag, $y_{i,t-1}$, captures the initial income level at the start of each growth spell, $\Delta y_{it}$. The explanatory variables include:
(i) $x_{1it}=\ln(inv_{it})$, the investment rate (share of GDP), capturing capital accumulation;
(ii) $x_{2it}=\ln(n_{it})=\ln(p_{it}+g+q)$, where $p_{it}$ is population growth, $g$ technological progress, and $q$ depreciation, with $g+q=0.05$ as in \citet{Islam1995};
(iii) $x_{3it}=\ln(educ_{it})$, the tertiary educational attainment rate, reflecting human capital; and
(iv) $x_{4it}=\ln(sci\&tech_{it})$, the share of employment in science and technology, representing innovation intensity.
The inclusion of human capital and innovation proxies, $x_{3it}$ and $x_{4it}$, follows endogenous growth theory, which treats these as internal drivers of long-run growth, in contrast to the exogeneity assumption in Solow's neoclassical framework.

The spatial weighting matrix is specified as:
\begin{equation}
w_{ii} = 0; \quad w_{ij} = \frac{exp\left(-\zeta d_{ij} \right)}{\sum_{j=1}^{N} exp\left(-\zeta d_{ij} \right)}; \quad j \neq i \text{,}\label{W_empirical}%
\end{equation}
where $d_{ij}$ represents the great-circle distance between regions $i$ and $j$,\footnote{This is measured in kilometers using the latitude and longitude coordinates of regional centroids.} and $\zeta$ is the distance decay parameter. Following \citet{ErturKoch2007}, $\zeta$ is set to $0.02$, which prioritises interactions among proximate regions. Alternative specifications for $\mathbf{W}$, such as inverse squared distances, are explored in Appendix C. The matrix $\mathbf{W}$ is then row-normalized so that each row sums to 1.

The set of instruments used is
\begin{equation}
\mathbf{\widehat{Z}}^{1}_{i}=\left(  {\mathbf{M}_{\widehat{\mathbf{F}}}}\mathbf{X}_{i}, \; \mathbf{M}_{\widehat{\mathbf{F}}} \mathbf{M}_{\widehat{\mathbf{F}}_{-1}}\mathbf{X}_{i,-1}, \; \mathbf{M}_{\widehat{\mathbf{F}}} \mathbf{M}_{\widehat{\mathbf{F}}_{-2}}\mathbf{X}_{i,-2}, \; \mathbf{M}_{\widehat{\mathbf{F}}} \sum_{j=1}^Nw_{ij}\mathbf{X}_{j}, \;  \mathbf{M}_{\widehat{\mathbf{F}}} \mathbf{M}_{\widehat{\mathbf{F}}_{-1}}\sum_{j=1}^Nw_{ij}\mathbf{X}_{j,-1} \right)  \text{,}
\label{Z_matrix_empirical}%
\end{equation}
where $\mathbf{X}_{i}=(\mathbf{x}_{i1},\cdots,\mathbf{x}_{iT})^{\prime }$ with $\mathbf{x}_{it}=\left(x_{1it},\dots,x_{4it}\right)^{\prime}$.
Thus, a total of 20 instruments is employed. In the baseline specification, this instrument set treats all covariates, including the investment rate, as strictly exogenous. While this assumption may appear restrictive, it should be interpreted as applying to the idiosyncratic component of the error term, rather than to the full disturbance that also includes latent common factors. In particular, investment decisions may depend on past output and other aggregate conditions through these common factors, which are explicitly controlled for in the estimation. Thus, the assumption does not preclude forward-looking behavior or feedback from macroeconomic conditions, but rules out correlation between investment and the idiosyncratic shocks. In Table \ref{tab:Robustness} in the Appendix, we also report robustness results that relax this assumption by treating investment as weakly exogenous (see Column [6]).

\subsection{Results}\label{subsec-empirical_results}

Table \ref{tab:MAIN} reports results for the model in Eq.~\eqref{eq:empirical} together with several nested specifications obtained by progressively relaxing parameter restrictions. Each column corresponds to a distinct model, illustrating how these restrictions affect parameter estimates and their interpretation.
Column [6] presents the most general specification, defined in Eq.~\eqref{eq:empirical}, which includes  spatial spillovers, interactive fixed effects, and slope heterogeneity, and is estimated using the Mean Group estimator in Eq.~\eqref{eq:MGest}.
At the other extreme, Column [1] reports the most restricted specification, imposing $u_{it}=\alpha_{i}+\tau_{t}+\varepsilon_{it}$, $\psi_{0,i}=\psi_{i}=\gamma_{\ell,i}=0$ for all $i$ and $\ell$, $\rho_{i}=\rho$, and $\beta_{\ell,i}=\beta_{\ell}$, thereby excluding spatial spillovers, enforcing slope homogeneity, and adopting a two-way error components structure. Columns [2]-[5] introduce intermediate relaxations of these constraints.
Specifically, Column [2] augments Column [1] by replacing the additive error structure with interactive fixed effects. Column [3] relaxes slope homogeneity while retaining the absence of spatial and interactive effects.
Column [4] introduces spatial spillovers to the baseline model in Column [1], maintaining slope homogeneity and a two-way error structure. Column [5] further extends this specification by incorporating interactive fixed effects, while still imposing parameter homogeneity.\footnote{Results in Columns [1]-[3] are obtained using the Stata command \texttt{xtivdfreg} \citep{KripfganzSarafidis2021}, and those in Columns [4]-[6] using the wrapper command \texttt{spxtivdfreg} \citep{KripfganzSarafidis2025}.}

We start with the most restrictive specification, Column [1]. The coefficient on $\ln\left(gdp_{i,t-1}\right)$ is negative and highly significant, supporting the hypothesis of conditional convergence in the regions. This finding aligns with the Solow growth model, which predicts that regions with higher initial GDP per capita grow more slowly as they approach their steady-state income levels. The coefficient on $\ln\left(inv_{it}\right)$ is positive and significant, indicating that higher investment rates contribute positively to regional growth. Conversely, the coefficient on $\ln\left(n_{it}\right)$ is negative and significant, consistent with neoclassical growth theory's prediction that higher population growth reduces per capita output growth by diluting capital per worker. The coefficients on both educational attainment and share of employment in science and technology are close to zero and statistically insignificant.
Allowing for interactive effects in the error process, as in Column [2], and slope-parameter heterogeneity, as in Column [3], leads to a more negative coefficient on $\ln \left(gdp_{i,t-1}\right)$, which implies faster conditional convergence. This suggests that accounting for unobserved common shocks/trends or parameter heterogeneity across regions is important.

Turning to specifications that incorporate spatial spillovers (i.e., Columns [4] and [5]), the estimated coefficients on $\sum_{j=1}^Nw_{ij}\Delta \ln \left(gdp_{jt}\right)$ are close to the spatial unit-root boundary, with confidence intervals that include one. Taken at face value, these estimates imply extremely persistent spatial feedback, with shocks to neighboring regions' growth rates dissipating only very slowly across the spatial network. %suggest the presence of a spatial unit root, as the confidence intervals include one in both cases. Taken at face value, this suggests that shocks to neighboring regions' growth rates persist indefinitely and propagate throughout the spatial network without decay, which is counterintuitive.
At the same time, the coefficient of $\ln\left(gdp_{i,t-1}\right)$ is close to zero, while the coefficient on $\sum_{j=1}^N w_{ij} \ln(gdp_{j,t-1})$ is positive, so that their combined effect exceeds zero. This points to the absence of conditional convergence under these specifications. Combined with rejection of the null of valid instruments based on the J-test (using the homogeneous 2SIV estimator of \citet{CuiSarafidisYamagata2023}), this provides evidence against slope homogeneity and suggests potential model misspecification.

Importantly, the most general specification presented in Column [6] yields a markedly different picture of growth dynamics. Allowing for regional heterogeneity reduces the estimated spatial feedback from $0.896$--$0.997$ in Columns [4]--[5] to $0.628$, while also restoring conditional convergence: the sum of the coefficients on $\ln(gdp_{i,t-1})$ and $\sum_{j=1}^N w_{ij}\ln(gdp_{j,t-1})$ is $-0.282$ and is statistically significant based on a one-sided test against the alternative that the sum is negative ($t=-4.87$, $p<0.001$).
This suggests that imposing slope homogeneity may be driving the apparent near-unit-root behavior in Columns [4] and [5], a particularly relevant concern given the inferential difficulties associated with strongly persistent and local-to-unity spatial processes highlighted by \citet{MullerWatson2024}. The coefficient on the investment rate of a region's neighbors is positive and statistically significant, highlighting substantial spillovers of regional investment activities. Notably, a similar pattern emerges for the coefficient on tertiary educational attainment, suggesting that higher education levels in neighboring regions contribute positively to a region's economic growth through knowledge diffusion and human capital externalities.

\begin{table}[htbp]
\centering
\caption{{Results for several empirical growth model specifications} \index{Results for Empirical Growth Model Table 5.1}}
\label{tab:MAIN}
\begin{tabular}{lcccccc}
%\multicolumn{7}{c}{\textbf{Table 5.1: Results for Empirical Growth Model.}}                                                                                                                                   %\\ \hline
\hline
                        & [1]          & [2]          & [3]          & [4]          & [5]          & [6]          \\
\hline
$\ln \left(gdp_{i,t-1}\right)$ & -0.137***    & -0.267***    & -0.216***    & -0.011       & -0.010       & -0.473***    \\
                         & (0.032)      & (0.034)      & (0.017)      & (0.034)      & (0.042)      & (0.028)      \\

$\ln\left(inv_{it}\right)$ & 0.035***     & 0.013        & 0.016**      & 0.003        & -0.004       & 0.027**      \\
                         & (0.006)      & (0.008)      & (0.007)      & (0.004)      & (0.005)      & (0.013)      \\

$\ln\left(n_{it}\right)$   & -0.629***    & -0.636***    & -0.271       & -0.617***    & -0.679***    & -0.582**     \\
                         & (0.153)      & (0.122)      & (0.249)      & (0.125)      & (0.123)      & (0.281)      \\

$\ln\left(educ_{it}\right)$ & -0.006       & 0.013        & 0.002        & -0.004       & 0.002        & 0.007        \\
                         & (0.007)      & (0.017)      & (0.015)      & (0.006)      & (0.008)      & (0.026)      \\

$\ln\left(sci \& tech_{it}\right)$ & -0.006       & -0.013       & -0.034***    & 0.008        & -0.009       & -0.003       \\
                         & (0.008)      & (0.011)      & (0.011)      & (0.006)      & (0.008)      & (0.016)      \\
\hline
\multicolumn{7}{l}{ } \\
$\sum_{j=1}^Nw_{ij}\Delta \ln \left(gdp_{jt}\right)$  &              &              &              & 0.896***     & 0.997***     & 0.628***     \\
                                &              &              &              & (0.145)      & (0.048)      & (0.033)      \\

$\sum_{j=1}^Nw_{ij} \ln\left(gdp_{j,t-1}\right)$       &              &              &              & 0.075        & 0.039        & 0.192***     \\
                                &              &              &              & (0.081)      & (0.050)      & (0.052)      \\

$\sum_{j=1}^Nw_{ij}\ln\left(inv_{jt}\right)$        &              &              &              & 0.001        & 0.003        & 0.071***     \\
                                &              &              &              & (0.012)      & (0.011)      & (0.021)      \\

$\sum_{j=1}^Nw_{ij}\ln\left(n_{jt}\right)$          &              &              &              & 0.334*       & 0.568***     & 0.643        \\
                                &              &              &              & (0.198)      & (0.135)      & (0.549)      \\

$\sum_{j=1}^Nw_{ij}\ln\left(educ_{jt}\right)$       &              &              &              & -0.000       & 0.007        & 0.150***     \\
                                &              &              &              & (0.011)      & (0.017)      & (0.035)      \\

$\sum_{j=1}^Nw_{ij}\ln\left(sci \& tech_{jt}\right)$ &              &              &              & -0.017*      & 0.004        & 0.022        \\
                                &              &              &              & (0.009)      & (0.011)      & (0.050)      \\
\hline
$N_g$                  & 266          & 266          & 266          & 266          & 266          & 266          \\
$\chi^2_J$             & 29.949       & 39.201       &              & 12.966       & 16.002       &              \\
$p_J$                  & 0.000        & 0.000        &              & 0.024        & 0.067        &              \\
$r_1$                  & 0            & 1            & 0            & 0            & 1            & 1            \\
$r_1+r_2$                  & 0            & 2            &              & 0            & 1            &              \\
\hline
\end{tabular}\smallskip

\begin{minipage}{\textwidth}
\small
\textit{Notes:} Standard errors are in parentheses. * denotes significance at the 10\% level, ** at the 5\% level, and *** at the 1\% level. The reported J-statistics are based on the over-identifying restrictions test associated with the homogeneous IV estimator of \citet{CuiSarafidisYamagata2023}; rejection of the null provides evidence against slope parameter homogeneity. Specification [1] imposes $u_{it}=\alpha_{i}+\tau_t+\varepsilon_{it}$, $\psi_{0,i}=\psi_{i}=\gamma_{\ell,i}=0$ for all $i$ and $\ell$, $\rho_{i}=\rho$ and $\beta_{\ell,i}=\beta_{\ell}$ for all $i$ and $\ell$ in Eq. \eqref{eq:empirical}. Specification [2] is as [1] except that it relaxes the two-way error components model restriction $u_{it}=\alpha_{i}+\tau_t+\varepsilon_{it}$ in [1]. Specification [3] is  as [1] except that it relaxes the slope parameter homogeneity restriction. Specification [4] imposes $u_{it}=\alpha_{i}+\tau_t+\varepsilon_{it}$ and slope parameter homogeneity but includes spatial spillovers. Specification [5] is as [4] except that it relaxes the two-way error components restriction by allowing for interactive effects. Finally, Specification [6] is as [5] but relaxes the slope parameter homogeneity restriction, which makes it the most general model.
\end{minipage}
\end{table}

\begin{figure}[h!]
    \centering
    \includegraphics[scale=0.65]{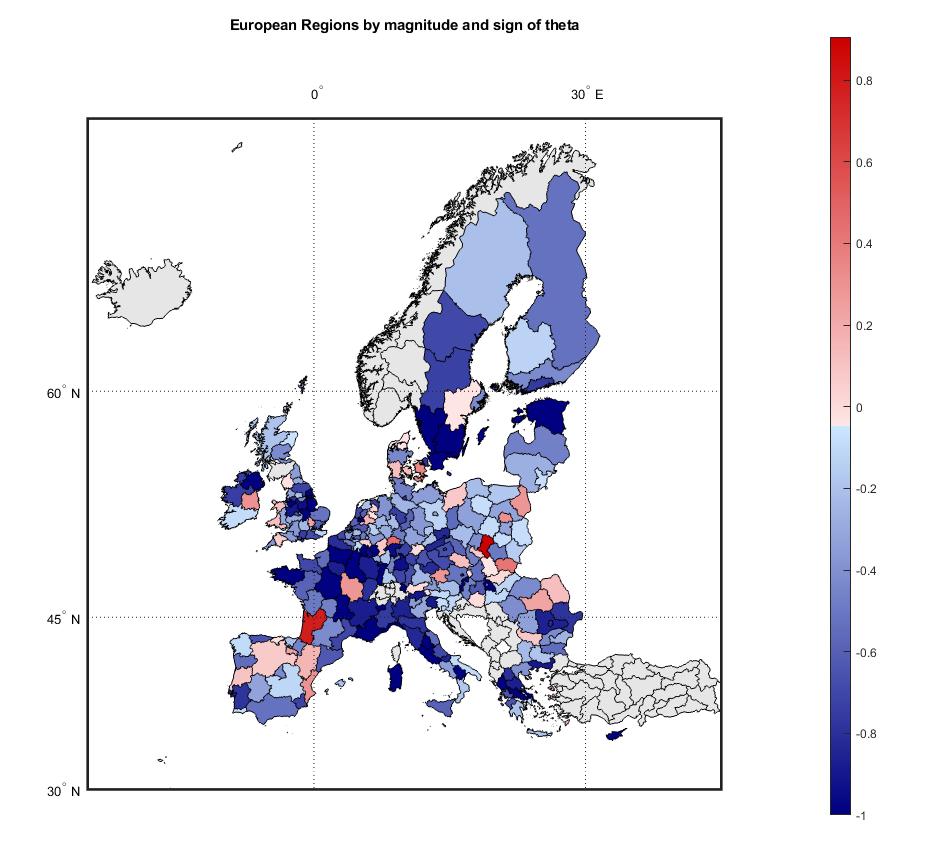}
    \caption{Convergence patterns across regional Europe}
    \label{fig:convergence}
\end{figure}

Figure \ref{fig:convergence} illustrates the individual-specific estimates of $\theta_{i}(=\delta_{i}+\psi_{i})$, the sum of the coefficients on the initial level of GDP per capita of region $i$ and its neighbors, using a map of European regions shaded according to their magnitude and sign. Regions shaded in dark blue correspond to more negative estimates of $\theta_i$, while those in dark red correspond to estimates well above zero. Regions shaded in pale colours indicate estimates close to zero and statistically insignificant. As shown, most high-income regions are shaded in dark blue, consistent with negative coefficients on initial GDP and hence evidence of conditional convergence, whereby higher initial income levels are associated with slower subsequent growth. By contrast, several regions in southern Europe, which are relatively poorer, also exhibit strongly negative estimates, indicating faster growth consistent with higher marginal returns to investment. In contrast, regions such as central Spain and Portugal, Wales, and northern Romania display estimates close to zero, providing little evidence of conditional convergence. There is also considerable heterogeneity in the magnitude of $\theta_i$ across regions, suggesting variation in the speed of convergence.

Table \ref{tab:Impacts} reports the Mean Group estimates of the direct, indirect (spillover), and total effects of $\ln(inv_{it})$, $\ln(n_{it})$, $\ln(educ_{it})$, and $\ln(sci\&tech_{it})$. Following \citet{LesagePace2009} and \citet{DebarsyEtal2012}, the direct effect measures the average impact of a change in a region's own covariate on its own growth rate, while the indirect effect captures the average impact of that change on the growth rates of other regions through spatial linkages. The total effect combines these two components, reflecting both local and spillover influences on regional growth.

The Mean Group direct effect of $\ln\left(inv_{it}\right)$ is small but statistically significant at the $10\%$ level when employing a one-tailed test, indicating that on average a region's own investment rate has a modest but discernible impact on its economic growth. On the other hand, the indirect effect, capturing the influence of neighboring regions' investment rates, is much larger and highly significant. This finding is consistent with the conclusions of \citet{ElhorstEtal2024}, who document that while the local impact of the investment rate tends to be small, its spillover effects are substantial. Overall, a $1\%$ increase in the investment rate is associated with a $0.264\%$ total increase in the regional GDP per capita growth rate, of which $83.3\%$ is attributable to spillovers from neighboring regions. This substantial contribution of spillovers highlights the pivotal role of cross-regional investment linkages, such as shared infrastructure projects, inter-regional supply chains, and trade networks, in driving economic growth.

For $\ln(n_{it})$, the Mean Group direct and spillover effects differ in both magnitude and sign. A $1\%$ increase in a region's population growth rate is associated with a $0.531\%$ decrease in its GDP per capita growth rate, consistent with capital dilution effects predicted by neoclassical growth theory. By contrast, the same increase in neighboring regions' population growth rates is associated with a $0.696\%$ increase in the focal region's growth rate, possibly reflecting labor market integration, demand spillovers, and knowledge diffusion. Overall, these effects largely offset each other, indicating that regional growth dynamics are shaped by both local and spatial demographic factors.
Turning to $\ln\left(educ_{it}\right)$ and $\ln\left(sci \& tech_{it}\right)$, the total effects for both variables are positive and statistically significant at the $10\%$ level under a one-tailed test.\footnote{Note that the Mean Group direct effect of $\ln\left(sci \& tech_{it}\right)$ is positive even if the corresponding coefficient in model \eqref{eq:empirical} is negative. This arises because a negative coefficient on a spatial lag of a covariate does not necessarily imply a negative spillover effect; see \citet{LesagePace2009} (p. 71).} These effects are primarily driven by their spillover components. In particular, the Mean Group indirect effect of $\ln\left(educ_{it}\right)$ is sizable, indicating that regions benefit from higher levels of tertiary education in neighboring areas, consistent with human capital spillovers emphasized in endogenous growth theory. Similarly, the indirect effect of $\ln\left(sci \& tech_{it}\right)$, although smaller in magnitude, suggests that regions also benefit from innovation and technology diffusion originating in adjacent areas.

\begin{table}[htbp]\centering
\caption{{Mean Group direct, indirect and total effects} \index{Impacts Table 5.2}}
\label{tab:Impacts}
\begin{tabular}{lcccccc}
\hline
%\multicolumn{6}{c}{\textbf{Table 5.1: Results for Empirical Growth Model.}}                                                                                                                                   %\\ \hline
             & Impact      & Std. Err.   & z score     & P>|z|       & [95\% Conf. Interval] \\
\hline
\textbf{Direct} \\
$\ln\left(inv_{it}\right)$  & 0.044     & 0.032     & 1.344      & 0.179      & [-0.020, 0.107] \\
$\ln\left(n_{it}\right)$    & -0.531    & 0.313     & -1.693     & 0.090      & [-1.144, 0.083] \\
$\ln\left(educ_{it}\right)$ & 0.036     & 0.314     & 0.114      & 0.909      & [-0.580, 0.651] \\
$\ln\left(sci \& tech_{it}\right)$ & 0.001     & 0.035     & 0.016      & 0.987      & [-0.068, 0.069] \\
\hline
\textbf{Indirect} \\
$\ln\left(inv_{it}\right)$
 & 0.220 & 0.034 & 6.471 & 0.000 & [0.153, 0.287] \\
$\ln\left(n_{it}\right)$
 & 0.696 & 0.318 & 2.189 & 0.029 & [0.073, 1.319] \\
$\ln\left(educ_{it}\right)$
 & 0.386 & 0.309 & 1.249 & 0.212 & [-0.220, 0.992] \\
$\ln\left(sci \& tech_{it}\right)$
 & 0.050 & 0.037 & 1.351 & 0.177 & [-0.023, 0.123] \\
\hline
\textbf{Total} \\
$\ln\left(inv_{it}\right)$
 & 0.264 & 0.036 & 7.333 & 0.000 & [0.193, 0.335] \\
$\ln\left(n_{it}\right)$
 & 0.165 & 0.321 & 0.514 & 0.607 & [-0.464, 0.794] \\
$\ln\left(educ_{it}\right)$
 & 0.422 & 0.317 & 1.331 & 0.183 & [-0.199, 1.043] \\
$\ln\left(sci \& tech_{it}\right)$
 & 0.051 & 0.039 & 1.308 & 0.191 & [-0.025, 0.127] \\
\hline
\end{tabular}
\end{table}

\begin{figure}[h!]
    \centering
    \includegraphics[scale=0.65]{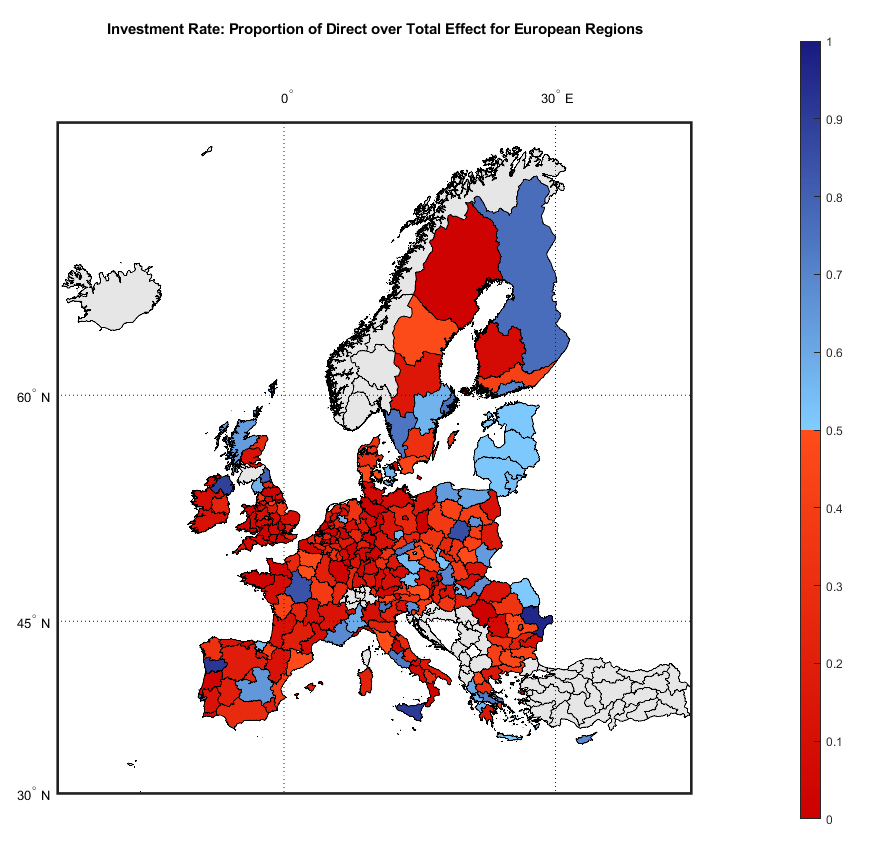}
    \caption{Investment rate effects on regional growth in Europe: proportion of direct over total effect}
    \label{fig:investment}
\end{figure}

Figure \ref{fig:investment} displays a map of European regions, shaded according to the share of the total effect of the investment rate, $\ln(inv_{it})$, attributable to the direct effect. Regions shaded in dark red correspond to shares close to zero, indicating that most of the effect operates through spillovers from neighboring regions. In contrast, regions shaded in dark blue correspond to shares close to one, suggesting that the effect is primarily driven by the region's own investment. As shown, many regions across Europe exhibit substantial spillover contributions relative to direct effects. Notable exceptions include Mediterranean islands such as Crete, Cyprus, and Sicily, where the direct effect is more pronounced. Similar patterns are observed in northern Scotland, Northern Ireland, Thuringia in Germany (often referred to as ``the green heart of Germany'' due to its broad and dense forests), and several capital regions, including Lisbon and Rome.

Taken together, the results highlight the importance of accounting for both spatial spillovers and slope heterogeneity in regional growth models. Ignoring these features can lead to misleading inference on the magnitude and transmission of growth effects, particularly in the presence of common shocks. From a policy perspective, the results highlight the importance of inter-regional linkages. Strong spillover effects suggest that enhancing connectivity and coordination in education and innovation can generate substantial aggregate benefits.

%From a policy perspective, the results emphasize the necessity of fostering inter-regional cooperation and investment. Policies that enhance connectivity, whether through infrastructure, trade facilitation, or collaborative innovation networks, are likely to yield substantial aggregate benefits. Moreover, the prominence of spillover effects of education and technology underscores the value of region-wide initiatives to elevate human capital and technological capabilities.

\section{Concluding Remarks}\label{sec-conclusion}

This paper develops a Mean Group Instrumental Variables (MGIV) estimator tailored for spatial dynamic panel data models with interactive fixed effects. In contrast to existing methods that assume slope parameter homogeneity, the MGIV estimator explicitly accommodates heterogeneity across cross-sectional units. Theoretical results establish its consistency and asymptotic normality under large $N$ and $T$ asymptotics. Furthermore, the estimator is asymptotically unbiased, enabling valid inferences without bias correction. Its linear structure also renders it computationally efficient, making it suitable for large-scale applications.

\bibliographystyle{apalike}
\bibliography{spatial_MGIV.bib}

@article{AhnHorenstein2013,
  author  = {Ahn, S. C. and Horenstein, A. R.},
  title   = {Eigenvalue Ratio Test for the Number of Factors},
  journal = {Econometrica},
  year    = {2013},
  volume  = {81},
  number  = {3},
  pages   = {1203--1227}
}

@article{MullerWatson2024,
  author  = {M{\"u}ller, U. K. and Watson, M. W.},
  title   = {Spatial Unit Roots and Spurious Regression},
  journal = {Econometrica},
  year    = {2024},
  volume  = {92},
  number  = {5},
  pages   = {1661--1695}
}

@article{BaltagiShu2025,
  author  = {Baltagi, Badi H. and Shu, Jin},
  title   = {Spatial dynamic panel data model with interactive fixed effects and time-variant endogenous spatial weight matrices},
  journal = {Econometrics and Statistics},
  year    = {2025},
  note    = {In press}
}

@article{AquaroEtal2021,
  author  = {Aquaro, M. and Bailey, N. and Pesaran, M. H.},
  title   = {Estimation and inference for spatial models with heterogeneous coefficients: An application to {US} house prices},
  journal = {Journal of Applied Econometrics},
  year    = {2021},
  volume  = {36},
  number  = {1},
  pages   = {18--44}
}

@book{ArbiaEtal2021,
  author    = {Arbia, G. and Espa, G. and Giuliani, D.},
  title     = {Spatial Microeconometrics},
  publisher = {Routledge},
  address   = {London and New York},
  year      = {2021}
}

@article{ArbiaEtal2010,
  author  = {Arbia, G. and Basile, R. and Salvatore, M.},
  title   = {Institutions and geography: Empirical test of spatial growth models for {E}uropean regions},
  journal = {Economic Modelling},
  year    = {2010},
  volume  = {27},
  number  = {1},
  pages   = {12--21}
}

@article{Bai2003,
  author  = {Bai, J.},
  title   = {Inferential theory for factor models of large dimensions},
  journal = {Econometrica},
  year    = {2003},
  volume  = {71},
  number  = {1},
  pages   = {135--171}
}

@article{Bai2009,
  author  = {Bai, J.},
  title   = {Panel data models with interactive fixed effects},
  journal = {Econometrica},
  year    = {2009},
  volume  = {77},
  number  = {4},
  pages   = {1229--1279}
}

@article{BaiNg2002,
  author  = {Bai, J. and Ng, S.},
  title   = {Determining the Number of Factors in Approximate Factor Models},
  journal = {Econometrica},
  year    = {2002},
  volume  = {70},
  number  = {1},
  pages   = {191--221}
}

@article{BaiNg2006,
  author  = {Bai, J. and Ng, S.},
  title   = {Confidence Intervals for Diffusion Index Forecasts and Inference for Factor-Augmented Regressions},
  journal = {Econometrica},
  year    = {2006},
  volume  = {74},
  number  = {4},
  pages   = {1133--1150}
}

@article{BaiLi2021,
  author  = {Bai, J. and Li, K. P.},
  title   = {Dynamic spatial panel data models with common shocks},
  journal = {Journal of Econometrics},
  year    = {2021},
  volume  = {224},
  number  = {1},
  pages   = {134--160}
}

@article{BelloniChenChernozhukovHansen2012,
  author  = {Belloni, A. and Chen, D. and Chernozhukov, V. and Hansen, C.},
  title   = {Sparse Models and Methods for Optimal Instruments with an Application to Eminent Domain},
  journal = {Econometrica},
  year    = {2012},
  volume  = {80},
  number  = {6},
  pages   = {2369--2429}
}

@article{BramulleEtal2009,
  author  = {Bramoull{\'e}, Y. and Djebbari, H. and Fortin, B.},
  title   = {Identification of Peer Effects Through Social Networks},
  journal = {Journal of Econometrics},
  year    = {2009},
  volume  = {150},
  number  = {1},
  pages   = {41--55}
}

@article{CaoEtAl2024,
author  = {Cao, Y. and Jin, S. and Lu, X. and Su, L.},
  title   = {Oracle Efficient Estimation of Heterogeneous Dynamic Panel Data Models with
Interactive Fixed Effects},
  journal = {Journal of Business \& Economic Statistics},
  year    = {2024},
  volume  = {42},
  number  = {4},
  pages   = { 1169--1184}}

@article{ChenEtAl2022,
  author  = {Chen, J. and Shin, Y. and Zheng, C.},
  title   = {Estimation and inference in heterogeneous spatial panel data models with a multifactor error structure},
  journal = {Journal of Econometrics},
  year    = {2022},
  volume  = {229},
  number  = {1},
  pages   = {55--79}
}

@article{Case1991,
  author  = {Case, A. C.},
  title   = {Spatial patterns in household demand},
  journal = {Econometrica},
  year    = {1991},
  volume  = {59},
  pages   = {953--965}
}

@article{ChudikPesaran2015,
  author  = {Chudik, A. and Pesaran, M. H.},
  title   = {Common correlated effects estimation of heterogeneous dynamic panel data models with weakly exogenous regressors},
  journal = {Journal of Econometrics},
  year    = {2015},
  volume  = {188},
  number  = {2},
  pages   = {393--420}
}

@article{CuiSarafidisYamagata2023,
  author  = {Cui, G. and Sarafidis, V. and Yamagata, T.},
  title   = {I{V} Estimation of Spatial Dynamic Panels with Interactive Effects: Large Sample Theory and an Application on Bank Attitude Toward Risk},
  journal = {Econometrics Journal},
  year    = {2023},
  volume  = {26},
  number  = {2},
  pages   = {124--146}
}

@article{DebarsyEtal2012,
  author  = {Debarsy, N. and LeSage, J. P. and Pace, R. K.},
  title   = {The interpretation and presentation of results in spatial econometrics},
  journal = {The Review of Regional Studies},
  year    = {2012},
  volume  = {42},
  number  = {2},
  pages   = {107--142}
}

@article{HigginsMartellosio2023,
  author  = {Higgins, A. and Martellosio, F.},
  title   = {Shrinkage estimation of network spillovers with factor structured errors},
  journal = {Journal of Econometrics},
  year    = {2023},
  volume  = {233},
  number  = {1},
  pages   = {66--87}
}

@incollection{LeeYu2015,
  author    = {Lee, L.-F. and Yu, J.},
  title     = {Spatial Panel Data Models},
  booktitle = {The Oxford Handbook of Panel Data},
  editor    = {Baltagi, B.},
  publisher = {Oxford University Press},
  year      = {2015},
  chapter   = {12},
  pages     = {363--401}
}

@book{Elhorst2014,
  author    = {Elhorst, P.},
  title     = {Spatial Econometrics: From Cross-Sectional Data to Spatial Panels},
  publisher = {Springer},
  year      = {2014}
}

@incollection{Elhorst2021,
  author    = {Elhorst, P.},
  title     = {Spatial panel models and common factors},
  booktitle = {Handbook of Regional Science},
  editor    = {Fischer, M. and Nijkamp, P.},
  publisher = {Springer},
  address   = {Berlin, Heidelberg},
  year      = {2021},
  chapter   = {12},
  pages     = {2141--2159}
}

@article{ElhorstEtal2024,
  author  = {Elhorst, P. and Tziolas, I. and Tan, C. and Milionis, P.},
  title   = {The distance decay effect and spatial reach of spillovers},
  journal = {Journal of Geographical Systems},
  year    = {2024},
  volume  = {26},
  pages   = {265--289}
}

@article{ErturKoch2007,
  author  = {Ertur, C. and Koch, W.},
  title   = {Growth, technological interdependence and spatial externalities: theory and evidence},
  journal = {Journal of Applied Econometrics},
  year    = {2007},
  volume  = {22},
  number  = {6},
  pages   = {1033--1062}
}

@article{HardingLamarche2011,
  author  = {Harding, M. and Lamarche, C.},
  title   = {Least squares estimation of a panel data model with multifactor error structure and endogenous covariates},
  journal = {Economics Letters},
  year    = {2011},
  volume  = {111},
  number  = {3},
  pages   = {197--199}
}

@article{Hamermesh1995,
  author  = {Hamermesh, D. S.},
  title   = {Labor Demand and the Source of Adjustment Costs},
  journal = {The Economic Journal},
  year    = {1995},
  volume  = {105},
  number  = {430},
  pages   = {620--634}
}

@incollection{JappelliPistaferri2017,
  author    = {Jappelli, T. and Pistaferri, L.},
  title     = {Time, Habits, and Consumer Durables},
  booktitle = {The Economics of Consumption: Theory and Evidence},
  publisher = {Oxford University Press},
  year      = {2017},
  chapter   = {13},
  pages     = {234--256}
}

@article{JingEtAl2018,
  author  = {Jing, Z. and Elhorst, J. P. and Jakobs, J. and de Haan, J.},
  title   = {The propagation of financial turbulence: Interdependence, spillovers, and direct and indirect effects},
  journal = {Empirical Economics},
  year    = {2018},
  volume  = {55},
  pages   = {169--192}
}

@article{JuodisSarafidis2018ER,
  author  = {Juodis, A. and Sarafidis, V.},
  title   = {Fixed T dynamic panel data estimators with multifactor errors},
  journal = {Econometric Reviews},
  year    = {2018},
  volume  = {37},
  number  = {8},
  pages   = {893--929}
}

@article{Islam1995,
  author  = {Islam, N.},
  title   = {Growth Empirics: A Panel Data Approach},
  journal = {The Quarterly Journal of Economics},
  year    = {1995},
  volume  = {110},
  number  = {4},
  pages   = {1127--1170}
}

@article{KapoorKelejianPrucha2007,
  author  = {Kapoor, M. and Kelejian, H. H. and Prucha, I. R.},
  title   = {Panel data models with spatially correlated error components},
  journal = {Journal of Econometrics},
  year    = {2007},
  volume  = {140},
  number  = {1},
  pages   = {97--130}
}

@article{KelejianPrucha2001,
  author  = {Kelejian, H. H. and Prucha, I. R.},
  title   = {On the asymptotic distribution of the {M}oran {I} test statistic with applications},
  journal = {Journal of Econometrics},
  year    = {2001},
  volume  = {104},
  number  = {2},
  pages   = {219--257}
}

@book{KelejianPiras2017,
  author    = {Kelejian, H. and Piras, G.},
  title     = {Spatial Econometrics},
  publisher = {Academic Press},
  address   = {London},
  year      = {2017},
  edition   = {1st}
}

@article{Korniotis2010,
  author  = {Korniotis, G. M.},
  title   = {Estimating Panel Models With Internal and External Habit Formation},
  journal = {Journal of Business \& Economic Statistics},
  year    = {2010},
  volume  = {28},
  number  = {1},
  pages   = {145--158}
}

@article{KripfganzSarafidis2021,
  author  = {Kripfganz, S. and Sarafidis, V.},
  title   = {Instrumental Variable Estimation of Large-{T} Panel Data Models with Common Factors},
  journal = {Stata Journal},
  year    = {2021},
  volume  = {21},
  number  = {3},
  pages   = {1--28}
}

@article{KripfganzSarafidis2025,
  author  = {Kripfganz, S. and Sarafidis, V.},
  title   = {Estimating Spatial Dynamic Panel Data Models with Unobserved Common Factors in {S}tata},
  journal = {Journal of Statistical Software},
  year    = {2025},
  volume  = {113},
  number  = {6},
  pages   = {1--27}
}

@article{LeeYu2014,
  author  = {Lee, L.-F. and Yu, J.},
  title   = {Efficient {GMM} estimation of spatial dynamic panel data models with fixed effects},
  journal = {Journal of Econometrics},
  year    = {2014},
  volume  = {180},
  number  = {2},
  pages   = {174--197}
}

@book{LesagePace2009,
  author    = {LeSage, J. P. and Pace, R. K.},
  title     = {Introduction to Spatial Econometrics},
  publisher = {CRC Press},
  year      = {2009}
}

@article{LesageChih2016,
  author  = {LeSage, J. P. and Chih, Y.-Y.},
  title   = {Interpreting Heterogeneous Coefficient Spatial Autoregressive Panel Models},
  journal = {Economics Letters},
  year    = {2016},
  volume  = {142},
  pages   = {1--5}
}

@article{Manski1993,
  author  = {Manski, C.},
  title   = {Identification of endogenous social effects: The reflection problem},
  journal = {Review of Economic Studies},
  year    = {1993},
  volume  = {60},
  number  = {3},
  pages   = {531--542}
}

@article{MoonWeidner2017,
  author  = {Moon, R. H. and Weidner, M.},
  title   = {Dynamic Linear Panel Regression Models With Interactive Fixed Effects},
  journal = {Econometric Theory},
  year    = {2017},
  volume  = {33},
  number  = {1},
  pages   = {158--195}
}

@article{NeweyWest1987,
  author = {Newey, W. K. and West, K. D.},
  title = {A Simple, Positive Semi-Definite, Heteroskedasticity
and Autocorrelation Consistent Covariance Matrix},
  journal = {Econometrica},
  year = {1987},
  volume = {55},
  number = {3},
  pages = {703--708}
  }

@article{Nickell1981,
  author  = {Nickell, S.},
  title   = {Biases in Dynamic Models with Fixed Effects},
  journal = {Econometrica},
  year    = {1981},
  volume  = {49},
  number  = {6},
  pages   = {1417--1426}
}

@article{NorkuteEtal2021,
  author  = {Norkut{\.{e}}, M. and Sarafidis, V. and Yamagata, T. and Cui, G.},
  title   = {Instrumental variable estimation of dynamic linear panel data models with defactored regressors and a multifactor error structure},
  journal = {Journal of Econometrics},
  year    = {2021},
  volume  = {220},
  number  = {2},
  pages   = {416--446}
}

@article{Pesaran2006,
  author  = {Pesaran, M. H.},
  title   = {Estimation and Inference in Large Heterogeneous Panels with a Multifactor Error Structure},
  journal = {Econometrica},
  year    = {2006},
  volume  = {74},
  number  = {4},
  pages   = {967--1012}
}

@article{PesaranSmith1995,
  author  = {Pesaran, M. H. and Smith, R.},
  title   = {Estimating long-run relationships from dynamic heterogeneous panels},
  journal = {Journal of Econometrics},
  year    = {1995},
  volume  = {68},
  number  = {1},
  pages   = {79--113}
}

@article{PhillipsSull2007,
  author  = {Phillips, P. C. B. and Sul, D.},
  title   = {Bias in Dynamic Panel Estimation with Fixed Effects, Incidental Trends and Cross Section Dependence},
  journal = {Journal of Econometrics},
  year    = {2007},
  volume  = {137},
  number  = {1},
  pages   = {162--188}
}

@article{RobertsonSymons1992,
  author  = {Robertson, D. and Symons, J.},
  title   = {Some Strange Properties of Panel Data Estimators},
  journal = {Journal of Applied Econometrics},
  year    = {1992},
  volume  = {7},
  number  = {2},
  pages   = {175--189}
}

@misc{Sargent_Sims_1977,
  author = {Sargent, T. and Sims, C.},
  title  = {Business cycle modeling without pretending to have too much a priori economic theory},
  year   = {1977},
  note   = {MIMEO}
}

@article{SarafidisWansbeek2012,
  author  = {Sarafidis, V. and Wansbeek, T.},
  title   = {Cross-Sectional Dependence in Panel Data Analysis},
  journal = {Econometric Reviews},
  year    = {2012},
  volume  = {31},
  number  = {5},
  pages   = {483--531}
}

@article{SarafidisWansbeek2021,
  author  = {Sarafidis, V. and Wansbeek, T.},
  title   = {Celebrating 40 years of panel data analysis: Past, present and future},
  journal = {Journal of Econometrics},
  year    = {2021},
  volume  = {220},
  number  = {2},
  pages   = {215--226}
}

@article{ShiLee2017,
  author  = {Shi, W. and Lee, L.-F.},
  title   = {Spatial dynamic panel data models with interactive fixed effects},
  journal = {Journal of Econometrics},
  year    = {2017},
  volume  = {197},
  number  = {2},
  pages   = {323--347}
}

@article{YuEtAl2012,
  author  = {Yu, J. and de Jong, R. and Lee, L.-F.},
  title   = {Estimation for spatial dynamic panel data with fixed effects: the case of spatial cointegration},
  journal = {Journal of Econometrics},
  year    = {2012},
  volume  = {167},
  number  = {1},
  pages   = {16--37}
}

@article{YuEtAl2008,
  author  = {Yu, J. and de Jong, R. and Lee, L.-F.},
  title   = {Quasi-maximum likelihood estimators for spatial dynamic panel data with fixed effects when both n and {T} are large},
  journal = {Journal of Econometrics},
  year    = {2008},
  volume  = {146},
  number  = {1},
  pages   = {118--134}
}

@incollection{AnselinGalloJayet2008,
  author    = {Luc Anselin and Julie Le Gallo and H{\'e}l{\`e}ne Jayet},
  title     = {Spatial Panel Econometrics},
  booktitle = {The Econometrics of Panel Data},
  editor    = {L{\'a}szl{\'o} M{\'a}ty{\'a}s and Patrick Sevestre},
  publisher = {Springer},
  address   = {Berlin, Heidelberg},
  year      = {2008},
  edition   = {3rd},
  pages     = {625--660}
}

\newpage

\section*{Appendix A: Proofs of the Main Theoretical Results}\label{AppendixA}

\renewcommand{\thelemma}{A.\arabic{lemma}} \setcounter{lemma}{0}
\makeatletter
\renewcommand{\thetheorem}{A.\arabic{theorem}} \setcounter{theorem}{0}
\makeatletter
\renewcommand{\theproposition}{A.\arabic{proposition}} %
\setcounter{proposition}{0} \makeatletter
\renewcommand{\thecorollary}{A.\arabic{corollary}} \setcounter{corollary}{0}
\makeatletter
\renewcommand\theequation{A.\@arabic\c@equation } \setcounter{equation}{0}
\makeatother
%Throughout the appendix, for a matrix $\mathbf{A}$, we define $\mathbf{P}_{\mathbf{A}}=%
%\mathbf{A}(\mathbf{A}^{\prime }\mathbf{A})^{-1}\mathbf{A}^{\prime }$ and $%
%\mathbf{M}_{\mathbf{A}}=\mathbf{I}-\mathbf{P}_{\mathbf{A}}$. We denote a
%generic finite constant large enough as $C$, which needs not to be the same
%at each appearance.

In this appendix, we provide the proofs of the main theoretical results. To save space, we relegate the proofs of Propositions \ref{prop:asymrep_zu} and \ref{prop:asymrep_pzu} presented below to Section~S1 of the Online Supplement. %The lemmas used  and the lemmas used in the proofs as well as their proofs are relegated to Section S1 of the Online Supplement.
\smallskip

We first present Proposition \ref{prop:asymrep_zu}, which plays an important role in the proof of Theorem \ref{individual-asydist}.

\begin{proposition}
\label{prop:asymrep_zu}  Under Assumptions \ref{assumption-idioiny}--\ref{assumption-weight} and \ref{assumption-ident}\ref{assumption-ident-fullrank}, we have, for all $i$,
\begin{align}
&(a)~~T^{-1}\widehat{\mathbf{Z}}_{i}^{\prime }\mathbf{u}_{i}=T^{-1}\boldsymbol{%
\mathbf{Z}}_{i}^{\prime }\mathbf{u}_{i}+O_p(\delta_{N\!T}^{-2})=O_p(T^{-1/2})+O_p(\delta_{N\!T}^{-2}), \notag\\
&(b)~~\widehat{\mathbf{A}}_i=\mathbf{A}_i+O_p(\delta_{N\!T}^{-1}), \notag\\
&(c)~~\widehat{\mathbf{B}}_i=\mathbf{B}_i+O_p(\delta_{N\!T}^{-1}), \notag\\
&(d)~~\mathbf{A}_i=O_p(1),\ \ \mathbf{B}_i=O_p(1),
\end{align}
as $N,T\rightarrow\infty$.
%where $\boldsymbol{\mathbf{Z}}_{i}=\left(\sum_{j=1}^Nw_{ij}\mathbf{M}_{
%\mathbf{F}}\boldsymbol{\mathbf{X}}_{j},\mathbf{M}_{\mathbf{F}}\mathbf{M}_{
%\mathbf{F}_{-1}}\boldsymbol{\mathbf{X}}_{i,-1},\mathbf{M}_{\mathbf{F}}%
%\boldsymbol{\mathbf{X}}_{i}\right)$.
\end{proposition}

\medskip

\noindent \textbf{Proof of Theorem \ref{individual-asydist}.} By (\ref{individual-est}) and Proposition \ref{prop:asymrep_zu}, we have, for all $i$,
\begin{align}
\widehat{\boldsymbol{\theta}}_i-\boldsymbol{\theta}_i=&\left(\widehat{\mathbf{A}}_i^{\prime }%
\widehat{\mathbf{B}}^{-1}_i\widehat{\mathbf{A}}_i\right)^{-1}\widehat{%
\mathbf{A}}_i^{\prime }\widehat{\mathbf{B}}_i^{-1}\widehat{\mathbf{c}}_{y,i}-\boldsymbol{\theta}_i \nonumber\\
=&\left(\widehat{\mathbf{A}}_i^{\prime }%
\widehat{\mathbf{B}}^{-1}_i\widehat{\mathbf{A}}_i\right)^{-1}\widehat{%
\mathbf{A}}_i^{\prime }\widehat{\mathbf{B}}_i^{-1}\big[T^{-1}\widehat{\mathbf{Z}}_{i}^{\prime }\big(\mathbf{y}_i-\mathbf{C}_i\boldsymbol{\theta}_i\big)\big]\nonumber\\
=&\left(\widehat{\mathbf{A}}_i^{\prime }%
\widehat{\mathbf{B}}^{-1}_i\widehat{\mathbf{A}}_i\right)^{-1}\widehat{%
\mathbf{A}}_i^{\prime }\widehat{\mathbf{B}}_i^{-1}\big(T^{-1}\widehat{\mathbf{Z}}_{i}^{\prime }\mathbf{u}_i\big)\notag\\
=&\left({\mathbf{A}}_i^{\prime }%
{\mathbf{B}}^{-1}_i{\mathbf{A}}_i\right)^{-1}{%
\mathbf{A}}_i^{\prime }{\mathbf{B}}_i^{-1}\big(T^{-1}{\mathbf{Z}}_{i}^{\prime }\mathbf{u}_i\big)+O_p(T^{-1/2}\delta_{NT}^{-1}+\delta_{NT}^{-2}),\label{eq-theorem1proof1}
\end{align}
as $N,T\rightarrow\infty$. By Assumption \ref{assumption-ident}\ref{assumption-ident-fullrank} and Proposition \ref{prop:asymrep_zu}(a)\&(d), we can see that the RHS of the last equality in (\ref{eq-theorem1proof1}) tends to zero in probability as $N,T\rightarrow\infty$. This completes the proof of the consistency of $\widehat{\boldsymbol{\theta}}_i$.

For the asymptotic distribution, by Assumption
\ref{assumption-ident}\ref{assumption-ident-clt},
\begin{align}
T^{-1/2}\mathbf{Z}_{i}^{\prime }\mathbf{u}_i
\stackrel{d}{\longrightarrow}
N(\mathbf{0},\boldsymbol{\Phi}_{i,0}),
\qquad
\text{as }N,T\rightarrow\infty.
\label{eq-theorem1proof2}
\end{align}
Furthermore, by Assumption
\ref{assumption-ident}\ref{assumption-ident-limit},
$\left(\mathbf A_i^\prime\mathbf B_i^{-1}\mathbf A_i\right)^{-1}
\mathbf A_i^\prime\mathbf B_i^{-1}
\stackrel{p}{\longrightarrow}
\left(\mathbf A_{i,0}^\prime\mathbf B_{i,0}^{-1}\mathbf A_{i,0}\right)^{-1}
\mathbf A_{i,0}^\prime\mathbf B_{i,0}^{-1}$.
When $T/N^2\rightarrow0$,
$T^{1/2}
\left(T^{-1/2}\delta_{NT}^{-1}+\delta_{NT}^{-2}\right)
=
\delta_{NT}^{-1}
+T^{1/2}\delta_{NT}^{-2}
\rightarrow0$.
The limiting normal distribution then follows from
(\ref{eq-theorem1proof1}) and (\ref{eq-theorem1proof2}).
\hfill $\Box$

% ORIGINAL CLT

% For the asymptotic distribution, by (\ref{eq-IV}) and Assumptions \ref{assumption-idioiny}--\ref{assumption-weight} and \ref{assumption-ident}\ref{assumption-ident-limit}, we can show that
% \begin{align}
% %&\mathbb{E}\big(T^{-1/2}{\mathbf{Z}}_{i}^{\prime }\mathbf{u}_i\big)=\mathbf{0},\ \ \ \ \ {\rm Var}\big(T^{-1/2}{\mathbf{Z}}_{i}^{\prime }\mathbf{u}_i\big)=\boldsymbol{\Phi}_i, \notag  \\
% T^{-1/2}{\mathbf{Z}}_{i}^{\prime }\mathbf{u}_i\stackrel{d}{\longrightarrow} N(\mathbf{0},\ \boldsymbol{\Phi}_{i,0}),\ \ \ {\rm as}\ \ T\rightarrow\infty. \label{eq-theorem1proof2}
% \end{align}
% {\it Furthermore, when $T/N^2\rightarrow0$, $T^{1/2}\cdot (T^{-1/2}\delta_{NT}^{-1}+\delta_{NT}^{-2})=\delta_{NT}^{-1}+T^{1/2}\delta_{NT}^{-2}\rightarrow0$}. Combining (\ref{eq-theorem1proof1}) and (\ref{eq-theorem1proof2}), we can easily obtain the desired asymptotic normality for $\sqrt{T}\big(\widehat{\boldsymbol{\theta}}_i-\boldsymbol{\theta}_i\big)$.  \hfill $\Box$

\bigskip

\noindent \textbf{Proof of Proposition \ref{individual-Cov-Consistency}.} First, note that by Proposition \ref{prop:asymrep_zu}(b)\&(c) and Assumption \ref{assumption-ident}\ref{assumption-ident-limit}, we have
\begin{align}
\widehat{\mathbf{A}}_i\stackrel{p}\longrightarrow\mathbf{A}_{i,0}\ \ \mathrm{and}\ \ \widehat{\mathbf{B}}_i\stackrel{p}{\longrightarrow}\mathbf{B}_{i,0}\notag
\end{align}
as $N,T\rightarrow\infty$. By the Continuous Mapping Theorem and the full-rankness of the matrices, we have
\begin{align}
\big(\widehat{\mathbf{A}}_{i}^\prime \widehat{\mathbf{B}}_{i}^{-1}\widehat{\mathbf{A}}_{i}\big)^{-1}\widehat{\mathbf{A}}_{i}^\prime \widehat{\mathbf{B}}_{i}^{-1}\stackrel{p}{\longrightarrow} \big(\mathbf{A}_{i,0}^\prime\mathbf{B}_{i,0}^{-1}\mathbf{A}_{i,0}\big)^{-1}\mathbf{A}_{i,0}^\prime\mathbf{B}_{i,0}^{-1}.\notag
\end{align}
Then it remains to prove that
\begin{align}
\widehat{\boldsymbol{\Phi}}_{i}-\boldsymbol{\Phi}_{i,0}=o_p(1).\label{Prop31proof1}
\end{align}
To this end, define the infeasible Newey-West estimator
\begin{align}
\widetilde{\boldsymbol{\Phi}}_{i}=\widetilde{\boldsymbol{\Phi}}_{i}(0)+\sum_{h=1}^{L_T}K\left(\frac{h}{L_T+1}\right)\left[\widetilde{\boldsymbol{\Phi}}_{i}(h)+\widetilde{\boldsymbol{\Phi}}_{i}(h)^\prime\right],\notag
\end{align}
where $\widetilde{\boldsymbol{\Phi}}_{i}(h)=\frac{1}{T}\sum_{t=h+1}^T{\tilde{u}}_{it}{\tilde{u}}_{i,t-h}{\mathbf{z}}_{it}{\mathbf{z}}_{i,t-h}^\prime=\frac{1}{T}\sum_{t=h+1}^T\mathbf{s}_{it}\mathbf{s}_{i,t-h}^\prime$. Then,
\begin{align}
\widehat{\boldsymbol{\Phi}}_{i}-\boldsymbol{\Phi}_{i,0}=&\ \left(\widehat{\boldsymbol{\Phi}}_{i}-\widetilde{\boldsymbol{\Phi}}_{i}\right) + \left(\widetilde{\boldsymbol{\Phi}}_{i}-\mathbb{E}\big(\widetilde{\boldsymbol{\Phi}}_i\big)\right) + \left(\mathbb{E}\big(\widetilde{\boldsymbol{\Phi}}_i\big)-\boldsymbol{\Phi}_{i}\right)+\left(\boldsymbol{\Phi}_{i}-\boldsymbol{\Phi}_{i,0}\right)\notag\\
=:&\ \mathcal{D}_{i,1}+\mathcal{D}_{i,2}+\mathcal{D}_{i,3}+\mathcal{D}_{i,4}.\label{Prop31proof2}
\end{align}
By Assumption \ref{assumption-ident}\ref{assumption-ident-limit}, $\mathcal{D}_{i,4}=o(1)$. Hence, we only need to show that $\mathcal{D}_{i,k}=o_p(1)$ for $k=1,2$, and $\mathcal{D}_{i,3}=o(1)$. First, consider $\mathcal{D}_{i,1}$. Denoting $\hat{\mathbf{s}}_{it}=\hat{\mathbf{z}}_{it}\hat{\tilde{u}}_{it}$, we have $\widehat{\boldsymbol{\Phi}}_{i}(h)=\frac{1}{T}\sum_{t=h+1}^T\hat{\mathbf{s}}_{it}\hat{\mathbf{s}}_{i,t-h}^\prime$ and
\begin{align}
&\widehat{\boldsymbol{\Phi}}_{i}(h)-\widetilde{\boldsymbol{\Phi}}_{i}(h)=\frac{1}{T}\sum_{t=h+1}^T\big(\hat{\mathbf{s}}_{it}\hat{\mathbf{s}}_{i,t-h}^\prime-\mathbf{s}_{it}\mathbf{s}_{i,t-h}^\prime\big)\notag\\
=&\frac{1}{T}\sum_{t=h+1}^T\big(\hat{\mathbf{s}}_{it}-\mathbf{s}_{it}\big)\big(\hat{\mathbf{s}}_{i,t-h}-\mathbf{s}_{i,t-h}\big)^\prime + \frac{1}{T}\sum_{t=h+1}^T\big(\hat{\mathbf{s}}_{it}-\mathbf{s}_{it}\big)\mathbf{s}_{i,t-h}^\prime + \frac{1}{T}\sum_{t=h+1}^T{\mathbf{s}}_{it}\big(\hat{\mathbf{s}}_{i,t-h}-\mathbf{s}_{i,t-h}\big)^\prime.\notag
\end{align}
Therefore, for all $0\leq h\leq L_T$, we have
\begin{align}
&\left\Vert\widehat{\boldsymbol{\Phi}}_{i}(h)-\widetilde{\boldsymbol{\Phi}}_{i}(h)\right\Vert
\leq \frac{1}{T}\sum_{t=h+1}^T\Vert\hat{\mathbf{s}}_{it}-\mathbf{s}_{it}\Vert\Vert\hat{\mathbf{s}}_{i,t-h}-\mathbf{s}_{i,t-h}\Vert + \frac{1}{T}\sum_{t=h+1}^T\Vert\hat{\mathbf{s}}_{it}-\mathbf{s}_{it}\Vert \Vert\mathbf{s}_{i,t-h}\Vert\notag\\
& + \frac{1}{T}\sum_{t=h+1}^T \Vert{\mathbf{s}}_{it}\Vert \Vert\hat{\mathbf{s}}_{i,t-h}-\mathbf{s}_{i,t-h}\Vert
\leq  \frac{1}{T}\left(\sum_{t=1}^T\Vert\hat{\mathbf{s}}_{it}-\mathbf{s}_{it}\Vert^2\right)  + \frac{2}{T}\sqrt{\left(\sum_{t=1}^T\Vert\hat{\mathbf{s}}_{it}-\mathbf{s}_{it}\Vert^2\right) \left(\sum_{t=1}^T\Vert\mathbf{s}_{it}\Vert^2\right)}.\label{Prop3proof3}
\end{align}

On the other hand,
\begin{align}
\hat{\mathbf{s}}_{it}-\mathbf{s}_{it}=\hat{\mathbf{z}}_{it}\hat{\tilde{u}}_{it}-\mathbf{z}_{it}\tilde{u}_{it}=\big(\hat{\mathbf{z}}_{it}-\mathbf{z}_{it}\big)\big(\hat{\tilde{u}}_{it}-\tilde{u}_{it}\big) + \big(\hat{\mathbf{z}}_{it}-\mathbf{z}_{it}\big)\tilde{u}_{it} + \mathbf{z}_{it}\big(\hat{\tilde{u}}_{it}-\tilde{u}_{it}\big)\notag
\end{align}
and hence,
\begin{align}
\sum_{t=1}^T\Vert\hat{\mathbf{s}}_{it}-\mathbf{s}_{it}\Vert^2\leq C\sum_{t=1}^T\Vert\hat{\mathbf{z}}_{it}-\mathbf{z}_{it}\Vert^2 \vert\hat{\tilde{u}}_{it}-\tilde{u}_{it}\vert^2 + C\sum_{t=1}^T\Vert\hat{\mathbf{z}}_{it}-\mathbf{z}_{it}\Vert^2\vert\tilde{u}_{it}\vert^2 + C\sum_{t=1}^T\Vert\mathbf{z}_{it}\Vert^2\vert\hat{\tilde{u}}_{it}-\tilde{u}_{it}\vert^2.\label{Prop3proof4}
\end{align}
In addition, note that
\begin{align}
&\sum_{t=1}^T\Vert\hat{\mathbf{z}}_{it}-\mathbf{z}_{it}\Vert^2=\Vert\widehat{\mathbf{Z}}_i-\mathbf{Z}_i\Vert^2\notag\\
=&\left\Vert\left(\sum_{j=1}^Nw_{ij}\big(\mathbf{M}_{\widehat{%
\mathbf{F}}}-\mathbf{M}_{\mathbf{F}}\big)\mathbf{X}_{j},\ \big(\mathbf{M}_{\widehat{\mathbf{F}}}\mathbf{M}_{%
\widehat{\mathbf{F}}_{-1}}-\mathbf{M}_{\mathbf{F}}\mathbf{M}_{\mathbf{F}_{-1}}\big)\mathbf{X}_{i,-1},\ \big(\mathbf{M}_{\widehat{\mathbf{F}}}-\mathbf{M}_{\mathbf{F}}\big)
\mathbf{X}_{i}\right)\right\Vert^2\notag\\
=&\left\Vert\sum_{j=1}^Nw_{ij}\big(\mathbf{M}_{\widehat{%
\mathbf{F}}}-\mathbf{M}_{\mathbf{F}}\big)\mathbf{X}_{j}\right\Vert^2+ \left\Vert\big(\mathbf{M}_{\widehat{\mathbf{F}}}\mathbf{M}_{%
\widehat{\mathbf{F}}_{-1}}-\mathbf{M}_{\mathbf{F}}\mathbf{M}_{\mathbf{F}_{-1}}\big)\mathbf{X}_{i,-1}\right\Vert^2 + \left\Vert\big(\mathbf{M}_{\widehat{\mathbf{F}}}-\mathbf{M}_{\mathbf{F}}\big)
\mathbf{X}_{i}\right\Vert^2.\notag
\end{align}
Using similar arguments to the proof of Proposition \ref{prop:asymrep_zu}(c) in the Online Supplement, we can show that each of the above three terms is of order $O_p(\delta_{NT}^{-2}T)$ and hence,
\begin{align}
\sum_{t=1}^T\Vert\hat{\mathbf{z}}_{it}-\mathbf{z}_{it}\Vert^2=O_p(\delta_{NT}^{-2}T).\label{Prop3proof5}
\end{align}
Analogously, we have
\begin{align}
&\sum_{t=1}^T\vert\hat{\tilde{u}}_{it}-\tilde{u}_{it}\vert^2=\Vert\widehat{\widetilde{\mathbf{u}}}_i-\widetilde{\mathbf{u}}_i\Vert^2=\left\Vert \mathbf{M}_{\widehat{\mathbf{F}}}\widehat{\mathbf{u}}_{i}-\mathbf{M}_{\mathbf{F}}\mathbf{u}_i \right\Vert^2\notag\\
\leq & C\left(\Vert\mathbf{M}_{\widehat{\mathbf{F}}}-\mathbf{M}_{\mathbf{F}}\Vert^2 \Vert\widehat{\mathbf{u}}_{i}-\mathbf{u}_i\Vert^2 + \Vert\mathbf{M}_{\widehat{\mathbf{F}}}-\mathbf{M}_{\mathbf{F}}\Vert^2 \Vert\mathbf{u}_i\Vert^2 + \Vert \mathbf{M}_{\mathbf{F}}\big(\widehat{\mathbf{u}}_{i}-\mathbf{u}_i\big)\Vert^2\right)\notag\\
\leq & C\left[\Vert\mathbf{M}_{\widehat{\mathbf{F}}}-\mathbf{M}_{\mathbf{F}}\Vert^2 \Vert\widehat{\mathbf{u}}_{i}-\mathbf{u}_i\Vert^2 + \Vert\mathbf{M}_{\widehat{\mathbf{F}}}-\mathbf{M}_{\mathbf{F}}\Vert^2 \Vert\mathbf{u}_i\Vert^2 + \Vert \widehat{\mathbf{u}}_{i}-\mathbf{u}_i\Vert^2 + \left(\frac{1}{T}\Vert\mathbf{F}\Vert^2\right)^2\Vert\widehat{\mathbf{u}}_{i}-\mathbf{u}_i\Vert^2\right],\label{Prop3proof6}
\end{align}
where we have used $\mathbf{M}_{\mathbf{F}}=\mathbf{I}_T-\mathbf{F}\mathbf{F}^\prime/T$ in deriving the last inequality.
For $\Vert\widehat{\mathbf{u}}_{i}-\mathbf{u}_i\Vert^2$, note that
\begin{align}
\Vert\widehat{\mathbf{u}}_{i}-\mathbf{u}_i\Vert^2=\Vert \mathbf{y}_{i}-\mathbf{C}_{i}\widehat{\boldsymbol{\theta}}_i-\big(\mathbf{y}_{i}-\mathbf{C}_{i}{\boldsymbol{\theta}}_i\big)\Vert^2\leq \Vert\mathbf{C}_{i}\Vert^2 \Vert\widehat{\boldsymbol{\theta}}_i-{\boldsymbol{\theta}}_i\Vert^2.\label{Prop3proof7}
\end{align}

Following similar arguments to the proof of Proposition
\ref{prop:asymrep_zu}(d) (i.e., the proof of $\mathbf{A}_i=O_p(1)$)
in the Online Supplement, we can obtain
$\Vert\mathbf{C}_{i}\Vert^2=O_p(T)$.
Furthermore, by (\ref{eq-theorem1proof1}) and Proposition
\ref{prop:asymrep_zu}(a)\&(d), we have
$\Vert\widehat{\boldsymbol{\theta}}_i-\boldsymbol{\theta}_i\Vert
=O_p\big(T^{-1/2}+\delta_{NT}^{-2}\big)$,
and hence
$\Vert\widehat{\boldsymbol{\theta}}_i-\boldsymbol{\theta}_i\Vert^2
=O_p\big(T^{-1}+\delta_{NT}^{-4}\big)$.
Plugging these results into (\ref{Prop3proof7}), we obtain
$\Vert\widehat{\mathbf{u}}_{i}-\mathbf{u}_i\Vert^2
=O_p\big(1+T\delta_{NT}^{-4}\big)
=O_p(T\delta_{NT}^{-2})$.

% Following similar arguments to the proof of Proposition \ref{prop:asymrep_zu}(d) (i.e., the proof of $\mathbf{A}_i=O_p(1)$) in the Online Supplement, we can obtain $\Vert\mathbf{C}_{i}\Vert^2=O_p(T)$. Furthermore, by the asymptotic distributional result in Theorem \ref{individual-asydist}, we have $\Vert\widehat{\boldsymbol{\theta}}_i-{\boldsymbol{\theta}}_i\Vert^2=O_p(T^{-1})$. Plugging these results into (\ref{Prop3proof7}), we obtain $\Vert\widehat{\mathbf{u}}_{i}-\mathbf{u}_i\Vert^2=O_p(1)$.
By Lemma~1(f) in the Online Supplement, we have $\Vert\mathbf{M}_{\widehat{\mathbf{F}}}-\mathbf{M}_{\mathbf{F}}\Vert^2=O_p(\delta_{NT}^{-2})$ Moreover, as established in the proof
of Proposition~\ref{prop:asymrep_zu} in the Online Supplement,
$\left(\frac{1}{T}\Vert\mathbf{F}\Vert^2\right)^2=O_p(1)$. Plugging these results into (\ref{Prop3proof6}) and noticing that $\Vert\mathbf{u}_i\Vert^2\leq C\left(\Vert\mathbf{F}\Vert^2\Vert\boldsymbol{\lambda}_i\Vert^2+\Vert\mathbf{G}\Vert^2\Vert\boldsymbol{\phi}_i\Vert^2+\Vert\boldsymbol{\varepsilon}_{i}\Vert^2\right)=O_p(T)$, we obtain
\begin{align}
\sum_{t=1}^T\vert\hat{\tilde{u}}_{it}-\tilde{u}_{it}\vert^2=O_p(\delta_{NT}^{-2}\,T).\label{Prop3proof8}
\end{align}
Furthermore, under Assumptions \ref{assumption-idioiny}--\ref{assumption-weight}, we can show that $\vert\hat{\tilde{u}}_{it}-\tilde{u}_{it}\vert^2\leq C$, $\vert\tilde{u}_{it}\vert^2\leq C$, and $\Vert\mathbf{z}_{it}\Vert^2\leq C$ in probability for all $i$ and $t$. Combined with (\ref{Prop3proof4}), (\ref{Prop3proof5}) and (\ref{Prop3proof8}), this implies
\begin{align}
\sum_{t=1}^T\Vert\hat{\mathbf{s}}_{it}-\mathbf{s}_{it}\Vert^2=O_p(\delta_{NT}^{-2}\,T).\label{Prop3proof9}
\end{align}
On the other hand, note that
\begin{align}
\sum_{t=1}^T \Vert{\mathbf{s}}_{it}\Vert^2\leq \sum_{t=1}^T \Vert{\mathbf{z}}_{it}\Vert^2 \vert\tilde{u}_{it}\vert^2=O_p(T).\label{Prop3proof10}
\end{align}
Plugging (\ref{Prop3proof9}) and (\ref{Prop3proof10}) into (\ref{Prop3proof3}), we obtain
\begin{align}
\left\Vert\widehat{\boldsymbol{\Phi}}_{i}(h)-\widetilde{\boldsymbol{\Phi}}_{i}(h)\right\Vert=O_p(\delta_{NT}^{-1})\label{Prop3proof11}
\end{align}
for all $h=0,\ldots,L_T$. With the boundedness of the kernel $K(\cdot)$ and the condition $L_T\delta_{NT}^{-1}\rightarrow0$,  this implies that
\begin{align}
\Vert\mathcal{D}_{i,1}\Vert=&\ \Vert\widehat{\boldsymbol{\Phi}}_{i}-\widetilde{\boldsymbol{\Phi}}_{i}\Vert=\left\Vert \big(\widehat{\boldsymbol{\Phi}}_{i}(0)-\widetilde{\boldsymbol{\Phi}}_{i}(0)\big)+\sum_{h=1}^{L_T}K\left(\frac{h}{L_T+1}\right)\left[\big(\widehat{\boldsymbol{\Phi}}_{i}(h)-\widetilde{\boldsymbol{\Phi}}_{i}(h)\big)+\big(\widehat{\boldsymbol{\Phi}}_{i}(h)-\widetilde{\boldsymbol{\Phi}}_{i}(h)\big)^\prime\right]\right\Vert\notag\\
\leq &\ CL_T\cdot O_p(\delta_{NT}^{-1})=O_p(L_T\delta_{NT}^{-1})=o_p(1).\label{Prop3proof12}
\end{align}
Next, consider $\mathcal{D}_{i,2}$. First, note that by the assumption in (\ref{Sec3insert3}) of Proposition \ref{individual-Cov-Consistency}, we have
\begin{align}
&\mathbb{E}\Vert\widetilde{\boldsymbol{\Phi}}_{i}(h)-\mathbb{E}\big(\widetilde{\boldsymbol{\Phi}}_{i}(h)\big)\Vert^2=\mathbb{E}\left\Vert\frac{1}{T}\sum_{t=h+1}^T\big[\mathbf{s}_{it}\mathbf{s}_{i,t-h}^\prime-\mathbb{E}\big(\mathbf{s}_{it}\mathbf{s}_{i,t-h}^\prime\big)\big]\right\Vert^2\notag\\
=&\ \frac{1}{T^2}\sum_{t=h+1}^T\sum_{s=h+1}^T\mathbb{E}\left(\mathrm{Tr}\left\{\big[\mathrm{vec}\big(\mathbf{s}_{it}\mathbf{s}_{i,t-h}^\prime\big)-\mathbb{E}\big(\mathrm{vec}(\mathbf{s}_{it}\mathbf{s}_{i,t-h}^\prime)\big)\big]\big[\mathrm{vec}\big(\mathbf{s}_{is}\mathbf{s}_{i,s-h}^\prime\big)-\mathbb{E}\big(\mathrm{vec}(\mathbf{s}_{is}\mathbf{s}_{i,s-h}^\prime)\big)\big]^\prime\right\}\right)\notag\\
=&\ \frac{1}{T^2}\sum_{t=h+1}^T\sum_{s=h+1}^T\mathrm{Tr}\left(\mathbb{E}\left\{\big[\mathrm{vec}\big(\mathbf{s}_{it}\mathbf{s}_{i,t-h}^\prime\big)-\mathbb{E}\big(\mathrm{vec}(\mathbf{s}_{it}\mathbf{s}_{i,t-h}^\prime)\big)\big]\big[\mathrm{vec}\big(\mathbf{s}_{is}\mathbf{s}_{i,s-h}^\prime\big)-\mathbb{E}\big(\mathrm{vec}(\mathbf{s}_{is}\mathbf{s}_{i,s-h}^\prime)\big)\big]^\prime\right\}\right)\notag\\
=&\ \frac{1}{T^2}\sum_{t=h+1}^T\sum_{s=h+1}^T\mathrm{Tr}\left[\mathrm{Cov}\big(\mathrm{vec}(\mathbf{s}_{it}\mathbf{s}_{i,t-h}^\prime),\ \mathrm{vec}(\mathbf{s}_{is}\mathbf{s}_{i,s-h}^\prime)\big)\right]\notag\\
\leq&\ \frac{3k}{T^2}\sum_{t=h+1}^T\sum_{s=h+1}^T\left\Vert\mathrm{Cov}\big(\mathrm{vec}(\mathbf{s}_{it}\mathbf{s}_{i,t-h}^\prime),\ \mathrm{vec}(\mathbf{s}_{is}\mathbf{s}_{i,s-h}^\prime)\big)\right\Vert\leq C \frac{1}{T^2}\cdot T=\frac{C}{T}\label{Prop3proof13}
\end{align}
for all $h=0,1,\ldots, L_T$, where in the third equality, we used the fact that both the trace operator $\mathrm{Tr}(\cdot)$ and the expectation operator $\mathbb{E}(\cdot)$ are linear operators, and in the first inequality we used the result that $\mathrm{Tr}(\mathbf{D})\leq \sqrt{m}\cdot\Vert\mathbf{D}\Vert$ for any $m\times m$ square matrix $\mathbf{D}$. With the above result and the boundedness of $K(\cdot)$, we readily have
\begin{align}
\Vert\mathcal{D}_{i,2}\Vert =O_p\big(L_T\cdot T^{-1/2}\big)=o_p(1).\label{Prop3proof14}
\end{align}
Finally, consider $\mathcal{D}_{i,3}$. Under Assumptions \ref{assumption-idioiny}--\ref{assumption-weight}, we can show that $\mathbf{s}_{it}$ has zero mean and is stationary over $t$ with absolutely summable serial covariances. Denoting $\boldsymbol{\Omega}_{\mathbf{s},i,h}=\mathbb{E}\big(\mathbf{s}_{it}\mathbf{s}_{i,t-h}^\prime\big)$ for $h=0,1,\ldots$, then we have $\mathbb{E}\big(\widetilde{\boldsymbol{\Phi}}_i(h)\big)=\frac{1}{T}\sum_{t=h+1}^T\boldsymbol{\Omega}_{\mathbf{s},i,h}=\frac{T-h}{T}\boldsymbol{\Omega}_{\mathbf{s},i,h}$. Hence,
\begin{align}
\mathbb{E}\big(\widetilde{\boldsymbol{\Phi}}_i\big)=\boldsymbol{\Omega}_{\mathbf{s},i,0}+\sum_{h=1}^{L_T}K\left(\frac{h}{L_T+1}\right)\left(\frac{T-h}{T}\right)\big(\boldsymbol{\Omega}_{\mathbf{s},i,h}+\boldsymbol{\Omega}_{\mathbf{s},i,h}^\prime\big).\label{Prop3proof15}
\end{align}
Further note that
\begin{align}
\boldsymbol{\Phi}_i=\mathrm{Var}\left(\frac{1}{\sqrt{T}}\sum_{t=1}^T\mathbf{s}_{it}\right)=\frac{1}{T}\sum_{t=1}^T\sum_{s=1}^T\mathbb{E}\big(\mathbf{s}_{it}\mathbf{s}_{is}^\prime\big)=\boldsymbol{\Omega}_{\mathbf{s},i,0}+\sum_{h=1}^{T-1}\left(\frac{T-h}{T}\right)\big(\boldsymbol{\Omega}_{\mathbf{s},i,h}+\boldsymbol{\Omega}_{\mathbf{s},i,h}^\prime\big).\label{Prop3proof16}
\end{align}
Therefore,
\begin{align}
&\mathcal{D}_{i,3}= \mathbb{E}\big(\widetilde{\boldsymbol{\Phi}}_i\big)-\boldsymbol{\Phi}_i\notag\\
=&\ \sum_{h=1}^{L_T}\left[K\left(\frac{h}{L_T+1}\right)-1\right]\left(\frac{T-h}{T}\right)\big(\boldsymbol{\Omega}_{\mathbf{s},i,h}+\boldsymbol{\Omega}_{\mathbf{s},i,h}^\prime\big) - \sum_{h=L_T+1}^{T-1}\left(\frac{T-h}{T}\right)\big(\boldsymbol{\Omega}_{\mathbf{s},i,h}+\boldsymbol{\Omega}_{\mathbf{s},i,h}^\prime\big)\notag\\
=:&\ \mathcal{D}_{i,3,1} + \mathcal{D}_{i,3,2}.\label{Prop3proof17}
\end{align}
By the properties of $K(\cdot)$ and the absolute summability of $\boldsymbol{\Omega}_{\mathbf{s},i,h}$, for any $\epsilon>0$, there exists an integer $0<M<L_T$, such that for sufficiently large $T$, $\left\vert K\left(\frac{h}{L_T+1}\right)-1\right\vert<\epsilon/C^\ast$ for all $h=1,\ldots,M$ and $\sum_{h=M+1}^{L_T}\Vert\boldsymbol{\Omega}_{\mathbf{s},i,h}\Vert<\epsilon/C^\ast$, where $C^\ast=4\cdot\max\big\{\sum_{h=1}^\infty\Vert\boldsymbol{\Omega}_{\mathbf{s},i,h}\Vert,\  \big(1+\sup_{0<u<1}|K(u)|\big)\big\}$. Therefore,
\begin{align}
\Vert\mathcal{R}_{i,3,1}\Vert\leq&\ 2\sum_{h=1}^{M}\left\vert K\left(\frac{h}{L_T+1}\right)-1\right\vert\left(\frac{T-h}{T}\right)\Vert\boldsymbol{\Omega}_{\mathbf{s},i,h}\Vert+2\sum_{h=M+1}^{L_T}\left\vert K\left(\frac{h}{L_T+1}\right)-1\right\vert\left(\frac{T-h}{T}\right)\Vert\boldsymbol{\Omega}_{\mathbf{s},i,h}\Vert\notag\\
< &\ \frac{\epsilon}{C^\ast}\cdot \frac{C^\ast}{2}+ \frac{C^\ast}{2}\cdot \frac{\epsilon}{C^\ast}=\epsilon\label{Prop3proof18}
\end{align}
for sufficiently large $T$. This proves $\Vert\mathcal{D}_{i,3,1}\Vert=o(1)$. For $\Vert\mathcal{D}_{i,3,2}\Vert$, it is easy to see that
\begin{align}
\Vert\mathcal{D}_{i,3,2}\Vert\leq 2\sum_{h=L_T+1}^{T-1}\Vert\boldsymbol{\Omega}_{\mathbf{s},i,h}\Vert\rightarrow0\label{Prop3proof19}
\end{align}
as $T\rightarrow\infty$. Combining (\ref{Prop3proof17})--(\ref{Prop3proof19}), we obtain $\Vert\mathcal{D}_{i,3}\Vert=o(1)$. This completes the proof of Proposition \ref{individual-Cov-Consistency}. \hfill $\Box$

\bigskip

Next, we present Proposition \ref{prop:asymrep_pzu} before proving Theorem \ref{MG-asydist}.  \medskip

\begin{proposition}
\label{prop:asymrep_pzu}  Denote $\widehat{\mathbf{Q}}_i=\big(\widehat{\mathbf{A}}_{i}^\prime \widehat{\mathbf{B}}_{i}^{-1}\widehat{\mathbf{A}}_{i}\big)^{-1}\widehat{\mathbf{A}}_{i}^\prime \widehat{\mathbf{B}}_{i}^{-1}$ and ${\mathbf{Q}}_i=\big({\mathbf{A}}_{i}^\prime {\mathbf{B}}_{i}^{-1}{\mathbf{A}}_{i}\big)^{-1}{\mathbf{A}}_{i}^\prime {\mathbf{B}}_{i}^{-1}$. Under Assumptions \ref{assumption-idioiny}--\ref{assumption-loadings}, \ref{assumption-weight}\ref{assumption-weight-diag}-\ref{assumption-weight-bound}, \ref{assumption-random}, and \ref{assumption-uniident}\ref{assumption-uniident-unifullrank}, we have
\begin{align}
&(a)~~\frac{1}{N} \sum_{i=1}^N \left\Vert\frac{1}{T}\big(\widehat{\mathbf{Z}}_i-\mathbf{Z}_i\big)^\prime\mathbf{u}_i\right\Vert^2=O_p(\delta_{NT}^{-4})\,,  \notag \\
&(b)~~\frac{1}{N} \sum_{i=1}^N  \left\Vert\frac{1}{T}\mathbf{Z}_i^\prime\mathbf{u}_i\right\Vert^2=O_p(T^{-1})\,,  \notag\\
&(c)~~\frac{1}{N} \sum_{i=1}^N  \left\Vert\widehat{\mathbf{Q}}_i-{\mathbf{Q}}_i\right\Vert^2=O_p(\delta_{NT}^{-2})\,,  \notag \\
&(d)~~\frac{1}{N} \sum_{i=1}^N  \left\Vert{\mathbf{Q}}_i\right\Vert^2=O_p(1)\,,  \notag
\end{align}
as $N,T\rightarrow\infty$.
%where $\boldsymbol{\mathbf{Z}}_{i}=\left(\sum_{j=1}^Nw_{ij}\mathbf{M}_{
%\mathbf{F}}\boldsymbol{\mathbf{X}}_{j},\mathbf{M}_{\mathbf{F}}\mathbf{M}_{
%\mathbf{F}_{-1}}\boldsymbol{\mathbf{X}}_{i,-1},\mathbf{M}_{\mathbf{F}}%
%\boldsymbol{\mathbf{X}}_{i}\right)$.
\end{proposition}
\medskip

\noindent \textbf{Proof of Theorem \ref{MG-asydist}.} Note that
\begin{align}
\big(\widehat{\boldsymbol\theta}_{MG}-\boldsymbol{\theta}\big)=\frac{1}{N}\sum\limits_{i=1}^N\big(\widehat{\boldsymbol{\theta}}_i-\boldsymbol{\theta}_i\big)+\frac{1}{N}\sum\limits_{i=1}^N\big({\boldsymbol{\theta}}_i-\boldsymbol{\theta}\big)
=\frac{1}{N}\sum\limits_{i=1}^N\big(\widehat{\boldsymbol{\theta}}_i-\boldsymbol{\theta}_i\big)+\frac{1}{N}\sum\limits_{i=1}^N\mathbf{e}_i.\label{eq-theorem2proof1}
\end{align}
By (\ref{eq-theorem1proof1}), we have
 \begin{align}
\frac{1}{N}\sum\limits_{i=1}^N\big(\widehat{\boldsymbol{\theta}}_i-\boldsymbol{\theta}_i\big)=&\frac{1}{NT}\sum\limits_{i=1}^N\left(\widehat{\mathbf{A}}_i^{\prime }%
\widehat{\mathbf{B}}^{-1}_i\widehat{\mathbf{A}}_i\right)^{-1}\widehat{
\mathbf{A}}_i^{\prime }\widehat{\mathbf{B}}_i^{-1}\big(\widehat{\mathbf{Z}}_{i}^{\prime }\mathbf{u}_i\big)=\frac{1}{NT}\sum\limits_{i=1}^N\widehat{\mathbf{Q}}_i\big(\widehat{\mathbf{Z}}_{i}^{\prime }\mathbf{u}_i\big)\notag\\
=&\frac{1}{NT} \sum_{i=1}^{N}\left(\widehat{\mathbf{Q}}_i-\mathbf{Q}_i\right)  \left[\left(\widehat{\mathbf{Z}}_{i}-\mathbf{Z}
_{i}\right)^{\prime }\boldsymbol{u}_{i}\right]+\frac{1}{NT} \sum_{i=1}^{N}\left(\widehat{\mathbf{Q}}_i-\mathbf{Q}_i\right) \mathbf{Z}
_{i}^{\prime }\boldsymbol{u}_{i}\notag \\
&+\frac{1}{NT}\sum_{i=1}^{N} \mathbf{Q}_i \left[\left(\widehat{\mathbf{Z}}_{i}-\mathbf{Z}
_{i}\right)^{\prime }\boldsymbol{u}_{i}\right]+\frac{1}{NT}\sum_{i=1}^{N} \mathbf{Q}_i \mathbf{Z}
_{i}^{\prime }\boldsymbol{u}_{i}\notag\\
=:&\ \Theta_{NT,1}+\Theta_{NT,2}+\Theta_{NT,3}+\Theta_{NT,4}.\label{eq-theorem2proof2}
 \end{align}
First, consider $\Theta_{NT,1}$. By Proposition \ref{prop:asymrep_pzu}(a)\&(c), we obtain
\begin{align}
\left\Vert\Theta_{NT,1}\right\Vert\leq \sqrt{\left(\frac{1}{N}\sum_{i=1}^N\left\Vert\widehat{\mathbf{Q}}_i-{\mathbf{Q}}_i\right\Vert^2\right)\left(\frac{1}{N} \sum_{i=1}^N \left\Vert\frac{1}{T}\big(\widehat{\mathbf{Z}}_i-\mathbf{Z}_i\big)^\prime\mathbf{u}_i\right\Vert^2\right)}=O_p(\delta_{NT}^{-1}\cdot\delta_{NT}^{-2})=O_p(\delta_{NT}^{-3}).\label{eq-theorem2proof3}
\end{align}
Analogously, for the remaining terms in (\ref{eq-theorem2proof2}), we have
\begin{align}
&\left\Vert\Theta_{NT,2}\right\Vert\leq \sqrt{\left(\frac{1}{N}\sum_{i=1}^N\left\Vert\widehat{\mathbf{Q}}_i-{\mathbf{Q}}_i\right\Vert^2\right)\left(\frac{1}{N} \sum_{i=1}^N  \left\Vert\frac{1}{T}\mathbf{Z}_i^\prime\mathbf{u}_i\right\Vert^2\right)}=O_p(\delta_{NT}^{-1}\cdot T^{-1/2})=O_p(\delta_{NT}^{-1} T^{-1/2}),\label{eq-theorem2proof4}\\
&\left\Vert\Theta_{NT,3}\right\Vert\leq \sqrt{\left(\frac{1}{N} \sum_{i=1}^N  \left\Vert{\mathbf{Q}}_i\right\Vert^2\right)\left(\frac{1}{N} \sum_{i=1}^N \left\Vert\frac{1}{T}\big(\widehat{\mathbf{Z}}_i-\mathbf{Z}_i\big)^\prime\mathbf{u}_i\right\Vert^2\right)}=O_p(1)\cdot O_p(\delta_{NT}^{-2})=O_p(\delta_{NT}^{-2}),\label{eq-theorem2proof5}\\
&\left\Vert\Theta_{NT,4}\right\Vert\leq \sqrt{\left(\frac{1}{N} \sum_{i=1}^N  \left\Vert\mathbf{Q}_i\right\Vert^2\right)\left(\frac{1}{N} \sum_{i=1}^N  \left\Vert\frac{1}{T}\mathbf{Z}_i^\prime\mathbf{u}_i\right\Vert^2\right)}=O_p(1)\cdot O_p(T^{-1/2})=O_p(T^{-1/2}).\label{eq-theorem2proof6}
\end{align}
From (\ref{eq-theorem2proof2})--(\ref{eq-theorem2proof6}) we obtain
\begin{equation}\label{eq-theorem2proof11}
\frac{1}{N}\sum\limits_{i=1}^N\big(\widehat{\boldsymbol{\theta}}_i-\boldsymbol{\theta}_i\big)=O_p(\delta_{NT}^{-2}+T^{-1/2})=o_p(1).
\end{equation}
Meanwhile, by Assumption \ref{assumption-random}\ref{assumption-random-dist} and the Law of Large Numbers, we have
\begin{align}
\frac{1}{N}\sum\limits_{i=1}^N\mathbf{e}_i\stackrel{p}{\longrightarrow}\mathbf{0},\ \ \ {\rm as\ }N\rightarrow\infty.\label{eq-theorem2proofinsert1}
\end{align}
In view of (\ref{eq-theorem2proof1}), (\ref{eq-theorem2proof11}) and (\ref{eq-theorem2proofinsert1}), we have completed the proof of the consistency of the MGIV estimator.\medskip

To prove the asymptotic distribution for the MGIV estimator, note that under Assumption \ref{assumption-random}\ref{assumption-random-dist}, we have
 \begin{equation}
 \frac{1}{N^{1/2}}\sum\limits_{i=1}^N\mathbf{e}_i \stackrel{d}{\longrightarrow} N\big(\mathbf{0}, \boldsymbol{\Sigma}_{\boldsymbol\theta}\big),\label{eq-theorem2proofinsert2}
 \end{equation}
 as $N\rightarrow\infty$. In view of (\ref{eq-theorem2proof1}), it remains to show that
 \begin{equation}\label{eq-theorem2proof12}
N^{1/2}\cdot \frac{1}{N}\sum\limits_{i=1}^N\big(\widehat{\boldsymbol{\theta}}_i-\boldsymbol{\theta}_i\big)=o_p(1),\ \ \ {\rm as\ }N,T\rightarrow\infty.
 \end{equation}
 To prove this, {\it we first note that when $N/T^{2}\rightarrow0$, we have $N^{1/2}\cdot\delta_{NT}^{-1}\cdot T^{-1/2}\rightarrow0$ and $N^{1/2}\cdot \delta_{NT}^{-2}\rightarrow0$}.  Hence, from (\ref{eq-theorem2proof3})--(\ref{eq-theorem2proof5}), we have
 \begin{equation}
N^{1/2}\cdot \Theta_{NT,1}=o_p(1),\ \ \ N^{1/2}\cdot \Theta_{NT,2}=o_p(1),\ \ \ N^{1/2}\cdot \Theta_{NT,3}=o_p(1).\label{eq-theorem2proof13}
 \end{equation}
 For $\Theta_{NT,4}$, by Assumption \ref{assumption-uniident}\ref{assumption-uniident-uniorder}, we have
 \begin{align}
\Theta_{NT,4}=\frac{1}{NT}\sum_{i=1}^{N} \mathbf{Q}_i \mathbf{Z}
_{i}^{\prime }\boldsymbol{u}_{i}=\frac{1}{NT}\sum_{i=1}^{N}(\mathbf{A}_i^\prime\mathbf{B}_i^{-1}\mathbf{A}_i)^{-1}\mathbf{A}_i^\prime\mathbf{B}_i^{-1}\mathbf{Z}_i^\prime\big(\mathbf{G}\boldsymbol{%
\phi}_i+\boldsymbol{\varepsilon}_{i}\big)=O_p(N^{-1/2}T^{-1/2}).\notag
 \end{align}
This implies
\begin{align}
N^{1/2}\cdot \Theta_{NT,4}=O_p(T^{-1/2})=o_p(1).\label{eq-theorem2proofinsert3}
\end{align}
Combining (\ref{eq-theorem2proof2}) with (\ref{eq-theorem2proof13}) and (\ref{eq-theorem2proofinsert3}), we obtain (\ref{eq-theorem2proof12}). This completes the proof of the asymptotic distributional result.

Next, we prove the consistency of $\widehat{\boldsymbol{\Sigma}}_{\boldsymbol\theta}$. Note that $\widehat{\boldsymbol{\theta}}_i-\boldsymbol{\theta}=\widehat{\boldsymbol{\theta}}_i-\boldsymbol{\theta}_i+\boldsymbol{\theta}_i-\boldsymbol{\theta}=\widehat{\boldsymbol{\theta}}_i-\boldsymbol{\theta}_i+\mathbf{e}_i$. Hence,
\begin{align}
\widehat{\boldsymbol{\Sigma}}_{\boldsymbol\theta}-{\boldsymbol{\Sigma}}_{\boldsymbol\theta}=&\frac{1}{N-1}\sum\limits_{i=1}^N \left(\widehat{\boldsymbol{\theta}}_i-\widehat{\boldsymbol\theta}_{MG}\right)\left(\widehat{\boldsymbol{\theta}}_i-\widehat{\boldsymbol\theta}_{MG}\right)^\prime -{\boldsymbol{\Sigma}}_{\boldsymbol\theta}\notag\\
=&\frac{1}{N-1}\sum\limits_{i=1}^N \left(\widehat{\boldsymbol{\theta}}_i-\boldsymbol{\theta}\right)\left(\widehat{\boldsymbol{\theta}}_i-\boldsymbol{\theta}\right)^\prime-\frac{1}{N-1}\sum\limits_{i=1}^N \left(\widehat{\boldsymbol{\theta}}_i-\boldsymbol{\theta}\right)\left(\widehat{\boldsymbol{\theta}}_{MG}-\boldsymbol{\theta}\right)^\prime\notag\\
&-\frac{1}{N-1}\sum\limits_{i=1}^N \left(\widehat{\boldsymbol{\theta}}_{MG}-\boldsymbol{\theta}\right)\left(\widehat{\boldsymbol{\theta}}_i-\boldsymbol{\theta}\right)^\prime+\frac{N}{N-1}\left(\widehat{\boldsymbol{\theta}}_{MG}-\boldsymbol{\theta}\right)\left(\widehat{\boldsymbol{\theta}}_{MG}-\boldsymbol{\theta}\right)^\prime-{\boldsymbol{\Sigma}}_{\boldsymbol\theta}\notag\\
=&\frac{1}{N-1}\sum\limits_{i=1}^N \left(\widehat{\boldsymbol{\theta}}_i-\boldsymbol{\theta}_i\right)\left(\widehat{\boldsymbol{\theta}}_i-\boldsymbol{\theta}_i\right)^\prime + \frac{1}{N-1}\sum\limits_{i=1}^N \left(\widehat{\boldsymbol{\theta}}_i-\boldsymbol{\theta}_i\right)\mathbf{e}_i^\prime + \frac{1}{N-1}\sum\limits_{i=1}^N \mathbf{e}_i\left(\widehat{\boldsymbol{\theta}}_i-\boldsymbol{\theta}_i\right)^\prime \notag\\
&-\frac{1}{N-1}\sum\limits_{i=1}^N \left(\widehat{\boldsymbol{\theta}}_i-\boldsymbol{\theta}_i\right)\left(\widehat{\boldsymbol{\theta}}_{MG}-\boldsymbol{\theta}\right)^\prime -\frac{1}{N-1}\sum\limits_{i=1}^N \mathbf{e}_i\left(\widehat{\boldsymbol{\theta}}_{MG}-\boldsymbol{\theta}\right)^\prime \notag\\
&-\left(\widehat{\boldsymbol{\theta}}_{MG}-\boldsymbol{\theta}\right)\left[\frac{1}{N-1}\sum\limits_{i=1}^N \left(\widehat{\boldsymbol{\theta}}_i-\boldsymbol{\theta}_i\right)^\prime\right]-\left(\widehat{\boldsymbol{\theta}}_{MG}-\boldsymbol{\theta}\right)\left[\frac{1}{N-1}\sum\limits_{i=1}^N \mathbf{e}_i^\prime\right] \notag\\
&+\frac{N}{N-1}\left(\widehat{\boldsymbol{\theta}}_{MG}-\boldsymbol{\theta}\right)\left(\widehat{\boldsymbol{\theta}}_{MG}-\boldsymbol{\theta}\right)^\prime +\left[\frac{1}{N-1}\sum\limits_{i=1}^N\mathbf{e}_i\mathbf{e}_i^\prime - {\boldsymbol{\Sigma}}_{\boldsymbol\theta}\right]\notag\\
=:& \sum\limits_{k=1}^9\Theta_{NT,k}^\ast. \label{eq-theorem2proof15}
\end{align}
By Assumption \ref{assumption-random}\ref{assumption-random-dist}, the consistency of $\widehat{\boldsymbol{\theta}}_{MG}$, and the result in (\ref{eq-theorem2proof11}), we can easily show that
\begin{align}
\Theta_{NT,k}^\ast=o_p(1),\ \ \ k=4,5,\ldots,9.\label{eq-theorem2proof16}
\end{align}
%Following the proof of Theorem 5(ii) in \citet{NorkuteEtal2021} we can also show that
Furthermore, by (\ref{eq-theorem1proof1}) and Proposition \ref{prop:asymrep_zu} (a)\&(d),
we have
\begin{align}
\Theta_{NT,1}^\ast=O_p(T^{-1})=o_p(1).\label{eq-theorem2proof17}
\end{align}
For $\Theta_{NT,2}^\ast$, using Assumption \ref{assumption-random}\ref{assumption-random-dist}, we have
\begin{align}
\left\Vert\Theta_{NT,2}^\ast\right\Vert\leq \frac{1}{N-1}\sum\limits_{i=1}^N \left\Vert\left(\widehat{\boldsymbol{\theta}}_i-\boldsymbol{\theta}_i\right)\right\Vert \left\Vert\mathbf{e}_i\right\Vert\leq \frac{1}{N-1}\left(\sum\limits_{i=1}^N \left\Vert\left(\widehat{\boldsymbol{\theta}}_i-\boldsymbol{\theta}_i\right)\right\Vert^2\right)^{1/2} \left(\sum\limits_{i=1}^N\left\Vert\mathbf{e}_i\right\Vert^2\right)^{1/2}.\label{eq-theorem2proof18}
\end{align}
By Assumption \ref{assumption-random}\ref{assumption-random-dist}, $\mathbb{E}\left(\sum\limits_{i=1}^N\left\Vert\mathbf{e}_i\right\Vert^2\right)\leq CN$ and hence $\sum\limits_{i=1}^N\left\Vert\mathbf{e}_i\right\Vert^2=O_p(N)$. On the other hand, again by (\ref{eq-theorem1proof1}) and Proposition \ref{prop:asymrep_zu} (a)\&(d), we have $\sum\limits_{i=1}^N \left\Vert\left(\widehat{\boldsymbol{\theta}}_i-\boldsymbol{\theta}_i\right)\right\Vert^2=O_p(NT^{-1})$. Plugging these results into (\ref{eq-theorem2proof18}) we obtain
\begin{align}
\left\Vert\Theta_{NT,2}^\ast\right\Vert=O_p\left(\frac{N}{N-1}\cdot T^{-1/2}\right)=O_p(T^{-1/2})=o_p(1).\label{eq-theorem2proof19}
\end{align}
Similarly, we have $\left\Vert\Theta_{NT,3}^\ast\right\Vert=O_p(T^{-1/2})=o_p(1)$. Combining this with (\ref{eq-theorem2proof15})--(\ref{eq-theorem2proof17}) and (\ref{eq-theorem2proof19}), we obtain $\widehat{\boldsymbol{\Sigma}}_{\boldsymbol\theta}-{\boldsymbol{\Sigma}}_{\boldsymbol\theta}=o_p(1)$. This completes the proof of the consistency of $\widehat{\boldsymbol{\Sigma}}_{\boldsymbol\theta}$. \hfill $\Box$

\bigskip

\section*{Appendix B: Additional Simulation Results}

\renewcommand{\thetable}{B.\arabic{table}}
\setcounter{table}{0}
\makeatletter

\begin{landscape}
\begin{table}[ht]
\centering
\begin{threeparttable}
\footnotesize
\footscriptsize
\setlength{\tabcolsep}{2.7pt} % tighten if needed
\renewcommand{\arraystretch}{1.15}
\caption{Mean Group IV results for $\pi_{u}=1/4$}
\label{tab:sim_results_rho_psi_pi1/4_MGIV}
\begin{tabular}{l*{27}{c}}
%\toprule
\addlinespace[2pt]
\multicolumn{28}{c}{\small \textbf{Panel A ($\rho=0.4$)}}\\

& \multicolumn{6}{c}{\small \textbf{MGIV$^{[1]}$}} &  & \multicolumn{6}{c}{\small \textbf{MGIV$^{[2]}$}} &  & \multicolumn{6}{c}{\small \textbf{MGIV$^{[3]}$}} &  & \multicolumn{6}{c}{\small \textbf{MGIV$^{[L]}$}} \\
%\addlinespace[2pt]
\cmidrule(lr){2-7} \cmidrule(lr){9-14} \cmidrule(lr){16-21} \cmidrule(lr){23-28}

\multicolumn{28}{c}{\small \textbf{Case I:} $N=25\tau$, $T=100\tau$}\\

$\tau$ & Mean & RMSE & ARB & Size & Power & J-test &  & Mean & RMSE & ARB & Size & Power & J-test &  & Mean & RMSE & ARB & Size & Power & J-test &  & Mean & RMSE & ARB & Size & Power & J-test \\
\cmidrule(lr){2-7} \cmidrule(lr){9-14} \cmidrule(lr){16-21} \cmidrule(lr){23-28}

1 & 0.405 & 0.026 & 1.275 & 0.073 & 0.971 & 0.079 &  & 0.404 & 0.026 & 0.919 & 0.065 & 0.970 & 0.129 &  & 0.409 & 0.027 & 2.258 & 0.080 & 0.973 & 0.083 &  & 0.404 & 0.027 & 1.043 & 0.065 & 0.957 & 0.149 \\
2 & 0.403 & 0.018 & 0.816 & 0.058 & 1.000 & 0.072 &  & 0.402 & 0.018 & 0.570 & 0.058 & 1.000 & 0.292 &  & 0.406 & 0.019 & 1.540 & 0.074 & 1.000 & 0.065 &  & 0.403 & 0.018 & 0.620 & 0.057 & 1.000 & 0.123 \\
4 & 0.401 & 0.012 & 0.309 & 0.045 & 1.000 & 0.081 &  & 0.401 & 0.012 & 0.190 & 0.048 & 1.000 & 0.866 &  & 0.403 & 0.012 & 0.790 & 0.050 & 1.000 & 0.094 &  & 0.401 & 0.011 & 0.175 & 0.048 & 1.000 & 0.868 \\

\multicolumn{28}{c}{\small \textbf{Case II:} $N=100\tau$, $T=25\tau$}\\

$\tau$ & Mean & RMSE & ARB & Size & Power & J-test &  & Mean & RMSE & ARB & Size & Power & J-test &  & Mean & RMSE & ARB & Size & Power & J-test &  & Mean & RMSE & ARB & Size & Power & J-test \\
\cmidrule(lr){2-7} \cmidrule(lr){9-14} \cmidrule(lr){16-21} \cmidrule(lr){23-28}

1 & 0.406 & 0.020 & 1.485 & 0.106 & 1.000 & 0.058 &  & 0.405 & 0.020 & 1.261 & 0.104 & 1.000 & 0.101 &  & 0.408 & 0.021 & 2.045 & 0.138 & 1.000 & 0.060 &  & 0.407 & 0.022 & 1.696 & 0.091 & 0.993 & 0.151 \\
2 & 0.408 & 0.014 & 1.257 & 0.089 & 1.000 & 0.052 &  & 0.407 & 0.013 & 1.113 & 0.079 & 1.000 & 0.290 &  & 0.412 & 0.017 & 2.928 & 0.253 & 1.000 & 0.059 &  & 0.407 & 0.014 & 1.253 & 0.075 & 1.000 & 0.143 \\
4 & 0.405 & 0.009 & 1.342 & 0.082 & 1.000 & 0.094 &  & 0.404 & 0.008 & 1.019 & 0.068 & 1.000 & 0.889 &  & 0.409 & 0.012 & 2.346 & 0.319 & 1.000 & 0.081 &  & 0.400 & 0.006 & 0.025 & 0.067 & 1.000 & 0.891 \\

\multicolumn{28}{c}{\small \textbf{Case III:} $N=50\tau$, $T=50\tau$}\\

$\tau$ & Mean & RMSE & ARB & Size & Power & J-test &  & Mean & RMSE & ARB & Size & Power & J-test &  & Mean & RMSE & ARB & Size & Power & J-test &  & Mean & RMSE & ARB & Size & Power & J-test \\
\cmidrule(lr){2-7} \cmidrule(lr){9-14} \cmidrule(lr){16-21} \cmidrule(lr){23-28}
1 & 0.408 & 0.023 & 1.969 & 0.086 & 0.997 & 0.068 &  & 0.407 & 0.022 & 1.616 & 0.086 & 0.998 & 0.121 &  & 0.412 & 0.024 & 2.879 & 0.111 & 0.998 & 0.073 &  & 0.407 & 0.024 & 1.648 & 0.079 & 0.986 & 0.155 \\
2 & 0.405 & 0.014 & 1.224 & 0.078 & 1.000 & 0.069 &  & 0.404 & 0.014 & 0.908 & 0.069 & 1.000 & 0.300 &  & 0.409 & 0.016 & 2.203 & 0.122 & 1.000 & 0.069 &  & 0.404 & 0.014 & 1.002 & 0.070 & 1.000 & 0.140 \\
4 & 0.403 & 0.009 & 0.708 & 0.069 & 1.000 & 0.087 &  & 0.402 & 0.009 & 0.506 & 0.064 & 1.000 & 0.875 &  & 0.406 & 0.011 & 1.460 & 0.123 & 1.000 & 0.087 &  & 0.398 & 0.009 & .500 & 0.063 & 1.000 & 0.882 \\
\midrule
% ===== Panel B =====
\multicolumn{28}{c}{\small \textbf{Panel B ($\psi=0.25$)}}\\

& \multicolumn{6}{c}{\small \textbf{MGIV$^{[1]}$}} &  & \multicolumn{6}{c}{\small \textbf{MGIV$^{[2]}$}} &  & \multicolumn{6}{c}{\small \textbf{MGIV$^{[3]}$}} &  & \multicolumn{6}{c}{\small \textbf{MGIV$^{[L]}$}} \\
%\addlinespace[2pt]
\cmidrule(lr){2-7} \cmidrule(lr){9-14} \cmidrule(lr){16-21} \cmidrule(lr){23-28}

\multicolumn{28}{c}{\small \textbf{Case I:} $N=25\tau$, $T=100\tau$}\\

$\tau$ & Mean & RMSE & ARB & Size & Power & J-test &  & Mean & RMSE & ARB & Size & Power & J-test &  & Mean & RMSE & ARB & Size & Power & J-test &  & Mean & RMSE & ARB & Size & Power & J-test \\
\cmidrule(lr){2-7} \cmidrule(lr){9-14} \cmidrule(lr){16-21} \cmidrule(lr){23-28}

1 & 0.251 & 0.028 & 0.566 & 0.060 & 0.940 & 0.079 &  & 0.252 & 0.028 & 0.797 & 0.058 & 0.939 & 0.129 &  & 0.251 & 0.027 & 0.354 & 0.061 & 0.950 & 0.083 &  & 0.251 & 0.031 & 0.269 & 0.055 & 0.919 & 0.149 \\
2 & 0.250 & 0.016 & 0.205 & 0.061 & 1.000 & 0.072 &  & 0.250 & 0.016 & 0.124 & 0.059 & 1.000 & 0.292 &  & 0.249 & 0.016 & 0.359 & 0.058 & 1.000 & 0.065 &  & 0.250 & 0.016 & 0.192 & 0.062 & 0.999 & 0.123 \\
4 & 0.250 & 0.010 & 0.133 & 0.056 & 1.000 & 0.081 &  & 0.250 & 0.010 & 0.049 & 0.058 & 1.000 & 0.866 &  & 0.249 & 0.010 & 0.232 & 0.056 & 1.000 & 0.094 &  & 0.250 & 0.010 & 0.120 & 0.058 & 0.999 & 0.868 \\

\multicolumn{28}{c}{\small \textbf{Case II:} $N=100\tau$, $T=25\tau$}\\

$\tau$ & Mean & RMSE & ARB & Size & Power & J-test &  & Mean & RMSE & ARB & Size & Power & J-test &  & Mean & RMSE & ARB & Size & Power & J-test &  & Mean & RMSE & ARB & Size & Power & J-test \\
\cmidrule(lr){2-7} \cmidrule(lr){9-14} \cmidrule(lr){16-21} \cmidrule(lr){23-28}

1 & 0.251 & 0.028 & 0.563 & 0.053 & 0.941 & 0.058 &  & 0.252 & 0.027 & 0.615 & 0.043 & 0.955 & 0.101 &  & 0.251 & 0.026 & 0.272 & 0.054 & 0.955 & 0.060 &  & 0.251 & 0.033 & 0.208 & 0.045 & 0.868 & 0.151 \\
2 & 0.251 & 0.013 & 0.227 & 0.053 & 1.000 & 0.052 &  & 0.251 & 0.013 & 0.344 & 0.046 & 1.000 & 0.290 &  & 0.250 & 0.013 & 0.205 & 0.055 & 1.000 & 0.059 &  & 0.250 & 0.015 & 0.112 & 0.045 & 0.998 & 0.143 \\
4 & 0.249 & 0.007 & 0.250 & 0.053 & 1.000 & 0.094 &  & 0.250 & 0.007 & 0.092 & 0.050 & 1.000 & 0.889 &  & 0.249 & 0.007 & 0.508 & 0.057 & 1.000 & 0.081 &  & 0.250 & 0.007 & 0.102 & 0.050 & 0.999 & 0.891 \\

\multicolumn{28}{c}{\small \textbf{Case III:} $N=50\tau$, $T=50\tau$}\\

$\tau$ & Mean & RMSE & ARB & Size & Power & J-test &  & Mean & RMSE & ARB & Size & Power & J-test &  & Mean & RMSE & ARB & Size & Power & J-test &  & Mean & RMSE & ARB & Size & Power & J-test \\
\cmidrule(lr){2-7} \cmidrule(lr){9-14} \cmidrule(lr){16-21} \cmidrule(lr){23-28}

1 & 0.251 & 0.027 & 0.202 & 0.060 & 0.946 & 0.068 &  & 0.251 & 0.027 & 0.411 & 0.066 & 0.938 & 0.121 &  & 0.250 & 0.026 & 0.138 & 0.074 & 0.944 & 0.073 &  & 0.250 & 0.032 & 0.095 & 0.061 & 0.892 & 0.155 \\
2 & 0.250 & 0.014 & 0.010 & 0.061 & 1.000 & 0.069 &  & 0.250 & 0.014 & 0.122 & 0.059 & 1.000 & 0.300 &  & 0.249 & 0.014 & 0.266 & 0.059 & 1.000 & 0.069 &  & 0.250 & 0.015 & 0.102 & 0.066 & 1.000 & 0.140 \\
4 & 0.250 & 0.008 & 0.187 & 0.050 & 1.000 & 0.087 &  & 0.250 & 0.008 & 0.068 & 0.049 & 1.000 & 0.875 &  & 0.249 & 0.008 & 0.336 & 0.050 & 1.000 & 0.087 &  & 0.250 & 0.008 & 0.160 & 0.049 & 1.000 & 0.882 \\
% Row 36: (optional) final spacer or notes row
\multicolumn{28}{c}{}\\
\bottomrule
\end{tabular}
\begin{tablenotes}[para,flushleft]
\footnotesize
\textbf{Notes:} MGIV$^{[\kappa]}$ refers to the Mean Group IV estimator in Eq. \eqref{eq:MGest} based on the instrument matrix $\mathbf{\widehat{Z}}^{[\kappa]}_{i}$, for $\kappa \in
\{1,2,3,L \}$, as explained in Section \ref{subsec:Design}.
\end{tablenotes}
\end{threeparttable}
\end{table}
\end{landscape}

\begin{landscape}
\begin{table}[ht]
\centering
\begin{threeparttable}
\footscriptsize
\setlength{\tabcolsep}{2.7pt} % tighten if needed
\renewcommand{\arraystretch}{1.15}
\caption{Mean Group IV results for $\beta_{1}=3$}
\label{tab:sim_results_beta1_MGIV}
\begin{tabular}{l*{27}{c}}
%\toprule
\addlinespace[2pt]

\multicolumn{28}{c}{\small \textbf{Panel A ($\pi_{u}=3/4$)}}\\

& \multicolumn{6}{c}{\small \textbf{MGIV$^{[1]}$}} &  & \multicolumn{6}{c}{\small \textbf{MGIV$^{[2]}$}} &  & \multicolumn{6}{c}{\small \textbf{MGIV$^{[3]}$}} &  & \multicolumn{6}{c}{\small \textbf{MGIV$^{[L]}$}} \\
\cmidrule(lr){2-7} \cmidrule(lr){9-14} \cmidrule(lr){16-21} \cmidrule(lr){23-28}

\multicolumn{28}{c}{\small \textbf{Case I:} $N=25\tau$, $T=100\tau$}\\

$\tau$ & Mean & RMSE & ARB & Size & Power & J-test &  & Mean & RMSE & ARB & Size & Power & J-test &  & Mean & RMSE & ARB & Size & Power & J-test &  & Mean & RMSE & ARB & Size & Power & J-test \\
\cmidrule(lr){2-7} \cmidrule(lr){9-14} \cmidrule(lr){16-21} \cmidrule(lr){23-28}

1 & 2.994 & 0.063 & 0.207 & 0.048 & 0.332 & 0.0905 &  & 2.996 & 0.063 & 0.120 & 0.048 & 0.366 & 0.121 &  & 2.985 & 0.065 & 0.500 & 0.050 & 0.297 & 0.098 &  & 2.996 & 0.062 & 0.144 & 0.045 & 0.365 & 0.2165 \\
2 & 2.996 & 0.032 & 0.120 & 0.049 & 0.836 & 0.068 &  & 2.998 & 0.032 & 0.065 & 0.044 & 0.862 & 0.235 &  & 2.991 & 0.034 & 0.313 & 0.061 & 0.778 & 0.0745 &  & 2.998 & 0.032 & 0.061 & 0.044 & 0.861 & 0.321 \\
4 & 2.999 & 0.018 & 0.048 & 0.047 & 1.000 & 0.091 &  & 2.999 & 0.017 & 0.021 & 0.045 & 1.000 & 0.7525 &  & 2.995 & 0.018 & 0.164 & 0.052 & 1.000 & 0.091 &  & 2.998 & 0.018 & 0.144 & 0.045 & 0.999 & 0.7555 \\

\multicolumn{28}{c}{\small \textbf{Case II:} $N=100\tau$, $T=25\tau$}\\

$\tau$ & Mean & RMSE & ARB & Size & Power & J-test &  & Mean & RMSE & ARB & Size & Power & J-test &  & Mean & RMSE & ARB & Size & Power & J-test &  & Mean & RMSE & ARB & Size & Power & J-test \\
\cmidrule(lr){2-7} \cmidrule(lr){9-14} \cmidrule(lr){16-21} \cmidrule(lr){23-28}

1 & 3.006 & 0.079 & 0.212 & 0.054 & 0.290 & 0.063 &  & 3.009 & 0.079 & 0.302 & 0.057 & 0.305 & 0.1095 &  & 3.004 & 0.079 & 0.145 & 0.058 & 0.288 & 0.068 &  & 3.008 & 0.084 & 0.262 & 0.053 & 0.276 & 0.058 \\
2 & 2.997 & 0.033 & 0.095 & 0.047 & 0.839 & 0.061 &  & 3.000 & 0.032 & 0.005 & 0.042 & 0.878 & 0.2305 &  & 2.989 & 0.034 & 0.360 & 0.059 & 0.762 & 0.0585 &  & 3.000 & 0.032 & 0.005 & 0.041 & 0.878 & 0.2615 \\
4 & 2.996 & 0.016 & 0.149 & 0.052 & 1.000 & 0.074 &  & 2.998 & 0.016 & 0.072 & 0.047 & 1.000 & 0.781 &  & 2.987 & 0.020 & 0.423 & 0.118 & 0.998 & 0.077 &  & 2.999 & 0.016 & 0.033 & 0.046 & 0.999 & 0.790 \\

\multicolumn{28}{c}{\small \textbf{Case III:} $N=50\tau$, $T=50\tau$}\\

$\tau$ & Mean & RMSE & ARB & Size & Power & J-test &  & Mean & RMSE & ARB & Size & Power & J-test &  & Mean & RMSE & ARB & Size & Power & J-test &  & Mean & RMSE & ARB & Size & Power & J-test \\
\cmidrule(lr){2-7} \cmidrule(lr){9-14} \cmidrule(lr){16-21} \cmidrule(lr){23-28}

1 & 2.991 & 0.066 & 0.294 & 0.047 & 0.303 & 0.061 &  & 2.994 & 0.065 & 0.192 & 0.046 & 0.325 & 0.113 &  & 2.984 & 0.067 & 0.543 & 0.049 & 0.262 & 0.0765 &  & 2.994 & 0.065 & 0.204 & 0.045 & 0.329 & 0.1385 \\
2 & 2.995 & 0.032 & 0.171 & 0.048 & 0.845 & 0.0545 &  & 2.998 & 0.031 & 0.082 & 0.046 & 0.877 & 0.233 &  & 2.987 & 0.034 & 0.439 & 0.068 & 0.736 & 0.0625 &  & 2.998 & 0.031 & 0.081 & 0.047 & 0.873 & 0.2815 \\
4 & 2.996 & 0.017 & 0.120 & 0.055 & 1.000 & 0.0875 &  & 2.998 & 0.016 & 0.063 & 0.051 & 1.000 & 0.769 &  & 2.991 & 0.019 & 0.312 & 0.092 & 1.000 & 0.0855 &  & 2.990 & 0.018 & 0.333 & 0.050 & 0.999 & 0.775 \\

\midrule
\multicolumn{28}{c}{\small \textbf{Panel B ($\pi_{u}=1/4$)}}\\

& \multicolumn{6}{c}{\small \textbf{MGIV$^{[1]}$}} &  & \multicolumn{6}{c}{\small \textbf{MGIV$^{[2]}$}} &  & \multicolumn{6}{c}{\small \textbf{MGIV$^{[3]}$}} &  & \multicolumn{6}{c}{\small \textbf{MGIV$^{[L]}$}} \\
\cmidrule(lr){2-7} \cmidrule(lr){9-14} \cmidrule(lr){16-21} \cmidrule(lr){23-28}

\multicolumn{28}{c}{\small \textbf{Case I:} $N=25\tau$, $T=100\tau$}\\

$\tau$ & Mean & RMSE & ARB & Size & Power & J-test &  & Mean & RMSE & ARB & Size & Power & J-test &  & Mean & RMSE & ARB & Size & Power & J-test &  & Mean & RMSE & ARB & Size & Power & J-test \\
\cmidrule(lr){2-7} \cmidrule(lr){9-14} \cmidrule(lr){16-21} \cmidrule(lr){23-28}

1 & 2.991 & 0.063 & 0.308 & 0.053 & 0.323 & 0.079 &  & 2.994 & 0.062 & 0.207 & 0.054 & 0.343 & 0.1285 &  & 2.986 & 0.063 & 0.470 & 0.054 & 0.304 & 0.083 &  & 2.994 & 0.064 & 0.196 & 0.049 & 0.339 & 0.1485 \\
2 & 2.993 & 0.032 & 0.230 & 0.053 & 0.820 & 0.0715 &  & 2.995 & 0.032 & 0.173 & 0.059 & 0.828 & 0.2915 &  & 2.988 & 0.033 & 0.386 & 0.063 & 0.765 & 0.065 &  & 2.994 & 0.032 & 0.184 & 0.054 & 0.818 & 0.123 \\
4 & 2.996 & 0.016 & 0.121 & 0.040 & 1.000 & 0.0805 &  & 2.997 & 0.016 & 0.092 & 0.036 & 1.000 & 0.866 &  & 2.993 & 0.017 & 0.232 & 0.054 & 1.000 & 0.0935 &  & 2.994 & 0.020 & 0.203 & 0.036 & 1.000 & 0.868 \\

\multicolumn{28}{c}{\small \textbf{Case II:} $N=100\tau$, $T=25\tau$}\\

$\tau$ & Mean & RMSE & ARB & Size & Power & J-test &  & Mean & RMSE & ARB & Size & Power & J-test &  & Mean & RMSE & ARB & Size & Power & J-test &  & Mean & RMSE & ARB & Size & Power & J-test \\
\cmidrule(lr){2-7} \cmidrule(lr){9-14} \cmidrule(lr){16-21} \cmidrule(lr){23-28}

1 & 3.000 & 0.071 & 0.003 & 0.050 & 0.307 & 0.058 &  & 3.000 & 0.070 & 0.010 & 0.047 & 0.333 & 0.1005 &  & 2.998 & 0.068 & 0.053 & 0.045 & 0.338 & 0.060 &  & 3.002 & 0.084 & 0.083 & 0.049 & 0.279 & 0.1505 \\
2 & 2.991 & 0.034 & 0.315 & 0.059 & 0.770 & 0.052 &  & 2.993 & 0.033 & 0.237 & 0.056 & 0.823 & 0.290 &  & 2.986 & 0.034 & 0.466 & 0.073 & 0.728 & 0.0585 &  & 2.992 & 0.036 & 0.256 & 0.055 & 0.753 & 0.1425 \\
4 & 2.991 & 0.018 & 0.299 & 0.081 & 1.000 & 0.094 &  & 2.993 & 0.017 & 0.222 & 0.069 & 0.999 & 0.8885 &  & 2.986 & 0.021 & 0.482 & 0.153 & 0.991 & 0.081 &  & 2.988 & 0.019 & 0.400 & 0.069 & 0.999 & 0.8905 \\

\multicolumn{28}{c}{\small \textbf{Case III:} $N=50\tau$, $T=50\tau$}\\

$\tau$ & Mean & RMSE & ARB & Size & Power & J-test &  & Mean & RMSE & ARB & Size & Power & J-test &  & Mean & RMSE & ARB & Size & Power & J-test &  & Mean & RMSE & ARB & Size & Power & J-test \\
\cmidrule(lr){2-7} \cmidrule(lr){9-14} \cmidrule(lr){16-21} \cmidrule(lr){23-28}

1 & 2.990 & 0.065 & 0.338 & 0.047 & 0.337 & 0.0675 &  & 2.991 & 0.064 & 0.286 & 0.046 & 0.344 & 0.1205 &  & 2.985 & 0.064 & 0.515 & 0.053 & 0.297 & 0.0725 &  & 2.991 & 0.072 & 0.312 & 0.052 & 0.303 & 0.155 \\
2 & 2.990 & 0.032 & 0.317 & 0.057 & 0.827 & 0.0685 &  & 2.992 & 0.031 & 0.253 & 0.052 & 0.845 & 0.300 &  & 2.985 & 0.034 & 0.513 & 0.072 & 0.722 & 0.0685 &  & 2.992 & 0.033 & 0.259 & 0.055 & 0.817 & 0.1395 \\
4 & 2.994 & 0.016 & 0.207 & 0.063 & 1.000 & 0.087 &  & 2.995 & 0.016 & 0.160 & 0.052 & 1.000 & 0.8745 &  & 2.989 & 0.019 & 0.368 & 0.097 & 1.000 & 0.087 &  & 2.994 & 0.018 & 0.200 & 0.052 & 0.999 & 0.8815 \\

\multicolumn{28}{c}{}\\
\bottomrule
\end{tabular}
\begin{tablenotes}[para,flushleft]
\footnotesize
\textbf{Notes:} MGIV$^{[\kappa]}$ refers to the Mean Group IV estimator in Eq. \eqref{eq:MGest} based on the instrument matrix $\mathbf{\widehat{Z}}^{[\kappa]}_{i}$, for $\kappa \in
\{1,2,3,L \}$, as explained in Section \ref{subsec:Design}.
\end{tablenotes}
\end{threeparttable}
\end{table}
\end{landscape}

\begin{landscape}
\begin{table}[ht]
\centering
\begin{threeparttable}
\footscriptsize
\setlength{\tabcolsep}{2.7pt} % tighten if needed
\renewcommand{\arraystretch}{1.15}
\caption{Mean Group IV results for $\beta_{2}=1$}
\label{tab:sim_results_beta2_MGIV}
\begin{tabular}{l*{27}{c}}
%\toprule
\addlinespace[2pt]

\multicolumn{28}{c}{\small \textbf{Panel A ($\pi_{u}=3/4$)}}\\

& \multicolumn{6}{c}{\small \textbf{MGIV$^{[1]}$}} &  & \multicolumn{6}{c}{\small \textbf{MGIV$^{[2]}$}} &  & \multicolumn{6}{c}{\small \textbf{MGIV$^{[3]}$}} &  & \multicolumn{6}{c}{\small \textbf{MGIV$^{[L]}$}} \\
%\addlinespace[2pt]
\cmidrule(lr){2-7} \cmidrule(lr){9-14} \cmidrule(lr){16-21} \cmidrule(lr){23-28}

\multicolumn{28}{c}{\small \textbf{Case I:} $N=25\tau$, $T=100\tau$}\\

$\tau$ & Mean & RMSE & ARB & Size & Power & J-test &  & Mean & RMSE & ARB & Size & Power & J-test &  & Mean & RMSE & ARB & Size & Power & J-test &  & Mean & RMSE & ARB & Size & Power & J-test \\
\cmidrule(lr){2-7} \cmidrule(lr){9-14} \cmidrule(lr){16-21} \cmidrule(lr){23-28}

1 & 0.998 & 0.063 & 0.254 & 0.057 & 0.366 & 0.0905 &  & 0.999 & 0.063 & 0.152 & 0.058 & 0.358 & 0.121 &  & 0.995 & 0.063 & 0.520 & 0.060 & 0.344 & 0.098 &  & 0.998 & 0.063 & 0.173 & 0.058 & 0.361 & 0.2165 \\
2 & 0.999 & 0.031 & 0.147 & 0.054 & 0.869 & 0.068 &  & 0.999 & 0.031 & 0.096 & 0.049 & 0.878 & 0.235 &  & 0.997 & 0.032 & 0.344 & 0.053 & 0.854 & 0.0745 &  & 0.999 & 0.031 & 0.088 & 0.049 & 0.877 & 0.321 \\
4 & 1.000 & 0.017 & 0.026 & 0.060 & 1.000 & 0.091 &  & 1.000 & 0.017 & 0.002 & 0.059 & 1.000 & 0.7525 &  & 0.999 & 0.017 & 0.142 & 0.059 & 1.000 & 0.091 &  & 1.000 & 0.017 & 0.005 & 0.059 & 0.999 & 0.7555 \\

\multicolumn{28}{c}{\small \textbf{Case II:} $N=100\tau$, $T=25\tau$}\\

$\tau$ & Mean & RMSE & ARB & Size & Power & J-test &  & Mean & RMSE & ARB & Size & Power & J-test &  & Mean & RMSE & ARB & Size & Power & J-test &  & Mean & RMSE & ARB & Size & Power & J-test \\
\cmidrule(lr){2-7} \cmidrule(lr){9-14} \cmidrule(lr){16-21} \cmidrule(lr){23-28}

1 & 1.006 & 0.080 & 0.602 & 0.065 & 0.264 & 0.063 &  & 1.006 & 0.081 & 0.617 & 0.068 & 0.273 & 0.1095 &  & 1.004 & 0.079 & 0.443 & 0.069 & 0.257 & 0.068 &  & 1.006 & 0.087 & 0.560 & 0.059 & 0.255 & 0.058 \\
2 & 0.999 & 0.033 & 0.096 & 0.059 & 0.834 & 0.061 &  & 1.000 & 0.033 & 0.006 & 0.055 & 0.857 & 0.2305 &  & 0.997 & 0.033 & 0.344 & 0.054 & 0.841 & 0.0585 &  & 1.000 & 0.033 & 0.009 & 0.055 & 0.856 & 0.2615 \\
4 & 0.999 & 0.015 & 0.059 & 0.049 & 1.000 & 0.074 &  & 1.000 & 0.015 & 0.017 & 0.048 & 1.000 & 0.781 &  & 0.997 & 0.015 & 0.331 & 0.052 & 1.000 & 0.077 &  & 0.999 & 0.015 & 0.110 & 0.048 & 0.999 & 0.79 \\

\multicolumn{28}{c}{\small \textbf{Case III:} $N=50\tau$, $T=50\tau$}\\

$\tau$ & Mean & RMSE & ARB & Size & Power & J-test &  & Mean & RMSE & ARB & Size & Power & J-test &  & Mean & RMSE & ARB & Size & Power & J-test &  & Mean & RMSE & ARB & Size & Power & J-test \\
\cmidrule(lr){2-7} \cmidrule(lr){9-14} \cmidrule(lr){16-21} \cmidrule(lr){23-28}

1 & 1.000 & 0.068 & 0.009 & 0.067 & 0.343 & 0.061 &  & 1.001 & 0.067 & 0.138 & 0.058 & 0.338 & 0.113 &  & 0.998 & 0.067 & 0.182 & 0.065 & 0.324 & 0.0765 &  & 1.001 & 0.067 & 0.127 & 0.056 & 0.334 & 0.1385 \\
2 & 0.999 & 0.030 & 0.152 & 0.052 & 0.889 & 0.0545 &  & 0.999 & 0.030 & 0.084 & 0.054 & 0.885 & 0.233 &  & 0.996 & 0.031 & 0.427 & 0.060 & 0.869 & 0.0625 &  & 0.999 & 0.030 & 0.080 & 0.054 & 0.885 & 0.2815 \\
4 & 0.999 & 0.015 & 0.136 & 0.044 & 1.000 & 0.0875 &  & 0.999 & 0.015 & 0.076 & 0.045 & 1.000 & 0.769 &  & 0.997 & 0.016 & 0.326 & 0.050 & 1.000 & 0.0855 &  & 0.999 & 0.015 & 0.140 & 0.044 & 1.000 & 0.775 \\
\midrule
% ===== Panel B =====
\multicolumn{28}{c}{\small \textbf{Panel B ($\pi_{u}=1/4$)}}\\

& \multicolumn{6}{c}{\small \textbf{MGIV$^{[1]}$}} &  & \multicolumn{6}{c}{\small \textbf{MGIV$^{[2]}$}} &  & \multicolumn{6}{c}{\small \textbf{MGIV$^{[3]}$}} &  & \multicolumn{6}{c}{\small \textbf{MGIV$^{[L]}$}} \\
\cmidrule(lr){2-7} \cmidrule(lr){9-14} \cmidrule(lr){16-21} \cmidrule(lr){23-28}

\multicolumn{28}{c}{\small \textbf{Case I:} $N=25\tau$, $T=100\tau$}\\

$\tau$ & Mean & RMSE & ARB & Size & Power & J-test &  & Mean & RMSE & ARB & Size & Power & J-test &  & Mean & RMSE & ARB & Size & Power & J-test &  & Mean & RMSE & ARB & Size & Power & J-test \\
\cmidrule(lr){2-7} \cmidrule(lr){9-14} \cmidrule(lr){16-21} \cmidrule(lr){23-28}

1 & 0.995 & 0.061 & 0.463 & 0.055 & 0.354 & 0.079 &  & 0.996 & 0.062 & 0.371 & 0.053 & 0.375 & 0.1285 &  & 0.994 & 0.061 & 0.653 & 0.056 & 0.350 & 0.083 &  & 0.996 & 0.064 & 0.404 & 0.054 & 0.343 & 0.1485 \\
2 & 0.997 & 0.031 & 0.279 & 0.048 & 0.858 & 0.0715 &  & 0.998 & 0.031 & 0.208 & 0.049 & 0.867 & 0.2915 &  & 0.996 & 0.031 & 0.434 & 0.052 & 0.847 & 0.065 &  & 0.998 & 0.032 & 0.222 & 0.048 & 0.866 & 0.123 \\
4 & 0.999 & 0.016 & 0.144 & 0.040 & 1.000 & 0.0805 &  & 0.999 & 0.016 & 0.111 & 0.038 & 1.000 & 0.866 &  & 0.998 & 0.016 & 0.253 & 0.040 & 1.000 & 0.0935 &  & 0.998 & 0.017 & 0.021 & 0.037 & 1.000 & 0.868 \\

\multicolumn{28}{c}{\small \textbf{Case II:} $N=100\tau$, $T=25\tau$}\\

$\tau$ & Mean & RMSE & ARB & Size & Power & J-test &  & Mean & RMSE & ARB & Size & Power & J-test &  & Mean & RMSE & ARB & Size & Power & J-test &  & Mean & RMSE & ARB & Size & Power & J-test \\
\cmidrule(lr){2-7} \cmidrule(lr){9-14} \cmidrule(lr){16-21} \cmidrule(lr){23-28}

1 & 1.000 & 0.071 & 0.054 & 0.051 & 0.318 & 0.058 &  & 1.000 & 0.069 & 0.041 & 0.045 & 0.343 & 0.1005 &  & 0.999 & 0.070 & 0.095 & 0.051 & 0.345 & 0.06 &  & 1.001 & 0.081 & 0.115 & 0.049 & 0.273 & 0.1505 \\
2 & 0.997 & 0.033 & 0.257 & 0.054 & 0.838 & 0.052 &  & 0.998 & 0.032 & 0.166 & 0.054 & 0.858 & 0.29 &  & 0.996 & 0.032 & 0.410 & 0.053 & 0.853 & 0.0585 &  & 0.998 & 0.036 & 0.167 & 0.058 & 0.791 & 0.1425 \\
4 & 0.997 & 0.016 & 0.297 & 0.054 & 1.000 & 0.094 &  & 0.998 & 0.016 & 0.219 & 0.048 & 1.000 & 0.8885 &  & 0.995 & 0.016 & 0.488 & 0.064 & 1.000 & 0.081 &  & 0.999 & 0.016 & 0.070 & 0.047 & 0.999 & 0.8905 \\

\multicolumn{28}{c}{\small \textbf{Case III:} $N=50\tau$, $T=50\tau$}\\

$\tau$ & Mean & RMSE & ARB & Size & Power & J-test &  & Mean & RMSE & ARB & Size & Power & J-test &  & Mean & RMSE & ARB & Size & Power & J-test &  & Mean & RMSE & ARB & Size & Power & J-test \\
\cmidrule(lr){2-7} \cmidrule(lr){9-14} \cmidrule(lr){16-21} \cmidrule(lr){23-28}

1 & 0.999 & 0.065 & 0.064 & 0.052 & 0.370 & 0.0675 &  & 1.000 & 0.065 & 0.030 & 0.051 & 0.382 & 0.1205 &  & 0.998 & 0.064 & 0.174 & 0.055 & 0.364 & 0.0725 &  & 1.000 & 0.074 & 0.030 & 0.047 & 0.367 & 0.155 \\
2 & 0.997 & 0.032 & 0.265 & 0.062 & 0.852 & 0.0685 &  & 0.998 & 0.032 & 0.229 & 0.058 & 0.864 & 0.3 &  & 0.995 & 0.031 & 0.458 & 0.058 & 0.857 & 0.0685 &  & 0.998 & 0.033 & 0.205 & 0.058 & 0.842 & 0.1395 \\
4 & 0.998 & 0.016 & 0.201 & 0.048 & 1.000 & 0.087 &  & 0.998 & 0.016 & 0.158 & 0.047 & 1.000 & 0.8745 &  & 0.996 & 0.016 & 0.368 & 0.051 & 1.000 & 0.087 &  & 0.998 & 0.017 & 0.191 & 0.047 & 0.999 & 0.8815 \\
% Row 36: (optional) final spacer or notes row
\multicolumn{28}{c}{}\\
\bottomrule
\end{tabular}
\begin{tablenotes}[para,flushleft]
\footnotesize
\textbf{Notes:} MGIV$^{[\kappa]}$ refers to the Mean Group IV estimator in Eq. \eqref{eq:MGest} based on the instrument matrix $\mathbf{\widehat{Z}}^{[\kappa]}_{i}$, for $\kappa \in
\{1,2,3,L \}$, as explained in Section \ref{subsec:Design}.
\end{tablenotes}
\end{threeparttable}
\end{table}
\end{landscape}

Tables \ref{tab:sim_results_rho_IVi}-\ref{tab:sim_results_beta2_IVi} below report summary results for individual-specific IV estimates averaged across $i$ in terms of Mean, Standard Deviation (St. Dev.) and RMSE. All the four individual IV estimations using different instrument sets deliver means tightly centered around the true value across designs, while dispersion shrinks monotonically with sample size.
Among them, IV$_{i}^{[2]}$ and IV$_{i}^{[L]}$ are marginally the most precise and exhibit the lowest St. Dev./RMSE in most cases.

\begin{table}[H]
\centering
\begin{threeparttable}
\footnotesize
\setlength{\tabcolsep}{3pt}
\renewcommand{\arraystretch}{1.15}
\caption{Averaged results across individual-specific IV estimates for $\rho=0.4$}
\label{tab:sim_results_rho_IVi}
\begin{tabular}{l*{15}{c}}
\addlinespace[2pt]
\multicolumn{16}{c}{\small \textbf{Panel A ($\pi_{u}=3/4$)}}\\

& \multicolumn{3}{c}{\small \textbf{IV$_{i}^{[1]}$}} &  & \multicolumn{3}{c}{\small \textbf{IV$_{i}^{[2]}$}} &  & \multicolumn{3}{c}{\small \textbf{IV$_{i}^{[3]}$}} &  & \multicolumn{3}{c}{\small \textbf{IV$_{i}^{[L]}$}} \\
\cmidrule(lr){2-4} \cmidrule(lr){6-8} \cmidrule(lr){10-12} \cmidrule(lr){14-16}

\multicolumn{16}{c}{\small \textbf{Case I:} $N=25\tau$, $T=100\tau$}\\

$\tau$ & Mean & St.\ Dev. & RMSE &  & Mean & St.\ Dev. & RMSE &  & Mean & St.\ Dev. & RMSE &  & Mean & St.\ Dev. & RMSE \\
\cmidrule(lr){2-4} \cmidrule(lr){6-8} \cmidrule(lr){10-12} \cmidrule(lr){14-16}

1 & 0.401 & 0.137 & 0.137 &  & 0.400 & 0.136 & 0.136 &  & 0.407 & 0.136 & 0.136 &  & 0.400 & 0.135 & 0.135 \\
2 & 0.402 & 0.126 & 0.126 &  & 0.401 & 0.125 & 0.125 &  & 0.405 & 0.125 & 0.126 &  & 0.401 & 0.125 & 0.125 \\
4 & 0.401 & 0.121 & 0.121 &  & 0.401 & 0.120 & 0.120 &  & 0.403 & 0.120 & 0.121 &  & 0.401 & 0.120 & 0.120 \\

\multicolumn{16}{c}{\small \textbf{Case II:} $N=100\tau$, $T=25\tau$}\\

$\tau$ & Mean & St.\ Dev. & RMSE &  & Mean & St.\ Dev. & RMSE &  & Mean & St.\ Dev. & RMSE &  & Mean & St.\ Dev. & RMSE \\
\cmidrule(lr){2-4} \cmidrule(lr){6-8} \cmidrule(lr){10-12} \cmidrule(lr){14-16}

1 & 0.397 & 0.194 & 0.194 &  & 0.397 & 0.188 & 0.188 &  & 0.403 & 0.186 & 0.186 &  & 0.401 & 0.221 & 0.221 \\
2 & 0.404 & 0.156 & 0.156 &  & 0.402 & 0.153 & 0.153 &  & 0.411 & 0.152 & 0.153 &  & 0.402 & 0.153 & 0.153 \\
4 & 0.403 & 0.136 & 0.136 &  & 0.402 & 0.134 & 0.134 &  & 0.409 & 0.134 & 0.135 &  & 0.402 & 0.134 & 0.134 \\

\multicolumn{16}{c}{\small \textbf{Case III:} $N=50\tau$, $T=50\tau$}\\

$\tau$ & Mean & St.\ Dev. & RMSE &  & Mean & St.\ Dev. & RMSE &  & Mean & St.\ Dev. & RMSE &  & Mean & St.\ Dev. & RMSE \\
\cmidrule(lr){2-4} \cmidrule(lr){6-8} \cmidrule(lr){10-12} \cmidrule(lr){14-16}

1 & 0.404 & 0.157 & 0.157 &  & 0.402 & 0.154 & 0.154 &  & 0.411 & 0.153 & 0.154 &  & 0.402 & 0.155 & 0.155 \\
2 & 0.403 & 0.136 & 0.136 &  & 0.401 & 0.135 & 0.135 &  & 0.408 & 0.135 & 0.135 &  & 0.401 & 0.134 & 0.135 \\
4 & 0.402 & 0.126 & 0.126 &  & 0.401 & 0.125 & 0.125 &  & 0.405 & 0.125 & 0.125 &  & 0.401 & 0.125 & 0.125 \\
\midrule
% ===== Panel B =====
\multicolumn{16}{c}{\small \textbf{Panel B ($\pi_{u}=1/4$)}}\\

& \multicolumn{3}{c}{\small \textbf{IV$_{i}^{[1]}$}} &  & \multicolumn{3}{c}{\small \textbf{IV$_{i}^{[2]}$}} &  & \multicolumn{3}{c}{\small \textbf{IV$_{i}^{[3]}$}} &  & \multicolumn{3}{c}{\small \textbf{IV$_{i}^{[L]}$}} \\
\cmidrule(lr){2-4} \cmidrule(lr){6-8} \cmidrule(lr){10-12} \cmidrule(lr){14-16}

\multicolumn{16}{c}{\small \textbf{Case I:} $N=25\tau$, $T=100\tau$}\\

$\tau$ & Mean & St.\ Dev. & RMSE &  & Mean & St.\ Dev. & RMSE &  & Mean & St.\ Dev. & RMSE &  & Mean & St.\ Dev. & RMSE \\
\cmidrule(lr){2-4} \cmidrule(lr){6-8} \cmidrule(lr){10-12} \cmidrule(lr){14-16}

1 & 0.405 & 0.132 & 0.132 &  & 0.404 & 0.131 & 0.131 &  & 0.409 & 0.130 & 0.130 &  & 0.404 & 0.136 & 0.136 \\
2 & 0.403 & 0.124 & 0.124 &  & 0.402 & 0.123 & 0.124 &  & 0.406 & 0.123 & 0.123 &  & 0.403 & 0.125 & 0.125 \\
4 & 0.401 & 0.120 & 0.120 &  & 0.401 & 0.120 & 0.120 &  & 0.403 & 0.119 & 0.119 &  & 0.401 & 0.120 & 0.120 \\

\multicolumn{16}{c}{\small \textbf{Case II:} $N=100\tau$, $T=25\tau$}\\

$\tau$ & Mean & St.\ Dev. & RMSE &  & Mean & St.\ Dev. & RMSE &  & Mean & St.\ Dev. & RMSE &  & Mean & St.\ Dev. & RMSE \\
\cmidrule(lr){2-4} \cmidrule(lr){6-8} \cmidrule(lr){10-12} \cmidrule(lr){14-16}

1 & 0.406 & 0.167 & 0.167 &  & 0.405 & 0.162 & 0.163 &  & 0.408 & 0.159 & 0.160 &  & 0.407 & 0.197 & 0.198 \\
2 & 0.408 & 0.144 & 0.144 &  & 0.407 & 0.142 & 0.143 &  & 0.412 & 0.141 & 0.141 &  & 0.407 & 0.156 & 0.156 \\
4 & 0.405 & 0.131 & 0.131 &  & 0.404 & 0.130 & 0.130 &  & 0.409 & 0.129 & 0.130 &  & 0.404 & 0.130 & 0.130 \\

\multicolumn{16}{c}{\small \textbf{Case III:} $N=50\tau$, $T=50\tau$}\\

$\tau$ & Mean & St.\ Dev. & RMSE &  & Mean & St.\ Dev. & RMSE &  & Mean & St.\ Dev. & RMSE &  & Mean & St.\ Dev. & RMSE \\
\cmidrule(lr){2-4} \cmidrule(lr){6-8} \cmidrule(lr){10-12} \cmidrule(lr){14-16}

1 & 0.408 & 0.144 & 0.144 &  & 0.407 & 0.142 & 0.142 &  & 0.412 & 0.140 & 0.141 &  & 0.407 & 0.156 & 0.157 \\
2 & 0.405 & 0.131 & 0.131 &  & 0.404 & 0.130 & 0.130 &  & 0.409 & 0.129 & 0.129 &  & 0.404 & 0.134 & 0.134 \\
4 & 0.403 & 0.123 & 0.123 &  & 0.402 & 0.123 & 0.123 &  & 0.406 & 0.123 & 0.123 &  & 0.402 & 0.123 & 0.123 \\
\multicolumn{16}{c}{}\\
\bottomrule
\end{tabular}
\begin{tablenotes}[para,flushleft]
\footnotesize
\textbf{Notes:} Let $\widehat{\theta}_{\ell,i}^{(r)}$ denote the estimate of population mean coefficient $\theta_\ell$, $\ell\in\{1,\ldots,k+2\}$ for unit $i\in\{1,\ldots,N\}$ in replication $r\in\{1,\ldots,R\}$. For each $\ell$ and $i$ we first summarise across replications:
\[
\overline{\theta}_{\ell,i}\equiv \frac{1}{R}\sum_{r=1}^{R}\widehat{\theta}_{\ell,i}^{(r)},\qquad
\text{St.Dev}_{\ell,i}\equiv \sqrt{\frac{1}{R}\sum_{r=1}^{R}\!\big(\widehat{\theta}_{\ell,i}^{(r)}-\overline{\theta}_{\ell,i}\big)^{2}},\qquad
\text{RMSE}_{\ell,i}\equiv \sqrt{\frac{1}{R}\sum_{r=1}^{R}\!\big(\widehat{\theta}_{\ell,i}^{(r)}-\theta_{\ell}\big)^{2}}.
\]
Then we average these individual-specific quantities across $i$:
\[
\overline{\overline{\theta}}_{\ell} \;=\; \frac{1}{N}\sum_{i=1}^{N}\overline{\theta}_{\ell,i},\qquad
\overline{\text{St.Dev}}_{\ell} \;=\; \frac{1}{N}\sum_{i=1}^{N}\text{St.Dev}_{\ell,i},\qquad
\overline{\text{RMSE}}_{\ell} \;=\; \frac{1}{N}\sum_{i=1}^{N}\text{RMSE}_{\ell,i}.
\]
Reported in Tables \ref{tab:sim_results_rho_IVi}--\ref{tab:sim_results_beta2_IVi} are $\overline{\overline{\theta}}_{\ell}$, $\overline{\text{St.Dev}}_{\ell}$,  and $\overline{\text{RMSE}}_{\ell}$, $\ell=1,\ldots,k+2$.
\end{tablenotes}
\end{threeparttable}
\end{table}

\begin{table}[H]
\centering
\begin{threeparttable}
\footnotesize
\setlength{\tabcolsep}{3pt}
\renewcommand{\arraystretch}{1.15}
\caption{Averaged results across individual-specific IV estimates for $\psi=0.25$}
\label{tab:sim_results_psi_IVi}
\begin{tabular}{l*{15}{c}}
\addlinespace[2pt]

\multicolumn{16}{c}{\small \textbf{Panel A ($\pi_{u}=3/4$)}}\\

& \multicolumn{3}{c}{\small \textbf{IV$_{i}^{[1]}$}} &  & \multicolumn{3}{c}{\small \textbf{IV$_{i}^{[2]}$}} &  & \multicolumn{3}{c}{\small \textbf{IV$_{i}^{[3]}$}} &  & \multicolumn{3}{c}{\small \textbf{IV$_{i}^{[L]}$}} \\
\cmidrule(lr){2-4} \cmidrule(lr){6-8} \cmidrule(lr){10-12} \cmidrule(lr){14-16}

\multicolumn{16}{c}{\small \textbf{Case I:} $N=25\tau$, $T=100\tau$}\\

$\tau$ & Mean & St.\ Dev. & RMSE &  & Mean & St.\ Dev. & RMSE &  & Mean & St.\ Dev. & RMSE &  & Mean & St.\ Dev. & RMSE \\
\cmidrule(lr){2-4} \cmidrule(lr){6-8} \cmidrule(lr){10-12} \cmidrule(lr){14-16}

1 & 0.251 & 0.136 & 0.136 &  & 0.252 & 0.138 & 0.138 &  & 0.250 & 0.134 & 0.134 &  & 0.252 & 0.137 & 0.137 \\
2 & 0.250 & 0.110 & 0.110 &  & 0.250 & 0.112 & 0.112 &  & 0.250 & 0.110 & 0.110 &  & 0.250 & 0.112 & 0.112 \\
4 & 0.250 & 0.098 & 0.098 &  & 0.250 & 0.099 & 0.099 &  & 0.249 & 0.098 & 0.098 &  & 0.250 & 0.099 & 0.099 \\

\multicolumn{16}{c}{\small \textbf{Case II:} $N=100\tau$, $T=25\tau$}\\

$\tau$ & Mean & St.\ Dev. & RMSE &  & Mean & St.\ Dev. & RMSE &  & Mean & St.\ Dev. & RMSE &  & Mean & St.\ Dev. & RMSE \\
\cmidrule(lr){2-4} \cmidrule(lr){6-8} \cmidrule(lr){10-12} \cmidrule(lr){14-16}

1 & 0.254 & 0.296 & 0.296 &  & 0.253 & 0.293 & 0.293 &  & 0.252 & 0.281 & 0.282 &  & 0.252 & 0.345 & 0.345 \\
2 & 0.250 & 0.186 & 0.186 &  & 0.250 & 0.188 & 0.188 &  & 0.249 & 0.181 & 0.181 &  & 0.250 & 0.188 & 0.188 \\
4 & 0.250 & 0.134 & 0.134 &  & 0.250 & 0.136 & 0.136 &  & 0.249 & 0.132 & 0.133 &  & 0.250 & 0.136 & 0.136 \\

\multicolumn{16}{c}{\small \textbf{Case III:} $N=50\tau$, $T=50\tau$}\\

$\tau$ & Mean & St.\ Dev. & RMSE &  & Mean & St.\ Dev. & RMSE &  & Mean & St.\ Dev. & RMSE &  & Mean & St.\ Dev. & RMSE \\
\cmidrule(lr){2-4} \cmidrule(lr){6-8} \cmidrule(lr){10-12} \cmidrule(lr){14-16}

1 & 0.253 & 0.187 & 0.188 &  & 0.253 & 0.189 & 0.189 &  & 0.251 & 0.182 & 0.182 &  & 0.253 & 0.190 & 0.190 \\
2 & 0.250 & 0.134 & 0.134 &  & 0.250 & 0.136 & 0.136 &  & 0.249 & 0.133 & 0.133 &  & 0.250 & 0.136 & 0.136 \\
4 & 0.250 & 0.110 & 0.110 &  & 0.250 & 0.111 & 0.111 &  & 0.249 & 0.109 & 0.109 &  & 0.250 & 0.111 & 0.111 \\
\midrule
% ===== Panel B =====
\multicolumn{16}{c}{\small \textbf{Panel B ($\pi_{u}=1/4$)}}\\

& \multicolumn{3}{c}{\small \textbf{IV$_{i}^{[1]}$}} &  & \multicolumn{3}{c}{\small \textbf{IV$_{i}^{[2]}$}} &  & \multicolumn{3}{c}{\small \textbf{IV$_{i}^{[3]}$}} &  & \multicolumn{3}{c}{\small \textbf{IV$_{i}^{[L]}$}} \\
\cmidrule(lr){2-4} \cmidrule(lr){6-8} \cmidrule(lr){10-12} \cmidrule(lr){14-16}

\multicolumn{16}{c}{\small \textbf{Case I:} $N=25\tau$, $T=100\tau$}\\

$\tau$ & Mean & St.\ Dev. & RMSE &  & Mean & St.\ Dev. & RMSE &  & Mean & St.\ Dev. & RMSE &  & Mean & St.\ Dev. & RMSE \\
\cmidrule(lr){2-4} \cmidrule(lr){6-8} \cmidrule(lr){10-12} \cmidrule(lr){14-16}

1 & 0.251 & 0.140 & 0.140 &  & 0.252 & 0.140 & 0.140 &  & 0.251 & 0.135 & 0.135 &  & 0.251 & 0.152 & 0.152 \\
2 & 0.250 & 0.112 & 0.112 &  & 0.250 & 0.113 & 0.113 &  & 0.249 & 0.110 & 0.111 &  & 0.250 & 0.114 & 0.114 \\
4 & 0.250 & 0.099 & 0.099 &  & 0.250 & 0.099 & 0.099 &  & 0.249 & 0.099 & 0.099 &  & 0.250 & 0.099 & 0.099 \\

\multicolumn{16}{c}{\small \textbf{Case II:} $N=100\tau$, $T=25\tau$}\\

$\tau$ & Mean & St.\ Dev. & RMSE &  & Mean & St.\ Dev. & RMSE &  & Mean & St.\ Dev. & RMSE &  & Mean & St.\ Dev. & RMSE \\
\cmidrule(lr){2-4} \cmidrule(lr){6-8} \cmidrule(lr){10-12} \cmidrule(lr){14-16}

1 & 0.251 & 0.273 & 0.273 &  & 0.252 & 0.268 & 0.268 &  & 0.251 & 0.259 & 0.259 &  & 0.251 & 0.334 & 0.334 \\
2 & 0.251 & 0.185 & 0.185 &  & 0.251 & 0.185 & 0.185 &  & 0.250 & 0.177 & 0.177 &  & 0.250 & 0.209 & 0.209 \\
4 & 0.249 & 0.137 & 0.137 &  & 0.250 & 0.137 & 0.137 &  & 0.249 & 0.133 & 0.133 &  & 0.250 & 0.137 & 0.137 \\

\multicolumn{16}{c}{\small \textbf{Case III:} $N=50\tau$, $T=50\tau$}\\

$\tau$ & Mean & St.\ Dev. & RMSE &  & Mean & St.\ Dev. & RMSE &  & Mean & St.\ Dev. & RMSE &  & Mean & St.\ Dev. & RMSE \\
\cmidrule(lr){2-4} \cmidrule(lr){6-8} \cmidrule(lr){10-12} \cmidrule(lr){14-16}

1 & 0.251 & 0.185 & 0.185 &  & 0.251 & 0.185 & 0.185 &  & 0.250 & 0.177 & 0.177 &  & 0.250 & 0.214 & 0.215 \\
2 & 0.250 & 0.138 & 0.138 &  & 0.250 & 0.138 & 0.138 &  & 0.249 & 0.134 & 0.134 &  & 0.250 & 0.144 & 0.144 \\
4 & 0.250 & 0.112 & 0.112 &  & 0.250 & 0.112 & 0.112 &  & 0.249 & 0.110 & 0.110 &  & 0.250 & 0.112 & 0.112 \\
\multicolumn{16}{c}{}\\
\bottomrule
\end{tabular}
\begin{tablenotes}[para,flushleft]
\footnotesize
\textbf{Notes:} As in Table \ref{tab:sim_results_rho_IVi}.
\end{tablenotes}
\end{threeparttable}
\end{table}

\begin{table}[H]
\centering
\begin{threeparttable}
\footnotesize
\setlength{\tabcolsep}{3pt}
\renewcommand{\arraystretch}{1.15}
\caption{Averaged results across individual-specific IV estimates for $\beta_{1}=3$}
\label{sim_results_beta1_IVi}
\begin{tabular}{l*{15}{c}}
\addlinespace[2pt]

\multicolumn{16}{c}{\small \textbf{Panel A ($\pi_{u}=3/4$)}}\\

& \multicolumn{3}{c}{\small \textbf{IV$_{i}^{[1]}$}} &  & \multicolumn{3}{c}{\small \textbf{IV$_{i}^{[2]}$}} &  & \multicolumn{3}{c}{\small \textbf{IV$_{i}^{[3]}$}} &  & \multicolumn{3}{c}{\small \textbf{IV$_{i}^{[L]}$}} \\
\cmidrule(lr){2-4} \cmidrule(lr){6-8} \cmidrule(lr){10-12} \cmidrule(lr){14-16}

\multicolumn{16}{c}{\small \textbf{Case I:} $N=25\tau$, $T=100\tau$}\\

$\tau$ & Mean & St.\ Dev. & RMSE &  & Mean & St.\ Dev. & RMSE &  & Mean & St.\ Dev. & RMSE &  & Mean & St.\ Dev. & RMSE \\
\cmidrule(lr){2-4} \cmidrule(lr){6-8} \cmidrule(lr){10-12} \cmidrule(lr){14-16}

1 & 2.994 & 0.327 & 0.327 &  & 2.996 & 0.324 & 0.324 &  & 2.985 & 0.326 & 0.326 &  & 2.996 & 0.323 & 0.324 \\
2 & 2.996 & 0.235 & 0.235 &  & 2.998 & 0.233 & 0.234 &  & 2.991 & 0.235 & 0.235 &  & 2.998 & 0.233 & 0.233 \\
4 & 2.999 & 0.182 & 0.182 &  & 2.999 & 0.181 & 0.181 &  & 2.995 & 0.182 & 0.182 &  & 2.999 & 0.181 & 0.181 \\

\multicolumn{16}{c}{\small \textbf{Case II:} $N=100\tau$, $T=25\tau$}\\

$\tau$ & Mean & St.\ Dev. & RMSE &  & Mean & St.\ Dev. & RMSE &  & Mean & St.\ Dev. & RMSE &  & Mean & St.\ Dev. & RMSE \\
\cmidrule(lr){2-4} \cmidrule(lr){6-8} \cmidrule(lr){10-12} \cmidrule(lr){14-16}

1 & 3.006 & 0.779 & 0.780 &  & 3.009 & 0.776 & 0.777 &  & 3.004 & 0.763 & 0.764 &  & 3.008 & 0.843 & 0.844 \\
2 & 2.997 & 0.470 & 0.470 &  & 3.000 & 0.465 & 0.466 &  & 2.989 & 0.466 & 0.466 &  & 3.000 & 0.466 & 0.466 \\
4 & 2.996 & 0.319 & 0.319 &  & 2.998 & 0.317 & 0.317 &  & 2.987 & 0.318 & 0.319 &  & 2.998 & 0.317 & 0.317 \\

\multicolumn{16}{c}{\small \textbf{Case III:} $N=50\tau$, $T=50\tau$}\\

$\tau$ & Mean & St.\ Dev. & RMSE &  & Mean & St.\ Dev. & RMSE &  & Mean & St.\ Dev. & RMSE &  & Mean & St.\ Dev. & RMSE \\
\cmidrule(lr){2-4} \cmidrule(lr){6-8} \cmidrule(lr){10-12} \cmidrule(lr){14-16}

1 & 2.991 & 0.480 & 0.480 &  & 2.994 & 0.476 & 0.477 &  & 2.984 & 0.475 & 0.476 &  & 2.994 & 0.478 & 0.478 \\
2 & 2.995 & 0.319 & 0.320 &  & 2.998 & 0.317 & 0.317 &  & 2.987 & 0.319 & 0.320 &  & 2.998 & 0.317 & 0.317 \\
4 & 2.996 & 0.233 & 0.233 &  & 2.998 & 0.231 & 0.231 &  & 2.991 & 0.233 & 0.233 &  & 2.998 & 0.231 & 0.231 \\
\midrule
% ===== Panel B =====
\multicolumn{16}{c}{\small \textbf{Panel B ($\pi_{u}=1/4$)}}\\

& \multicolumn{3}{c}{\small \textbf{IV$_{i}^{[1]}$}} &  & \multicolumn{3}{c}{\small \textbf{IV$_{i}^{[2]}$}} &  & \multicolumn{3}{c}{\small \textbf{IV$_{i}^{[3]}$}} &  & \multicolumn{3}{c}{\small \textbf{IV$_{i}^{[L]}$}} \\
\cmidrule(lr){2-4} \cmidrule(lr){6-8} \cmidrule(lr){10-12} \cmidrule(lr){14-16}

\multicolumn{16}{c}{\small \textbf{Case I:} $N=25\tau$, $T=100\tau$}\\

$\tau$ & Mean & St.\ Dev. & RMSE &  & Mean & St.\ Dev. & RMSE &  & Mean & St.\ Dev. & RMSE &  & Mean & St.\ Dev. & RMSE \\
\cmidrule(lr){2-4} \cmidrule(lr){6-8} \cmidrule(lr){10-12} \cmidrule(lr){14-16}

1 & 2.991 & 0.323 & 0.324 &  & 2.994 & 0.319 & 0.320 &  & 2.986 & 0.320 & 0.321 &  & 2.994 & 0.333 & 0.333 \\
2 & 2.993 & 0.228 & 0.228 &  & 2.995 & 0.228 & 0.228 &  & 2.988 & 0.228 & 0.228 &  & 2.994 & 0.230 & 0.230 \\
4 & 2.996 & 0.175 & 0.175 &  & 2.997 & 0.175 & 0.175 &  & 2.993 & 0.176 & 0.176 &  & 2.997 & 0.175 & 0.175 \\

\multicolumn{16}{c}{\small \textbf{Case II:} $N=100\tau$, $T=25\tau$}\\

$\tau$ & Mean & St.\ Dev. & RMSE &  & Mean & St.\ Dev. & RMSE &  & Mean & St.\ Dev. & RMSE &  & Mean & St.\ Dev. & RMSE \\
\cmidrule(lr){2-4} \cmidrule(lr){6-8} \cmidrule(lr){10-12} \cmidrule(lr){14-16}

1 & 3.000 & 0.721 & 0.721 &  & 3.000 & 0.708 & 0.708 &  & 2.998 & 0.697 & 0.698 &  & 3.002 & 0.841 & 0.842 \\
2 & 2.991 & 0.456 & 0.457 &  & 2.993 & 0.451 & 0.451 &  & 2.986 & 0.445 & 0.445 &  & 2.992 & 0.497 & 0.497 \\
4 & 2.991 & 0.310 & 0.310 &  & 2.993 & 0.308 & 0.308 &  & 2.986 & 0.307 & 0.307 &  & 2.993 & 0.308 & 0.308 \\

\multicolumn{16}{c}{\small \textbf{Case III:} $N=50\tau$, $T=50\tau$}\\

$\tau$ & Mean & St.\ Dev. & RMSE &  & Mean & St.\ Dev. & RMSE &  & Mean & St.\ Dev. & RMSE &  & Mean & St.\ Dev. & RMSE \\
\cmidrule(lr){2-4} \cmidrule(lr){6-8} \cmidrule(lr){10-12} \cmidrule(lr){14-16}

1 & 2.990 & 0.461 & 0.461 &  & 2.991 & 0.454 & 0.454 &  & 2.985 & 0.449 & 0.450 &  & 2.991 & 0.506 & 0.506 \\
2 & 2.990 & 0.312 & 0.312 &  & 2.992 & 0.310 & 0.310 &  & 2.985 & 0.309 & 0.309 &  & 2.992 & 0.323 & 0.323 \\
4 & 2.994 & 0.224 & 0.224 &  & 2.995 & 0.224 & 0.224 &  & 2.989 & 0.224 & 0.225 &  & 2.995 & 0.224 & 0.224 \\
\multicolumn{16}{c}{}\\
\bottomrule
\end{tabular}
\begin{tablenotes}[para,flushleft]
\footnotesize
\textbf{Notes:} As in Table \ref{tab:sim_results_rho_IVi}.
\end{tablenotes}
\end{threeparttable}
\end{table}

\begin{table}[H]
\centering
\begin{threeparttable}
\footnotesize
\setlength{\tabcolsep}{3pt}
\renewcommand{\arraystretch}{1.15}
\caption{Averaged results across individual-specific IV estimates for $\beta_{2}=1$}
\label{tab:sim_results_beta2_IVi}
\begin{tabular}{l*{15}{c}}
\addlinespace[2pt]

\multicolumn{16}{c}{\small \textbf{Panel A ($\pi_{u}=3/4$)}}\\

& \multicolumn{3}{c}{\small \textbf{IV$_{i}^{[1]}$}} &  & \multicolumn{3}{c}{\small \textbf{IV$_{i}^{[2]}$}} &  & \multicolumn{3}{c}{\small \textbf{IV$_{i}^{[3]}$}} &  & \multicolumn{3}{c}{\small \textbf{IV$_{i}^{[L]}$}} \\
\cmidrule(lr){2-4} \cmidrule(lr){6-8} \cmidrule(lr){10-12} \cmidrule(lr){14-16}

\multicolumn{16}{c}{\small \textbf{Case I:} $N=25\tau$, $T=100\tau$}\\

$\tau$ & Mean & St.\ Dev. & RMSE &  & Mean & St.\ Dev. & RMSE &  & Mean & St.\ Dev. & RMSE &  & Mean & St.\ Dev. & RMSE \\
\cmidrule(lr){2-4} \cmidrule(lr){6-8} \cmidrule(lr){10-12} \cmidrule(lr){14-16}

1 & 0.998 & 0.315 & 0.316 &  & 0.999 & 0.315 & 0.315 &  & 0.995 & 0.313 & 0.313 &  & 0.998 & 0.314 & 0.314 \\
2 & 0.999 & 0.225 & 0.225 &  & 0.999 & 0.225 & 0.225 &  & 0.997 & 0.224 & 0.224 &  & 0.999 & 0.225 & 0.225 \\
4 & 1.000 & 0.175 & 0.175 &  & 1.000 & 0.175 & 0.175 &  & 0.999 & 0.175 & 0.175 &  & 1.000 & 0.175 & 0.175 \\

\multicolumn{16}{c}{\small \textbf{Case II:} $N=100\tau$, $T=25\tau$}\\

$\tau$ & Mean & St.\ Dev. & RMSE &  & Mean & St.\ Dev. & RMSE &  & Mean & St.\ Dev. & RMSE &  & Mean & St.\ Dev. & RMSE \\
\cmidrule(lr){2-4} \cmidrule(lr){6-8} \cmidrule(lr){10-12} \cmidrule(lr){14-16}

1 & 1.006 & 0.782 & 0.782 &  & 1.006 & 0.776 & 0.776 &  & 1.004 & 0.765 & 0.766 &  & 1.006 & 0.849 & 0.849 \\
2 & 0.999 & 0.462 & 0.462 &  & 1.000 & 0.460 & 0.460 &  & 0.997 & 0.456 & 0.456 &  & 1.000 & 0.460 & 0.460 \\
4 & 0.999 & 0.307 & 0.307 &  & 1.000 & 0.306 & 0.306 &  & 0.997 & 0.305 & 0.305 &  & 1.000 & 0.306 & 0.306 \\

\multicolumn{16}{c}{\small \textbf{Case III:} $N=50\tau$, $T=50\tau$}\\

$\tau$ & Mean & St.\ Dev. & RMSE &  & Mean & St.\ Dev. & RMSE &  & Mean & St.\ Dev. & RMSE &  & Mean & St.\ Dev. & RMSE \\
\cmidrule(lr){2-4} \cmidrule(lr){6-8} \cmidrule(lr){10-12} \cmidrule(lr){14-16}

1 & 1.000 & 0.465 & 0.465 &  & 1.001 & 0.466 & 0.466 &  & 0.998 & 0.460 & 0.460 &  & 1.001 & 0.467 & 0.467 \\
2 & 0.999 & 0.307 & 0.307 &  & 0.999 & 0.307 & 0.307 &  & 0.996 & 0.305 & 0.305 &  & 0.999 & 0.307 & 0.307 \\
4 & 0.999 & 0.223 & 0.224 &  & 0.999 & 0.223 & 0.223 &  & 0.997 & 0.223 & 0.223 &  & 0.999 & 0.223 & 0.223 \\
\midrule
% ===== Panel B =====
\multicolumn{16}{c}{\small \textbf{Panel B ($\pi_{u}=1/4$)}}\\

& \multicolumn{3}{c}{\small \textbf{IV$_{i}^{[1]}$}} &  & \multicolumn{3}{c}{\small \textbf{IV$_{i}^{[2]}$}} &  & \multicolumn{3}{c}{\small \textbf{IV$_{i}^{[3]}$}} &  & \multicolumn{3}{c}{\small \textbf{IV$_{i}^{[L]}$}} \\
\cmidrule(lr){2-4} \cmidrule(lr){6-8} \cmidrule(lr){10-12} \cmidrule(lr){14-16}

\multicolumn{16}{c}{\small \textbf{Case I:} $N=25\tau$, $T=100\tau$}\\

$\tau$ & Mean & St.\ Dev. & RMSE &  & Mean & St.\ Dev. & RMSE &  & Mean & St.\ Dev. & RMSE &  & Mean & St.\ Dev. & RMSE \\
\cmidrule(lr){2-4} \cmidrule(lr){6-8} \cmidrule(lr){10-12} \cmidrule(lr){14-16}

1 & 0.995 & 0.318 & 0.318 &  & 0.996 & 0.317 & 0.318 &  & 0.994 & 0.313 & 0.313 &  & 0.996 & 0.331 & 0.331 \\
2 & 0.997 & 0.228 & 0.228 &  & 0.998 & 0.228 & 0.228 &  & 0.996 & 0.226 & 0.226 &  & 0.998 & 0.230 & 0.230 \\
4 & 0.999 & 0.176 & 0.176 &  & 0.999 & 0.176 & 0.177 &  & 0.998 & 0.176 & 0.176 &  & 0.999 & 0.176 & 0.177 \\

\multicolumn{16}{c}{\small \textbf{Case II:} $N=100\tau$, $T=25\tau$}\\

$\tau$ & Mean & St.\ Dev. & RMSE &  & Mean & St.\ Dev. & RMSE &  & Mean & St.\ Dev. & RMSE &  & Mean & St.\ Dev. & RMSE \\
\cmidrule(lr){2-4} \cmidrule(lr){6-8} \cmidrule(lr){10-12} \cmidrule(lr){14-16}

1 & 1.000 & 0.721 & 0.721 &  & 1.000 & 0.705 & 0.705 &  & 0.999 & 0.695 & 0.695 &  & 1.001 & 0.833 & 0.833 \\
2 & 0.997 & 0.455 & 0.455 &  & 0.998 & 0.450 & 0.450 &  & 0.996 & 0.443 & 0.443 &  & 0.998 & 0.496 & 0.496 \\
4 & 0.997 & 0.309 & 0.309 &  & 0.998 & 0.308 & 0.308 &  & 0.995 & 0.305 & 0.305 &  & 0.998 & 0.308 & 0.308 \\

\multicolumn{16}{c}{\small \textbf{Case III:} $N=50\tau$, $T=50\tau$}\\

$\tau$ & Mean & St.\ Dev. & RMSE &  & Mean & St.\ Dev. & RMSE &  & Mean & St.\ Dev. & RMSE &  & Mean & St.\ Dev. & RMSE \\
\cmidrule(lr){2-4} \cmidrule(lr){6-8} \cmidrule(lr){10-12} \cmidrule(lr){14-16}

1 & 0.999 & 0.459 & 0.459 &  & 1.000 & 0.454 & 0.454 &  & 0.998 & 0.447 & 0.447 &  & 1.000 & 0.513 & 0.513 \\
2 & 0.997 & 0.310 & 0.310 &  & 0.998 & 0.309 & 0.309 &  & 0.995 & 0.305 & 0.305 &  & 0.998 & 0.320 & 0.320 \\
4 & 0.998 & 0.226 & 0.226 &  & 0.998 & 0.225 & 0.225 &  & 0.996 & 0.224 & 0.224 &  & 0.998 & 0.225 & 0.225 \\
\multicolumn{16}{c}{}\\
\bottomrule
\end{tabular}
\begin{tablenotes}[para,flushleft]
\footnotesize
\textbf{Notes:} As in Table \ref{tab:sim_results_rho_IVi}.
\end{tablenotes}
\end{threeparttable}
\end{table}

Tables \ref{tab:sim_results_rho_2SIV}-\ref{tab:sim_results_beta2_2SIV} below report results on the 2SIV (pooled) estimator developed by \citet{CuiSarafidisYamagata2023} based on instrument matrix $\mathbf{\widehat{Z}}^{[\kappa]}_{i}$, for $\kappa \in \{1,2,3,L \}$, as explained in Section \ref{subsec:Design}. This is defined as:
\begin{equation}\label{eq:two-step_IV}
\widehat{\boldsymbol{\theta}}_{2SIV}=(\widetilde{\mathbf{A}}'\widetilde{\mathbf{B}}^{-1}\widetilde{\mathbf{A}})^{-1}\widetilde{\mathbf{A}}'\widetilde{\mathbf{B}}^{-1}\widetilde{\mathbf{c}}_y,
\end{equation}
where
 \[\widetilde{\mathbf{A}}=\frac{1}{NT}\sum_{i=1}^N  \widehat{\mathbf{Z}}_{i}'\mathbf{M}_{\widehat{\mathbf{H}}}\mathbf{C}_{i};\ \ \widetilde{\mathbf{B}}=\frac{1}{NT}\sum_{i=1}^N  \widehat{\mathbf{Z}}_{i}'\mathbf{M}_{\widehat{\mathbf{H}}}\widehat{\mathbf{Z}}_{i};\ \ \widetilde{\mathbf{c}}_y=\frac{1}{NT}\sum_{i=1}^N  \widehat{\mathbf{Z}}_{i}'\mathbf{M}_{\widehat{\mathbf{H}}}\mathbf{y}_{i},\]
and $\mathbf{M}_{\widehat{\mathbf{H}}}=\mathbf{I}-\widehat{\mathbf{H}}(\widehat{\mathbf{H}}'\widehat{\mathbf{H}})^{-1}\widehat{\mathbf{H}}'$ with $\widehat{\mathbf{H}}$ defined as $\sqrt{T}$ times the eigenvectors corresponding to the $r_{y}$ largest eigenvalues of the $T \times T$ matrix $(NT)^{-1}\sum_{i=1}^N  \widehat{\mathbf{u}}_{i}\widehat{\mathbf{u}}_{i}'$, where $\widehat{\mathbf{u}}_{i}$ is the $T\times 1$ vector of residuals from the first-stage homogeneous IV estimation.

The performance of these pooled IV estimators is unsatisfactory: absolute relative bias is substantial and increases with sample size, while empirical size is consistently and severely distorted. This outcome is expected, as pooled estimators are inconsistent under slope parameter heterogeneity, and therefore invalid moment conditions propagate to inference.

\begin{landscape}
\begin{table}[ht]
\centering
\begin{threeparttable}
\footscriptsize
\setlength{\tabcolsep}{2.7pt}
\renewcommand{\arraystretch}{1.15}
\caption{2SIV results for $\rho=0.4$}
\label{tab:sim_results_rho_2SIV}
\begin{tabular}{l*{27}{c}}
\addlinespace[2pt]

\multicolumn{28}{c}{\small \textbf{Panel A ($\pi_{u}=3/4$)}}\\

& \multicolumn{6}{c}{\small \textbf{2SIV$^{[1]}$}} &  & \multicolumn{6}{c}{\small \textbf{2SIV$^{[2]}$}} &  & \multicolumn{6}{c}{\small \textbf{2SIV$^{[3]}$}} &  & \multicolumn{6}{c}{\small \textbf{2SIV$^{[L]}$}} \\
\cmidrule(lr){2-7} \cmidrule(lr){9-14} \cmidrule(lr){16-21} \cmidrule(lr){23-28}

\multicolumn{28}{c}{\small \textbf{Case I:} $N=25\tau$, $T=100\tau$}\\

$\tau$ & Mean & RMSE & ARB & Size & Power & J-test &  & Mean & RMSE & ARB & Size & Power & J-test &  & Mean & RMSE & ARB & Size & Power & J-test &  & Mean & RMSE & ARB & Size & Power & J-test \\
\cmidrule(lr){2-7} \cmidrule(lr){9-14} \cmidrule(lr){16-21} \cmidrule(lr){23-28}

1 & 0.409 & 0.030 & 2.204 & 0.129 & 0.925 & 0.091  &  & 0.414 & 0.032 & 3.418 & 0.149 & 0.929 & 0.121  &  & 0.409 & 0.030 & 2.359 & 0.130 & 0.932 & 0.098  &  & 0.414 & 0.033 & 3.500 & 0.124 & 0.909 & 0.217 \\
2 & 0.411 & 0.022 & 2.842 & 0.132 & 1.000 & 0.068  &  & 0.416 & 0.024 & 4.019 & 0.188 & 1.000 & 0.235  &  & 0.412 & 0.022 & 2.902 & 0.137 & 1.000 & 0.075  &  & 0.416 & 0.025 & 4.088 & 0.174 & 0.999 & 0.321 \\
4 & 0.413 & 0.017 & 3.169 & 0.193 & 1.000 & 0.091  &  & 0.418 & 0.021 & 4.373 & 0.327 & 1.000 & 0.753  &  & 0.413 & 0.018 & 3.211 & 0.202 & 1.000 & 0.091  &  & 0.420 & 0.023 & 4.873 & 0.326 & 1.000 & 0.756 \\

\multicolumn{28}{c}{\small \textbf{Case II:} $N=100\tau$, $T=25\tau$}\\

$\tau$ & Mean & RMSE & ARB & Size & Power & J-test &  & Mean & RMSE & ARB & Size & Power & J-test &  & Mean & RMSE & ARB & Size & Power & J-test &  & Mean & RMSE & ARB & Size & Power & J-test \\
\cmidrule(lr){2-7} \cmidrule(lr){9-14} \cmidrule(lr){16-21} \cmidrule(lr){23-28}

1 & 0.409 & 0.023 & 2.174 & 0.091 & 0.998 & 0.063  &  & 0.413 & 0.025 & 3.353 & 0.133 & 0.998 & 0.110  &  & 0.409 & 0.023 & 2.318 & 0.098 & 0.997 & 0.068  &  & 0.410 & 0.024 & 2.550 & 0.101 & 0.997 & 0.058 \\
2 & 0.412 & 0.017 & 2.944 & 0.195 & 1.000 & 0.061  &  & 0.417 & 0.020 & 4.155 & 0.327 & 1.000 & 0.231  &  & 0.412 & 0.017 & 3.019 & 0.203 & 1.000 & 0.059  &  & 0.417 & 0.020 & 4.128 & 0.312 & 1.000 & 0.262 \\
4 & 0.413 & 0.015 & 3.207 & 0.459 & 1.000 & 0.074  &  & 0.418 & 0.019 & 4.424 & 0.739 & 1.000 & 0.781  &  & 0.413 & 0.015 & 3.250 & 0.465 & 1.000 & 0.077  &  & 0.425 & 0.025 & 6.176 & 0.738 & 0.999 & 0.790 \\

\multicolumn{28}{c}{\small \textbf{Case III:} $N=50\tau$, $T=50\tau$}\\

$\tau$ & Mean & RMSE & ARB & Size & Power & J-test &  & Mean & RMSE & ARB & Size & Power & J-test &  & Mean & RMSE & ARB & Size & Power & J-test &  & Mean & RMSE & ARB & Size & Power & J-test \\
\cmidrule(lr){2-7} \cmidrule(lr){9-14} \cmidrule(lr){16-21} \cmidrule(lr){23-28}

1 & 0.411 & 0.026 & 2.739 & 0.111 & 0.992 & 0.061  &  & 0.416 & 0.028 & 3.863 & 0.151 & 0.992 & 0.113  &  & 0.411 & 0.026 & 2.851 & 0.115 & 0.990 & 0.077  &  & 0.415 & 0.028 & 3.737 & 0.133 & 0.981 & 0.139 \\
2 & 0.412 & 0.018 & 3.032 & 0.159 & 1.000 & 0.055  &  & 0.417 & 0.022 & 4.228 & 0.261 & 1.000 & 0.233  &  & 0.412 & 0.019 & 3.098 & 0.160 & 1.000 & 0.063  &  & 0.417 & 0.022 & 4.251 & 0.245 & 1.000 & 0.282 \\
4 & 0.413 & 0.016 & 3.195 & 0.313 & 1.000 & 0.088  &  & 0.418 & 0.020 & 4.385 & 0.523 & 1.000 & 0.769  &  & 0.413 & 0.016 & 3.236 & 0.322 & 1.000 & 0.086  &  & 0.427 & 0.029 & 6.841 & 0.522 & 0.999 & 0.775 \\

\midrule
\multicolumn{28}{c}{\small \textbf{Panel B ($\pi_{u}=1/4$)}}\\

& \multicolumn{6}{c}{\small \textbf{2SIV$^{[1]}$}} &  & \multicolumn{6}{c}{\small \textbf{2SIV$^{[2]}$}} &  & \multicolumn{6}{c}{\small \textbf{2SIV$^{[3]}$}} &  & \multicolumn{6}{c}{\small \textbf{2SIV$^{[L]}$}} \\
\cmidrule(lr){2-7} \cmidrule(lr){9-14} \cmidrule(lr){16-21} \cmidrule(lr){23-28}

\multicolumn{28}{c}{\small \textbf{Case I:} $N=25\tau$, $T=100\tau$}\\

$\tau$ & Mean & RMSE & ARB & Size & Power & J-test &  & Mean & RMSE & ARB & Size & Power & J-test &  & Mean & RMSE & ARB & Size & Power & J-test &  & Mean & RMSE & ARB & Size & Power & J-test \\
\cmidrule(lr){2-7} \cmidrule(lr){9-14} \cmidrule(lr){16-21} \cmidrule(lr){23-28}

1 & 0.409 & 0.029 & 2.299 & 0.124 & 0.954 & 0.079  &  & 0.414 & 0.031 & 3.483 & 0.151 & 0.943 & 0.129  &  & 0.410 & 0.030 & 2.469 & 0.124 & 0.946 & 0.083  &  & 0.409 & 0.029 & 2.347 & 0.092 & 0.935 & 0.149 \\
2 & 0.411 & 0.021 & 2.860 & 0.127 & 1.000 & 0.072  &  & 0.416 & 0.024 & 4.052 & 0.186 & 1.000 & 0.292  &  & 0.412 & 0.021 & 2.924 & 0.130 & 1.000 & 0.065  &  & 0.412 & 0.021 & 2.889 & 0.115 & 1.000 & 0.123 \\
4 & 0.412 & 0.017 & 2.981 & 0.172 & 1.000 & 0.081  &  & 0.417 & 0.021 & 4.184 & 0.302 & 1.000 & 0.866  &  & 0.412 & 0.017 & 3.028 & 0.181 & 1.000 & 0.094  &  & 0.410 & 0.015 & 2.426 & 0.302 & 1.000 & 0.868 \\

\multicolumn{28}{c}{\small \textbf{Case II:} $N=100\tau$, $T=25\tau$}\\

$\tau$ & Mean & RMSE & ARB & Size & Power & J-test &  & Mean & RMSE & ARB & Size & Power & J-test &  & Mean & RMSE & ARB & Size & Power & J-test &  & Mean & RMSE & ARB & Size & Power & J-test \\
\cmidrule(lr){2-7} \cmidrule(lr){9-14} \cmidrule(lr){16-21} \cmidrule(lr){23-28}

1 & 0.409 & 0.021 & 2.210 & 0.094 & 0.999 & 0.058  &  & 0.414 & 0.024 & 3.421 & 0.135 & 0.999 & 0.101  &  & 0.410 & 0.022 & 2.378 & 0.106 & 0.999 & 0.060  &  & 0.409 & 0.022 & 2.265 & 0.072 & 0.997 & 0.151 \\
2 & 0.412 & 0.016 & 2.936 & 0.194 & 1.000 & 0.052  &  & 0.417 & 0.020 & 4.131 & 0.340 & 1.000 & 0.290  &  & 0.412 & 0.016 & 3.004 & 0.197 & 1.000 & 0.059  &  & 0.412 & 0.016 & 2.977 & 0.178 & 1.000 & 0.143 \\
4 & 0.413 & 0.014 & 3.158 & 0.480 & 1.000 & 0.094  &  & 0.418 & 0.019 & 4.368 & 0.730 & 1.000 & 0.889  &  & 0.413 & 0.014 & 3.201 & 0.487 & 1.000 & 0.081  &  & 0.411 & 0.013 & 2.854 & 0.729 & 0.999 & 0.891 \\

\multicolumn{28}{c}{\small \textbf{Case III:} $N=50\tau$, $T=50\tau$}\\

$\tau$ & Mean & RMSE & ARB & Size & Power & J-test &  & Mean & RMSE & ARB & Size & Power & J-test &  & Mean & RMSE & ARB & Size & Power & J-test &  & Mean & RMSE & ARB & Size & Power & J-test \\
\cmidrule(lr){2-7} \cmidrule(lr){9-14} \cmidrule(lr){16-21} \cmidrule(lr){23-28}

1 & 0.410 & 0.024 & 2.497 & 0.107 & 0.996 & 0.068  &  & 0.415 & 0.027 & 3.669 & 0.153 & 0.995 & 0.121  &  & 0.411 & 0.025 & 2.682 & 0.118 & 0.996 & 0.073  &  & 0.410 & 0.024 & 2.549 & 0.084 & 0.995 & 0.155 \\
2 & 0.412 & 0.018 & 3.004 & 0.170 & 1.000 & 0.069  &  & 0.417 & 0.022 & 4.202 & 0.254 & 1.000 & 0.300  &  & 0.412 & 0.018 & 3.078 & 0.175 & 1.000 & 0.069  &  & 0.412 & 0.018 & 3.040 & 0.153 & 1.000 & 0.140 \\
4 & 0.413 & 0.016 & 3.162 & 0.316 & 1.000 & 0.087  &  & 0.418 & 0.020 & 4.379 & 0.503 & 1.000 & 0.875  &  & 0.413 & 0.016 & 3.206 & 0.320 & 1.000 & 0.087  &  & 0.418 & 0.020 & 4.446 & 0.503 & 1.000 & 0.882 \\
\multicolumn{28}{c}{}\\
\bottomrule
\end{tabular}
\begin{tablenotes}[para,flushleft]
\footnotesize
\textbf{Notes:} 2SIV$^{[\kappa]}$ refers to the 2SIV (pooled) estimator  developed by \citet{CuiSarafidisYamagata2023} based on instrument matrix $\mathbf{\widehat{Z}}^{[\kappa]}_{i}$, for $\kappa \in \{1,2,3,L \}$, as explained in Section \ref{subsec:Design}.
\end{tablenotes}
\end{threeparttable}
\end{table}
\end{landscape}

\begin{landscape}
\begin{table}[ht]
\centering
\begin{threeparttable}
\footscriptsize
\setlength{\tabcolsep}{2.7pt}
\renewcommand{\arraystretch}{1.15}
\caption{2SIV results for $\psi=0.25$}
\label{tab:sim_results_psi_2SIV}
\begin{tabular}{l*{27}{c}}
\addlinespace[2pt]

\multicolumn{28}{c}{\small \textbf{Panel A ($\pi_{u}=3/4$)}}\\

& \multicolumn{6}{c}{\small \textbf{2SIV$^{[1]}$}} &  & \multicolumn{6}{c}{\small \textbf{2SIV$^{[2]}$}} &  & \multicolumn{6}{c}{\small \textbf{2SIV$^{[3]}$}} &  & \multicolumn{6}{c}{\small \textbf{2SIV$^{[L]}$}} \\
\cmidrule(lr){2-7} \cmidrule(lr){9-14} \cmidrule(lr){16-21} \cmidrule(lr){23-28}

\multicolumn{28}{c}{\small \textbf{Case I:} $N=25\tau$, $T=100\tau$}\\

$\tau$ & Mean & RMSE & ARB & Size & Power & J-test &  & Mean & RMSE & ARB & Size & Power & J-test &  & Mean & RMSE & ARB & Size & Power & J-test &  & Mean & RMSE & ARB & Size & Power & J-test \\
\cmidrule(lr){2-7} \cmidrule(lr){9-14} \cmidrule(lr){16-21} \cmidrule(lr){23-28}

1 & 0.248 & 0.027 & 0.643 & 0.103 & 0.944 & 0.091 &  & 0.248 & 0.027 & 0.692 & 0.097 & 0.934 & 0.121 &  & 0.248 & 0.027 & 0.624 & 0.097 & 0.939 & 0.098 &  & 0.248 & 0.027 & 0.687 & 0.054 & 0.932 & 0.217 \\
2 & 0.249 & 0.016 & 0.479 & 0.074 & 1.000 & 0.068 &  & 0.248 & 0.016 & 0.622 & 0.070 & 1.000 & 0.235 &  & 0.249 & 0.016 & 0.368 & 0.072 & 1.000 & 0.075 &  & 0.248 & 0.016 & 0.634 & 0.054 & 1.000 & 0.321 \\
4 & 0.249 & 0.010 & 0.594 & 0.064 & 1.000 & 0.091 &  & 0.248 & 0.010 & 0.730 & 0.067 & 1.000 & 0.753 &  & 0.249 & 0.010 & 0.464 & 0.063 & 1.000 & 0.091 &  & 0.256 & 0.011 & 2.440 & 0.067 & 0.999 & 0.756 \\

\multicolumn{28}{c}{\small \textbf{Case II:} $N=100\tau$, $T=25\tau$}\\

$\tau$ & Mean & RMSE & ARB & Size & Power & J-test &  & Mean & RMSE & ARB & Size & Power & J-test &  & Mean & RMSE & ARB & Size & Power & J-test &  & Mean & RMSE & ARB & Size & Power & J-test \\
\cmidrule(lr){2-7} \cmidrule(lr){9-14} \cmidrule(lr){16-21} \cmidrule(lr){23-28}

1 & 0.248 & 0.024 & 0.624 & 0.059 & 0.981 & 0.063 &  & 0.249 & 0.025 & 0.559 & 0.064 & 0.970 & 0.110 &  & 0.249 & 0.024 & 0.503 & 0.056 & 0.975 & 0.068 &  & 0.249 & 0.025 & 0.476 & 0.031 & 0.968 & 0.058 \\
2 & 0.248 & 0.012 & 0.706 & 0.061 & 1.000 & 0.061 &  & 0.248 & 0.013 & 0.706 & 0.063 & 1.000 & 0.231 &  & 0.249 & 0.012 & 0.540 & 0.061 & 1.000 & 0.059 &  & 0.248 & 0.013 & 0.701 & 0.042 & 1.000 & 0.262 \\
4 & 0.249 & 0.007 & 0.432 & 0.056 & 1.000 & 0.074 &  & 0.249 & 0.007 & 0.544 & 0.064 & 1.000 & 0.781 &  & 0.249 & 0.007 & 0.316 & 0.064 & 1.000 & 0.077 &  & 0.250 & 0.007 & 0.042 & 0.063 & 0.999 & 0.790 \\

\multicolumn{28}{c}{\small \textbf{Case III:} $N=50\tau$, $T=50\tau$}\\

$\tau$ & Mean & RMSE & ARB & Size & Power & J-test &  & Mean & RMSE & ARB & Size & Power & J-test &  & Mean & RMSE & ARB & Size & Power & J-test &  & Mean & RMSE & ARB & Size & Power & J-test \\
\cmidrule(lr){2-7} \cmidrule(lr){9-14} \cmidrule(lr){16-21} \cmidrule(lr){23-28}

1 & 0.249 & 0.024 & 0.453 & 0.072 & 0.968 & 0.061 &  & 0.249 & 0.025 & 0.581 & 0.068 & 0.969 & 0.113 &  & 0.249 & 0.024 & 0.491 & 0.064 & 0.972 & 0.077 &  & 0.249 & 0.025 & 0.564 & 0.044 & 0.968 & 0.139 \\
2 & 0.249 & 0.013 & 0.612 & 0.069 & 1.000 & 0.055 &  & 0.248 & 0.014 & 0.697 & 0.060 & 1.000 & 0.233 &  & 0.249 & 0.013 & 0.417 & 0.062 & 1.000 & 0.063 &  & 0.248 & 0.014 & 0.702 & 0.042 & 1.000 & 0.282 \\
4 & 0.249 & 0.008 & 0.509 & 0.058 & 1.000 & 0.088 &  & 0.248 & 0.008 & 0.646 & 0.062 & 1.000 & 0.769 &  & 0.249 & 0.008 & 0.382 & 0.055 & 1.000 & 0.086 &  & 0.256 & 0.010 & 2.209 & 0.061 & 1.000 & 0.775 \\
\midrule
% ===== Panel B =====
\multicolumn{28}{c}{\small \textbf{Panel B ($\pi_{u}=1/4$)}}\\

& \multicolumn{6}{c}{\small \textbf{2SIV$^{[1]}$}} &  & \multicolumn{6}{c}{\small \textbf{2SIV$^{[2]}$}} &  & \multicolumn{6}{c}{\small \textbf{2SIV$^{[3]}$}} &  & \multicolumn{6}{c}{\small \textbf{2SIV$^{[L]}$}} \\
\cmidrule(lr){2-7} \cmidrule(lr){9-14} \cmidrule(lr){16-21} \cmidrule(lr){23-28}

\multicolumn{28}{c}{\small \textbf{Case I:} $N=25\tau$, $T=100\tau$}\\

$\tau$ & Mean & RMSE & ARB & Size & Power & J-test &  & Mean & RMSE & ARB & Size & Power & J-test &  & Mean & RMSE & ARB & Size & Power & J-test &  & Mean & RMSE & ARB & Size & Power & J-test \\
\cmidrule(lr){2-7} \cmidrule(lr){9-14} \cmidrule(lr){16-21} \cmidrule(lr){23-28}

1 & 0.249 & 0.027 & 0.292 & 0.099 & 0.938 & 0.079 &  & 0.249 & 0.027 & 0.263 & 0.098 & 0.930 & 0.129 &  & 0.249 & 0.027 & 0.223 & 0.097 & 0.934 & 0.083 &  & 0.250 & 0.027 & 0.099 & 0.041 & 0.927 & 0.149 \\
2 & 0.249 & 0.016 & 0.474 & 0.071 & 1.000 & 0.072 &  & 0.248 & 0.016 & 0.630 & 0.076 & 1.000 & 0.292 &  & 0.249 & 0.016 & 0.418 & 0.076 & 1.000 & 0.065 &  & 0.249 & 0.016 & 0.369 & 0.041 & 1.000 & 0.123 \\
4 & 0.249 & 0.010 & 0.485 & 0.067 & 1.000 & 0.081 &  & 0.249 & 0.010 & 0.619 & 0.068 & 1.000 & 0.866 &  & 0.249 & 0.010 & 0.352 & 0.066 & 1.000 & 0.094 &  & 0.256 & 0.011 & 2.269 & 0.068 & 1.000 & 0.868 \\

\multicolumn{28}{c}{\small \textbf{Case II:} $N=100\tau$, $T=25\tau$}\\

$\tau$ & Mean & RMSE & ARB & Size & Power & J-test &  & Mean & RMSE & ARB & Size & Power & J-test &  & Mean & RMSE & ARB & Size & Power & J-test &  & Mean & RMSE & ARB & Size & Power & J-test \\
\cmidrule(lr){2-7} \cmidrule(lr){9-14} \cmidrule(lr){16-21} \cmidrule(lr){23-28}

1 & 0.249 & 0.023 & 0.574 & 0.058 & 0.981 & 0.058 &  & 0.249 & 0.024 & 0.456 & 0.065 & 0.978 & 0.101 &  & 0.249 & 0.023 & 0.411 & 0.066 & 0.981 & 0.060 &  & 0.249 & 0.024 & 0.343 & 0.022 & 0.970 & 0.151 \\
2 & 0.249 & 0.012 & 0.477 & 0.062 & 1.000 & 0.052 &  & 0.249 & 0.012 & 0.518 & 0.062 & 1.000 & 0.290 &  & 0.249 & 0.012 & 0.339 & 0.064 & 1.000 & 0.059 &  & 0.249 & 0.012 & 0.313 & 0.024 & 1.000 & 0.143 \\
4 & 0.249 & 0.007 & 0.548 & 0.061 & 1.000 & 0.094 &  & 0.248 & 0.007 & 0.665 & 0.061 & 1.000 & 0.889 &  & 0.249 & 0.007 & 0.412 & 0.057 & 1.000 & 0.081 &  & 0.256 & 0.008 & 2.310 & 0.060 & 0.999 & 0.891 \\

\multicolumn{28}{c}{\small \textbf{Case III:} $N=50\tau$, $T=50\tau$}\\

$\tau$ & Mean & RMSE & ARB & Size & Power & J-test &  & Mean & RMSE & ARB & Size & Power & J-test &  & Mean & RMSE & ARB & Size & Power & J-test &  & Mean & RMSE & ARB & Size & Power & J-test \\
\cmidrule(lr){2-7} \cmidrule(lr){9-14} \cmidrule(lr){16-21} \cmidrule(lr){23-28}

1 & 0.249 & 0.024 & 0.430 & 0.078 & 0.974 & 0.068 &  & 0.249 & 0.025 & 0.387 & 0.081 & 0.973 & 0.121 &  & 0.249 & 0.024 & 0.339 & 0.078 & 0.975 & 0.073 &  & 0.250 & 0.025 & 0.208 & 0.032 & 0.971 & 0.155 \\
2 & 0.249 & 0.013 & 0.390 & 0.063 & 1.000 & 0.069 &  & 0.249 & 0.013 & 0.482 & 0.061 & 1.000 & 0.300 &  & 0.249 & 0.013 & 0.270 & 0.062 & 1.000 & 0.069 &  & 0.249 & 0.013 & 0.249 & 0.036 & 1.000 & 0.140 \\
4 & 0.249 & 0.008 & 0.454 & 0.068 & 1.000 & 0.087 &  & 0.249 & 0.008 & 0.585 & 0.065 & 1.000 & 0.875 &  & 0.249 & 0.008 & 0.317 & 0.063 & 1.000 & 0.087 &  & 0.240 & 0.013 & 4.112 & 0.065 & 0.999 & 0.882 \\
\multicolumn{28}{c}{}\\
\bottomrule
\end{tabular}
\begin{tablenotes}[para,flushleft]
\footnotesize
\textbf{Notes:} 2SIV$^{[\kappa]}$ refers to the 2SIV (pooled) estimator in Eq. \eqref{eq:MGest} based on the instruments $\mathbf{\widehat{Z}}^{[\kappa]}_{i}$, for $\kappa \in \{1,2,3,L \}$, as explained in Section \ref{subsec:Design}.
\end{tablenotes}
\end{threeparttable}
\end{table}
\end{landscape}

\begin{landscape}
\begin{table}[ht]
\centering
\begin{threeparttable}
\footscriptsize
\setlength{\tabcolsep}{2.7pt}
\renewcommand{\arraystretch}{1.15}
\caption{2SIV results for $\beta_{1}=3$}
\label{tab:sim_results_beta1_2SIV}
\begin{tabular}{l*{27}{c}}
\addlinespace[2pt]

\multicolumn{28}{c}{\small \textbf{Panel A ($\pi_{u}=3/4$)}}\\

& \multicolumn{6}{c}{\small \textbf{2SIV$^{[1]}$}} &  & \multicolumn{6}{c}{\small \textbf{2SIV$^{[2]}$}} &  & \multicolumn{6}{c}{\small \textbf{2SIV$^{[3]}$}} &  & \multicolumn{6}{c}{\small \textbf{2SIV$^{[L]}$}} \\
\cmidrule(lr){2-7} \cmidrule(lr){9-14} \cmidrule(lr){16-21} \cmidrule(lr){23-28}

\multicolumn{28}{c}{\small \textbf{Case I:} $N=25\tau$, $T=100\tau$}\\

$\tau$ & Mean & RMSE & ARB & Size & Power & J-test &  & Mean & RMSE & ARB & Size & Power & J-test &  & Mean & RMSE & ARB & Size & Power & J-test &  & Mean & RMSE & ARB & Size & Power & J-test \\
\cmidrule(lr){2-7} \cmidrule(lr){9-14} \cmidrule(lr){16-21} \cmidrule(lr){23-28}

1 & 2.999 & 0.062 & 0.049 & 0.087 & 0.353 & 0.0905 &  & 2.990 & 0.062 & 0.319 & 0.087 & 0.290 & 0.121 &  & 2.997 & 0.062 & 0.085 & 0.085 & 0.337 & 0.098 &  & 2.990 & 0.062 & 0.341 & 0.054 & 0.279 & 0.2165 \\
2 & 3.000 & 0.033 & 0.005 & 0.057 & 0.858 & 0.068 &  & 2.992 & 0.033 & 0.283 & 0.066 & 0.780 & 0.235 &  & 2.999 & 0.033 & 0.020 & 0.061 & 0.846 & 0.0745 &  & 2.991 & 0.033 & 0.300 & 0.049 & 0.775 & 0.321 \\
4 & 3.001 & 0.019 & 0.023 & 0.055 & 1.000 & 0.091 &  & 2.992 & 0.020 & 0.271 & 0.080 & 0.998 & 0.7525 &  & 3.000 & 0.019 & 0.013 & 0.054 & 1.000 & 0.091 &  & 2.997 & 0.018 & 0.089 & 0.080 & 0.997 & 0.7555 \\

\multicolumn{28}{c}{\small \textbf{Case II:} $N=100\tau$, $T=25\tau$}\\

$\tau$ & Mean & RMSE & ARB & Size & Power & J-test &  & Mean & RMSE & ARB & Size & Power & J-test &  & Mean & RMSE & ARB & Size & Power & J-test &  & Mean & RMSE & ARB & Size & Power & J-test \\
\cmidrule(lr){2-7} \cmidrule(lr){9-14} \cmidrule(lr){16-21} \cmidrule(lr){23-28}

1 & 3.006 & 0.064 & 0.205 & 0.067 & 0.380 & 0.063 &  & 3.001 & 0.063 & 0.035 & 0.062 & 0.364 & 0.1095 &  & 3.005 & 0.064 & 0.181 & 0.063 & 0.390 & 0.068 &  & 3.004 & 0.064 & 0.147 & 0.040 & 0.365 & 0.058 \\
2 & 3.006 & 0.031 & 0.208 & 0.058 & 0.920 & 0.061 &  & 2.999 & 0.030 & 0.033 & 0.058 & 0.896 & 0.2305 &  & 3.006 & 0.031 & 0.193 & 0.059 & 0.918 & 0.0585 &  & 2.999 & 0.030 & 0.028 & 0.037 & 0.900 & 0.2615 \\
4 & 3.004 & 0.016 & 0.137 & 0.055 & 1.000 & 0.074 &  & 2.996 & 0.016 & 0.138 & 0.055 & 1.000 & 0.781 &  & 3.004 & 0.016 & 0.127 & 0.051 & 1.000 & 0.077 &  & 2.995 & 0.016 & 0.174 & 0.055 & 1.000 & 0.79 \\

\multicolumn{28}{c}{\small \textbf{Case III:} $N=50\tau$, $T=50\tau$}\\

$\tau$ & Mean & RMSE & ARB & Size & Power & J-test &  & Mean & RMSE & ARB & Size & Power & J-test &  & Mean & RMSE & ARB & Size & Power & J-test &  & Mean & RMSE & ARB & Size & Power & J-test \\
\cmidrule(lr){2-7} \cmidrule(lr){9-14} \cmidrule(lr){16-21} \cmidrule(lr){23-28}

1 & 3.003 & 0.062 & 0.098 & 0.067 & 0.378 & 0.061 &  & 2.996 & 0.061 & 0.121 & 0.062 & 0.358 & 0.113 &  & 3.002 & 0.062 & 0.074 & 0.065 & 0.374 & 0.0765 &  & 2.997 & 0.060 & 0.099 & 0.039 & 0.354 & 0.1385 \\
2 & 3.002 & 0.032 & 0.074 & 0.064 & 0.886 & 0.0545 &  & 2.994 & 0.032 & 0.194 & 0.062 & 0.833 & 0.233 &  & 3.002 & 0.032 & 0.058 & 0.062 & 0.885 & 0.0625 &  & 2.994 & 0.032 & 0.200 & 0.046 & 0.831 & 0.2815 \\
4 & 3.002 & 0.017 & 0.068 & 0.053 & 1.000 & 0.0875 &  & 2.994 & 0.018 & 0.216 & 0.077 & 1.000 & 0.769 &  & 3.002 & 0.017 & 0.058 & 0.057 & 1.000 & 0.0855 &  & 2.993 & 0.018 & 0.246 & 0.076 & 1.000 & 0.775 \\
\midrule
% ===== Panel B =====
\multicolumn{28}{c}{\small \textbf{Panel B ($\pi_{u}=1/4$)}}\\

& \multicolumn{6}{c}{\small \textbf{2SIV$^{[1]}$}} &  & \multicolumn{6}{c}{\small \textbf{2SIV$^{[2]}$}} &  & \multicolumn{6}{c}{\small \textbf{2SIV$^{[3]}$}} &  & \multicolumn{6}{c}{\small \textbf{2SIV$^{[L]}$}} \\
\cmidrule(lr){2-7} \cmidrule(lr){9-14} \cmidrule(lr){16-21} \cmidrule(lr){23-28}

\multicolumn{28}{c}{\small \textbf{Case I:} $N=25\tau$, $T=100\tau$}\\

$\tau$ & Mean & RMSE & ARB & Size & Power & J-test &  & Mean & RMSE & ARB & Size & Power & J-test &  & Mean & RMSE & ARB & Size & Power & J-test &  & Mean & RMSE & ARB & Size & Power & J-test \\
\cmidrule(lr){2-7} \cmidrule(lr){9-14} \cmidrule(lr){16-21} \cmidrule(lr){23-28}

1 & 3.003 & 0.060 & 0.087 & 0.082 & 0.383 & 0.079 &  & 2.995 & 0.060 & 0.173 & 0.088 & 0.346 & 0.1285 &  & 3.002 & 0.060 & 0.051 & 0.087 & 0.393 & 0.083 &  & 3.002 & 0.060 & 0.075 & 0.035 & 0.377 & 0.1485 \\
2 & 3.000 & 0.032 & 0.007 & 0.065 & 0.866 & 0.0715 &  & 2.992 & 0.033 & 0.276 & 0.075 & 0.785 & 0.2915 &  & 3.000 & 0.032 & 0.009 & 0.069 & 0.859 & 0.065 &  & 3.000 & 0.032 & 0.000 & 0.031 & 0.845 & 0.123 \\
4 & 3.000 & 0.017 & 0.007 & 0.046 & 1.000 & 0.0805 &  & 2.991 & 0.019 & 0.301 & 0.071 & 0.999 & 0.866 &  & 2.999 & 0.017 & 0.019 & 0.044 & 1.000 & 0.0935 &  & 2.999 & 0.016 & 0.030 & 0.071 & 0.998 & 0.868 \\

\multicolumn{28}{c}{\small \textbf{Case II:} $N=100\tau$, $T=25\tau$}\\

$\tau$ & Mean & RMSE & ARB & Size & Power & J-test &  & Mean & RMSE & ARB & Size & Power & J-test &  & Mean & RMSE & ARB & Size & Power & J-test &  & Mean & RMSE & ARB & Size & Power & J-test \\
\cmidrule(lr){2-7} \cmidrule(lr){9-14} \cmidrule(lr){16-21} \cmidrule(lr){23-28}

1 & 3.005 & 0.057 & 0.171 & 0.061 & 0.469 & 0.058 &  & 3.000 & 0.056 & 0.012 & 0.061 & 0.425 & 0.1005 &  & 3.004 & 0.057 & 0.144 & 0.061 & 0.465 & 0.060 &  & 3.005 & 0.057 & 0.165 & 0.022 & 0.444 & 0.1505 \\
2 & 3.006 & 0.029 & 0.213 & 0.056 & 0.954 & 0.052 &  & 2.999 & 0.028 & 0.028 & 0.048 & 0.940 & 0.290 &  & 3.006 & 0.029 & 0.198 & 0.053 & 0.956 & 0.0585 &  & 3.006 & 0.029 & 0.204 & 0.019 & 0.954 & 0.1425 \\
4 & 3.004 & 0.015 & 0.136 & 0.056 & 1.000 & 0.094 &  & 2.996 & 0.015 & 0.138 & 0.060 & 1.000 & 0.8885 &  & 3.004 & 0.015 & 0.126 & 0.058 & 1.000 & 0.081 &  & 2.993 & 0.016 & 0.233 & 0.060 & 0.999 & 0.8905 \\

\multicolumn{28}{c}{\small \textbf{Case III:} $N=50\tau$, $T=50\tau$}\\

$\tau$ & Mean & RMSE & ARB & Size & Power & J-test &  & Mean & RMSE & ARB & Size & Power & J-test &  & Mean & RMSE & ARB & Size & Power & J-test &  & Mean & RMSE & ARB & Size & Power & J-test \\
\cmidrule(lr){2-7} \cmidrule(lr){9-14} \cmidrule(lr){16-21} \cmidrule(lr){23-28}

1 & 3.004 & 0.056 & 0.147 & 0.065 & 0.446 & 0.0675 &  & 2.998 & 0.055 & 0.082 & 0.062 & 0.406 & 0.1205 &  & 3.003 & 0.056 & 0.113 & 0.065 & 0.448 & 0.0725 &  & 3.004 & 0.056 & 0.142 & 0.022 & 0.445 & 0.155 \\
2 & 3.003 & 0.030 & 0.101 & 0.054 & 0.938 & 0.0685 &  & 2.995 & 0.029 & 0.171 & 0.061 & 0.893 & 0.300 &  & 3.002 & 0.029 & 0.082 & 0.056 & 0.940 & 0.0685 &  & 3.003 & 0.030 & 0.092 & 0.028 & 0.926 & 0.1395 \\
4 & 3.002 & 0.016 & 0.069 & 0.048 & 1.000 & 0.087 &  & 2.993 & 0.017 & 0.222 & 0.065 & 1.000 & 0.8745 &  & 3.002 & 0.016 & 0.058 & 0.048 & 1.000 & 0.087 &  & 2.999 & 0.015 & 0.019 & 0.065 & 0.999 & 0.8815 \\
\multicolumn{28}{c}{}\\
\bottomrule
\end{tabular}
\begin{tablenotes}[para,flushleft]
\footnotesize
\textbf{Notes:} 2SIV$^{[\kappa]}$ refers to the 2SIV (pooled) estimator in Eq. \eqref{eq:MGest} based on the instruments $\mathbf{\widehat{Z}}^{[\kappa]}_{i}$, for $\kappa \in \{1,2,3,L \}$, as explained in Section \ref{subsec:Design}.
\end{tablenotes}
\end{threeparttable}
\end{table}
\end{landscape}

\begin{landscape}
\begin{table}[ht]
\centering
\begin{threeparttable}
\footscriptsize
\setlength{\tabcolsep}{2.7pt}
\renewcommand{\arraystretch}{1.15}
\caption{2SIV results for $\beta_{2}=1$}
\label{tab:sim_results_beta2_2SIV}
\begin{tabular}{l*{27}{c}}
\addlinespace[2pt]

\multicolumn{28}{c}{\small \textbf{Panel A ($\pi_{u}=3/4$)}}\\

& \multicolumn{6}{c}{\small \textbf{2SIV$^{[1]}$}} &  & \multicolumn{6}{c}{\small \textbf{2SIV$^{[2]}$}} &  & \multicolumn{6}{c}{\small \textbf{2SIV$^{[3]}$}} &  & \multicolumn{6}{c}{\small \textbf{2SIV$^{[L]}$}} \\
\cmidrule(lr){2-7} \cmidrule(lr){9-14} \cmidrule(lr){16-21} \cmidrule(lr){23-28}

\multicolumn{28}{c}{\small \textbf{Case I:} $N=25\tau$, $T=100\tau$}\\

$\tau$ & Mean & RMSE & ARB & Size & Power & J-test &  & Mean & RMSE & ARB & Size & Power & J-test &  & Mean & RMSE & ARB & Size & Power & J-test &  & Mean & RMSE & ARB & Size & Power & J-test \\
\cmidrule(lr){2-7} \cmidrule(lr){9-14} \cmidrule(lr){16-21} \cmidrule(lr){23-28}

1 & 1.010 & 0.063 & 0.997 & 0.099 & 0.408 & 0.0905 &  & 1.007 & 0.063 & 0.732 & 0.095 & 0.393 & 0.121 &  & 1.010 & 0.063 & 0.954 & 0.096 & 0.400 & 0.098 &  & 1.007 & 0.062 & 0.719 & 0.048 & 0.405 & 0.2165 \\
2 & 1.008 & 0.033 & 0.813 & 0.074 & 0.905 & 0.068 &  & 1.005 & 0.032 & 0.534 & 0.068 & 0.901 & 0.235 &  & 1.008 & 0.033 & 0.798 & 0.073 & 0.909 & 0.0745 &  & 1.005 & 0.032 & 0.517 & 0.046 & 0.895 & 0.321 \\
4 & 1.009 & 0.021 & 0.880 & 0.089 & 1.000 & 0.091 &  & 1.006 & 0.019 & 0.581 & 0.076 & 1.000 & 0.7525 &  & 1.009 & 0.020 & 0.868 & 0.089 & 1.000 & 0.091 &  & 1.006 & 0.019 & 0.568 & 0.075 & 0.999 & 0.7555 \\

\multicolumn{28}{c}{\small \textbf{Case II:} $N=100\tau$, $T=25\tau$}\\

$\tau$ & Mean & RMSE & ARB & Size & Power & J-test &  & Mean & RMSE & ARB & Size & Power & J-test &  & Mean & RMSE & ARB & Size & Power & J-test &  & Mean & RMSE & ARB & Size & Power & J-test \\
\cmidrule(lr){2-7} \cmidrule(lr){9-14} \cmidrule(lr){16-21} \cmidrule(lr){23-28}

1 & 1.027 & 0.074 & 2.691 & 0.107 & 0.348 & 0.063 &  & 1.025 & 0.073 & 2.499 & 0.106 & 0.345 & 0.1095 &  & 1.026 & 0.074 & 2.643 & 0.108 & 0.353 & 0.068 &  & 1.026 & 0.073 & 2.600 & 0.076 & 0.320 & 0.058 \\
2 & 1.015 & 0.034 & 1.545 & 0.094 & 0.949 & 0.061 &  & 1.013 & 0.033 & 1.299 & 0.082 & 0.947 & 0.2305 &  & 1.015 & 0.034 & 1.527 & 0.094 & 0.947 & 0.0585 &  & 1.013 & 0.033 & 1.303 & 0.055 & 0.945 & 0.2615 \\
4 & 1.013 & 0.020 & 1.312 & 0.142 & 1.000 & 0.074 &  & 1.010 & 0.019 & 1.032 & 0.109 & 1.000 & 0.781 &  & 1.013 & 0.020 & 1.302 & 0.138 & 1.000 & 0.077 &  & 1.008 & 0.017 & 0.809 & 0.109 & 0.999 & 0.79 \\

\multicolumn{28}{c}{\small \textbf{Case III:} $N=50\tau$, $T=50\tau$}\\

$\tau$ & Mean & RMSE & ARB & Size & Power & J-test &  & Mean & RMSE & ARB & Size & Power & J-test &  & Mean & RMSE & ARB & Size & Power & J-test &  & Mean & RMSE & ARB & Size & Power & J-test \\
\cmidrule(lr){2-7} \cmidrule(lr){9-14} \cmidrule(lr){16-21} \cmidrule(lr){23-28}

1 & 1.016 & 0.066 & 1.632 & 0.092 & 0.422 & 0.061 &  & 1.014 & 0.065 & 1.390 & 0.091 & 0.410 & 0.113 &  & 1.016 & 0.065 & 1.580 & 0.093 & 0.415 & 0.0765 &  & 1.014 & 0.065 & 1.413 & 0.056 & 0.412 & 0.1385 \\
2 & 1.012 & 0.033 & 1.215 & 0.084 & 0.919 & 0.0545 &  & 1.009 & 0.032 & 0.942 & 0.081 & 0.919 & 0.233 &  & 1.012 & 0.033 & 1.197 & 0.082 & 0.918 & 0.0625 &  & 1.009 & 0.032 & 0.936 & 0.055 & 0.918 & 0.2815 \\
4 & 1.010 & 0.018 & 0.965 & 0.086 & 1.000 & 0.0875 &  & 1.007 & 0.017 & 0.677 & 0.061 & 1.000 & 0.769 &  & 1.010 & 0.018 & 0.954 & 0.084 & 1.000 & 0.0855 &  & 1.011 & 0.018 & 1.102 & 0.061 & 0.999 & 0.775 \\
\midrule
% ===== Panel B =====
\multicolumn{28}{c}{\small \textbf{Panel B ($\pi_{u}=1/4$)}}\\

& \multicolumn{6}{c}{\small \textbf{2SIV$^{[1]}$}} &  & \multicolumn{6}{c}{\small \textbf{2SIV$^{[2]}$}} &  & \multicolumn{6}{c}{\small \textbf{2SIV$^{[3]}$}} &  & \multicolumn{6}{c}{\small \textbf{2SIV$^{[L]}$}} \\
\cmidrule(lr){2-7} \cmidrule(lr){9-14} \cmidrule(lr){16-21} \cmidrule(lr){23-28}

\multicolumn{28}{c}{\small \textbf{Case I:} $N=25\tau$, $T=100\tau$}\\

$\tau$ & Mean & RMSE & ARB & Size & Power & J-test &  & Mean & RMSE & ARB & Size & Power & J-test &  & Mean & RMSE & ARB & Size & Power & J-test &  & Mean & RMSE & ARB & Size & Power & J-test \\
\cmidrule(lr){2-7} \cmidrule(lr){9-14} \cmidrule(lr){16-21} \cmidrule(lr){23-28}

1 & 1.008 & 0.059 & 0.801 & 0.085 & 0.437 & 0.079 &  & 1.005 & 0.059 & 0.517 & 0.084 & 0.423 & 0.1285 &  & 1.007 & 0.059 & 0.721 & 0.084 & 0.430 & 0.083 &  & 1.007 & 0.059 & 0.720 & 0.027 & 0.427 & 0.1485 \\
2 & 1.008 & 0.032 & 0.841 & 0.077 & 0.907 & 0.0715 &  & 1.006 & 0.031 & 0.553 & 0.072 & 0.902 & 0.2915 &  & 1.008 & 0.032 & 0.819 & 0.077 & 0.908 & 0.065 &  & 1.008 & 0.032 & 0.821 & 0.027 & 0.903 & 0.123 \\
4 & 1.008 & 0.019 & 0.795 & 0.066 & 1.000 & 0.0805 &  & 1.005 & 0.018 & 0.496 & 0.053 & 1.000 & 0.866 &  & 1.008 & 0.019 & 0.782 & 0.064 & 1.000 & 0.0935 &  & 1.010 & 0.019 & 0.990 & 0.053 & 0.999 & 0.868 \\

\multicolumn{28}{c}{\small \textbf{Case II:} $N=100\tau$, $T=25\tau$}\\

$\tau$ & Mean & RMSE & ARB & Size & Power & J-test &  & Mean & RMSE & ARB & Size & Power & J-test &  & Mean & RMSE & ARB & Size & Power & J-test &  & Mean & RMSE & ARB & Size & Power & J-test \\
\cmidrule(lr){2-7} \cmidrule(lr){9-14} \cmidrule(lr){16-21} \cmidrule(lr){23-28}

1 & 1.018 & 0.060 & 1.757 & 0.076 & 0.511 & 0.058 &  & 1.016 & 0.060 & 1.553 & 0.073 & 0.499 & 0.1005 &  & 1.017 & 0.060 & 1.704 & 0.074 & 0.500 & 0.060 &  & 1.017 & 0.060 & 1.682 & 0.024 & 0.451 & 0.1505 \\
2 & 1.016 & 0.033 & 1.588 & 0.095 & 0.950 & 0.052 &  & 1.013 & 0.032 & 1.339 & 0.082 & 0.952 & 0.29 &  & 1.016 & 0.033 & 1.568 & 0.097 & 0.948 & 0.0585 &  & 1.016 & 0.033 & 1.563 & 0.035 & 0.958 & 0.1425 \\
4 & 1.013 & 0.020 & 1.265 & 0.142 & 1.000 & 0.094 &  & 1.010 & 0.018 & 0.987 & 0.108 & 1.000 & 0.8885 &  & 1.013 & 0.020 & 1.254 & 0.139 & 1.000 & 0.081 &  & 1.002 & 0.014 & 0.202 & 0.108 & 1.000 & 0.8905 \\

\multicolumn{28}{c}{\small \textbf{Case III:} $N=50\tau$, $T=50\tau$}\\

$\tau$ & Mean & RMSE & ARB & Size & Power & J-test &  & Mean & RMSE & ARB & Size & Power & J-test &  & Mean & RMSE & ARB & Size & Power & J-test &  & Mean & RMSE & ARB & Size & Power & J-test \\
\cmidrule(lr){2-7} \cmidrule(lr){9-14} \cmidrule(lr){16-21} \cmidrule(lr){23-28}

1 & 1.013 & 0.059 & 1.319 & 0.082 & 0.509 & 0.0675 &  & 1.011 & 0.058 & 1.052 & 0.073 & 0.501 & 0.1205 &  & 1.012 & 0.058 & 1.239 & 0.077 & 0.507 & 0.0725 &  & 1.012 & 0.058 & 1.236 & 0.027 & 0.459 & 0.155 \\
2 & 1.012 & 0.033 & 1.242 & 0.075 & 0.934 & 0.0685 &  & 1.010 & 0.032 & 0.960 & 0.069 & 0.935 & 0.300 &  & 1.012 & 0.033 & 1.220 & 0.075 & 0.936 & 0.0685 &  & 1.012 & 0.033 & 1.219 & 0.031 & 0.938 & 0.1395 \\
4 & 1.010 & 0.019 & 1.045 & 0.102 & 1.000 & 0.087 &  & 1.007 & 0.018 & 0.749 & 0.070 & 1.000 & 0.8745 &  & 1.010 & 0.019 & 1.033 & 0.101 & 1.000 & 0.087 &  & 1.001 & 0.016 & 0.078 & 0.069 & 1.000 & 0.8815 \\
\multicolumn{28}{c}{}\\
\bottomrule
\end{tabular}
\begin{tablenotes}[para,flushleft]
\footnotesize
\textbf{Notes:} 2SIV$^{[\kappa]}$ refers to the 2SIV (pooled) estimator in Eq. \eqref{eq:MGest} based on the instruments $\mathbf{\widehat{Z}}^{[\kappa]}_{i}$, for $\kappa \in \{1,2,3,L \}$, as explained in Section \ref{subsec:Design}.
\end{tablenotes}
\end{threeparttable}
\end{table}
\end{landscape}

\clearpage

Tables \ref{tab:sim_results_QMLE}-\ref{tab:sim_results_CCEXIV} report summary results for the QMLE estimator of \citet{ShiLee2017} and the CCEX-IV estimator of \citet{ChenEtAl2022}, respectively, in terms of Mean, RMSE, ARB, size and power. CCEX-IV includes contemporaneous and $p_T$ lagged cross-sectional averages of $y_{it}$ and the regressors, where $p_T=\lfloor T^{1/3}\rfloor$. The inclusion of these lags is required because, in dynamic panels cross-sectional averages proxy the unobserved common factors only through an infinite distributed-lag representation. Following \citet{ChudikPesaran2015}, this infinite distributed-lag representation is approximated by including $p_T$ lags.

\begin{landscape}
\begin{table}[ht]
\centering
\begin{threeparttable}
\footscriptsize
\setlength{\tabcolsep}{3.2pt}
\renewcommand{\arraystretch}{1.15}
\caption{Results for the QMLE estimator of \citet{ShiLee2017}}
\label{tab:sim_results_QMLE}
\begin{tabular}{l*{20}{c}}
\toprule

\multicolumn{21}{c}{\small \textbf{Panel A ($\pi_{u}=3/4$)}}\\

& \multicolumn{5}{c}{\small \textbf{$\rho=0.4$}}
& \multicolumn{5}{c}{\small \textbf{$\psi=0.25$}}
& \multicolumn{5}{c}{\small \textbf{$\beta_1=3$}}
& \multicolumn{5}{c}{\small \textbf{$\beta_2=1$}} \\
\cmidrule(lr){2-6} \cmidrule(lr){7-11} \cmidrule(lr){12-16} \cmidrule(lr){17-21}

\multicolumn{21}{c}{\small \textbf{Case I:} $N=25\tau$, $T=100\tau$}\\

$\tau$ & Mean & RMSE & ARB & Size & Power
      & Mean & RMSE & ARB & Size & Power
      & Mean & RMSE & ARB & Size & Power
      & Mean & RMSE & ARB & Size & Power \\
\cmidrule(lr){2-6} \cmidrule(lr){7-11} \cmidrule(lr){12-16} \cmidrule(lr){17-21}

1 & 0.429 & 0.042 & 7.139 & 0.755 & 0.895 & 0.246 & 0.044 & 1.571 & 0.535 & 0.590 & 2.958 & 0.110 & 1.405 & 0.500 & 0.050 & 1.095 & 0.139 & 9.496 & 0.610 & 0.245 \\
2 & 0.427 & 0.035 & 6.762 & 0.840 & 0.990 & 0.246 & 0.022 & 1.629 & 0.410 & 1.000 & 2.980 & 0.056 & 0.656 & 0.415 & 0.285 & 1.052 & 0.072 & 5.178 & 0.545 & 0.455 \\
4 & 0.430 & 0.033 & 7.551 & 0.985 & 1.000 & 0.250 & 0.014 & 0.090 & 0.520 & 1.000 & 2.971 & 0.039 & 0.955 & 0.600 & 0.450 & 1.025 & 0.035 & 2.456 & 0.580 & 0.980 \\

\multicolumn{21}{c}{\small \textbf{Case II:} $N=100\tau$, $T=25\tau$}\\

$\tau$ & Mean & RMSE & ARB & Size & Power
      & Mean & RMSE & ARB & Size & Power
      & Mean & RMSE & ARB & Size & Power
      & Mean & RMSE & ARB & Size & Power \\
\cmidrule(lr){2-6} \cmidrule(lr){7-11} \cmidrule(lr){12-16} \cmidrule(lr){17-21}

1 & 0.409 & 0.021 & 2.273 & 0.460 & 1.000 & 0.246 & 0.026 & 1.457 & 0.185 & 0.935 & 2.987 & 0.075 & 0.427 & 0.275 & 0.215 & 1.114 & 0.135 & 11.370 & 0.715 & 0.375 \\
2 & 0.420 & 0.023 & 4.935 & 0.855 & 1.000 & 0.249 & 0.013 & 0.349 & 0.140 & 1.000 & 2.992 & 0.035 & 0.258 & 0.185 & 0.720 & 1.058 & 0.068 & 5.807 & 0.650 & 0.845 \\
4 & 0.428 & 0.029 & 6.977 & 1.000 & 1.000 & 0.249 & 0.008 & 0.529 & 0.210 & 1.000 & 2.980 & 0.027 & 0.664 & 0.430 & 0.955 & 1.025 & 0.032 & 2.549 & 0.570 & 1.000 \\

\multicolumn{21}{c}{\small \textbf{Case III:} $N=50\tau$, $T=50\tau$}\\

$\tau$ & Mean & RMSE & ARB & Size & Power
      & Mean & RMSE & ARB & Size & Power
      & Mean & RMSE & ARB & Size & Power
      & Mean & RMSE & ARB & Size & Power \\
\cmidrule(lr){2-6} \cmidrule(lr){7-11} \cmidrule(lr){12-16} \cmidrule(lr){17-21}

1 & 0.418 & 0.031 & 4.420 & 0.690 & 0.970 & 0.248 & 0.030 & 0.931 & 0.245 & 0.920 & 2.980 & 0.078 & 0.666 & 0.310 & 0.170 & 1.089 & 0.121 & 8.889 & 0.565 & 0.310 \\
2 & 0.424 & 0.027 & 5.976 & 0.900 & 1.000 & 0.246 & 0.018 & 1.581 & 0.305 & 1.000 & 2.984 & 0.039 & 0.544 & 0.260 & 0.585 & 1.050 & 0.066 & 5.042 & 0.560 & 0.710 \\
4 & 0.430 & 0.032 & 7.534 & 0.995 & 1.000 & 0.250 & 0.010 & 0.204 & 0.290 & 1.000 & 2.976 & 0.032 & 0.808 & 0.560 & 0.820 & 1.024 & 0.031 & 2.407 & 0.565 & 0.995 \\

\midrule
\multicolumn{21}{c}{\small \textbf{Panel B ($\pi_{u}=1/4$)}}\\

& \multicolumn{5}{c}{\small \textbf{$\rho=0.4$}}
& \multicolumn{5}{c}{\small \textbf{$\psi=0.25$}}
& \multicolumn{5}{c}{\small \textbf{$\beta_1=3$}}
& \multicolumn{5}{c}{\small \textbf{$\beta_2=1$}} \\
\cmidrule(lr){2-6} \cmidrule(lr){7-11} \cmidrule(lr){12-16} \cmidrule(lr){17-21}

\multicolumn{21}{c}{\small \textbf{Case I:} $N=25\tau$, $T=100\tau$}\\

$\tau$ & Mean & RMSE & ARB & Size & Power
      & Mean & RMSE & ARB & Size & Power
      & Mean & RMSE & ARB & Size & Power
      & Mean & RMSE & ARB & Size & Power \\
\cmidrule(lr){2-6} \cmidrule(lr){7-11} \cmidrule(lr){12-16} \cmidrule(lr){17-21}

1 & 0.425 & 0.044 & 6.273 & 0.830 & 0.795 & 0.248 & 0.037 & 0.974 & 0.520 & 0.730 & 2.978 & 0.082 & 0.730 & 0.505 & 0.165 & 1.041 & 0.090 & 4.062 & 0.535 & 0.285 \\
2 & 0.425 & 0.034 & 6.268 & 0.915 & 0.980 & 0.251 & 0.022 & 0.258 & 0.610 & 0.995 & 2.987 & 0.052 & 0.441 & 0.550 & 0.395 & 1.050 & 0.078 & 4.960 & 0.615 & 0.270 \\
4 & 0.429 & 0.034 & 7.331 & 0.980 & 1.000 & 0.249 & 0.014 & 0.321 & 0.700 & 1.000 & 2.978 & 0.040 & 0.717 & 0.685 & 0.510 & 1.040 & 0.052 & 3.990 & 0.815 & 0.885 \\

\multicolumn{21}{c}{\small \textbf{Case II:} $N=100\tau$, $T=25\tau$}\\

$\tau$ & Mean & RMSE & ARB & Size & Power
      & Mean & RMSE & ARB & Size & Power
      & Mean & RMSE & ARB & Size & Power
      & Mean & RMSE & ARB & Size & Power \\
\cmidrule(lr){2-6} \cmidrule(lr){7-11} \cmidrule(lr){12-16} \cmidrule(lr){17-21}

1 & 0.416 & 0.026 & 3.867 & 0.680 & 0.995 & 0.248 & 0.024 & 0.997 & 0.285 & 0.975 & 2.995 & 0.047 & 0.157 & 0.195 & 0.540 & 1.049 & 0.077 & 4.855 & 0.335 & 0.350 \\
2 & 0.424 & 0.027 & 5.949 & 0.935 & 1.000 & 0.249 & 0.013 & 0.612 & 0.320 & 1.000 & 2.993 & 0.031 & 0.232 & 0.270 & 0.850 & 1.045 & 0.058 & 4.458 & 0.605 & 0.805 \\
4 & 0.426 & 0.028 & 6.611 & 0.995 & 1.000 & 0.249 & 0.008 & 0.344 & 0.410 & 1.000 & 2.986 & 0.026 & 0.482 & 0.545 & 0.950 & 1.042 & 0.048 & 4.173 & 0.890 & 0.995 \\

\multicolumn{21}{c}{\small \textbf{Case III:} $N=50\tau$, $T=50\tau$}\\

$\tau$ & Mean & RMSE & ARB & Size & Power
      & Mean & RMSE & ARB & Size & Power
      & Mean & RMSE & ARB & Size & Power
      & Mean & RMSE & ARB & Size & Power \\
\cmidrule(lr){2-6} \cmidrule(lr){7-11} \cmidrule(lr){12-16} \cmidrule(lr){17-21}

1 & 0.419 & 0.032 & 4.639 & 0.765 & 0.975 & 0.250 & 0.026 & 0.067 & 0.315 & 0.960 & 2.984 & 0.062 & 0.533 & 0.320 & 0.205 & 1.040 & 0.077 & 3.970 & 0.400 & 0.320 \\
2 & 0.425 & 0.031 & 6.275 & 0.920 & 1.000 & 0.248 & 0.016 & 0.891 & 0.430 & 1.000 & 2.982 & 0.043 & 0.611 & 0.500 & 0.445 & 1.040 & 0.057 & 4.007 & 0.585 & 0.675 \\
4 & 0.427 & 0.030 & 6.843 & 0.980 & 1.000 & 0.249 & 0.010 & 0.402 & 0.460 & 1.000 & 2.982 & 0.030 & 0.598 & 0.690 & 0.885 & 1.044 & 0.051 & 4.413 & 0.885 & 0.990 \\

\multicolumn{21}{c}{}\\
\bottomrule
\end{tabular}
\begin{tablenotes}[para,flushleft]
\footnotesize
\textbf{Notes:} QMLE refers to the quasi-maximum likelihood estimator of \citet{ShiLee2017}, designed for large spatial, dynamic panels with common factors and homogeneous slope coefficients.
\end{tablenotes}
\end{threeparttable}
\end{table}
\end{landscape}

\begin{landscape}
\begin{table}[ht]
\centering
\begin{threeparttable}
\footscriptsize
\setlength{\tabcolsep}{3.2pt}
\renewcommand{\arraystretch}{1.15}
\caption{Results for the CCEX-IV estimator of \citet{ChenEtAl2022}}
\label{tab:sim_results_CCEXIV}
\begin{tabular}{l*{20}{c}}
\toprule

\multicolumn{21}{c}{\small \textbf{Panel A ($\pi_{u}=3/4$)}}\\

& \multicolumn{5}{c}{\small \textbf{$\rho=0.4$}}
& \multicolumn{5}{c}{\small \textbf{$\psi=0.25$}}
& \multicolumn{5}{c}{\small \textbf{$\beta_1=3$}}
& \multicolumn{5}{c}{\small \textbf{$\beta_2=1$}} \\
\cmidrule(lr){2-6} \cmidrule(lr){7-11} \cmidrule(lr){12-16} \cmidrule(lr){17-21}

\multicolumn{21}{c}{\small \textbf{Case I:} $N=25\tau$, $T=100\tau$}\\

$\tau$ & Mean & RMSE & ARB & Size & Power
      & Mean & RMSE & ARB & Size & Power
      & Mean & RMSE & ARB & Size & Power
      & Mean & RMSE & ARB & Size & Power \\
\cmidrule(lr){2-6} \cmidrule(lr){7-11} \cmidrule(lr){12-16} \cmidrule(lr){17-21}

1 & 0.413 & 0.032 & 3.211 & 0.420 & 0.760 & 0.249 & 0.036 & 0.491 & 0.220 & 0.480 & 2.992 & 0.106 & 0.272 & 0.190 & 0.090 & 1.250 & 0.273 & 25.010 & 0.760 & 0.165 \\
2 & 0.410 & 0.023 & 2.546 & 0.465 & 0.930 & 0.244 & 0.025 & 2.391 & 0.235 & 0.625 & 2.992 & 0.069 & 0.283 & 0.205 & 0.185 & 1.244 & 0.256 & 24.370 & 0.915 & 0.195 \\
4 & 0.411 & 0.018 & 2.849 & 0.585 & 0.995 & 0.250 & 0.018 & 0.123 & 0.255 & 0.785 & 3.002 & 0.046 & 0.081 & 0.265 & 0.325 & 1.251 & 0.258 & 25.100 & 0.995 & 0.160 \\

\multicolumn{21}{c}{\small \textbf{Case II:} $N=100\tau$, $T=25\tau$}\\

$\tau$ & Mean & RMSE & ARB & Size & Power
      & Mean & RMSE & ARB & Size & Power
      & Mean & RMSE & ARB & Size & Power
      & Mean & RMSE & ARB & Size & Power \\
\cmidrule(lr){2-6} \cmidrule(lr){7-11} \cmidrule(lr){12-16} \cmidrule(lr){17-21}

1 & 0.411 & 0.030 & 2.632 & 0.335 & 0.810 & 0.248 & 0.028 & 0.668 & 0.150 & 0.795 & 2.997 & 0.085 & 0.103 & 0.220 & 0.160 & 1.238 & 0.255 & 23.810 & 0.855 & 0.130 \\
2 & 0.411 & 0.019 & 2.745 & 0.500 & 1.000 & 0.249 & 0.015 & 0.509 & 0.155 & 0.990 & 2.996 & 0.066 & 0.146 & 0.370 & 0.170 & 1.249 & 0.256 & 24.880 & 0.995 & 0.140 \\
4 & 0.411 & 0.016 & 2.851 & 0.665 & 1.000 & 0.249 & 0.010 & 0.418 & 0.210 & 0.995 & 2.999 & 0.037 & 0.032 & 0.350 & 0.425 & 1.250 & 0.256 & 25.010 & 1.000 & 0.185 \\

\multicolumn{21}{c}{\small \textbf{Case III:} $N=50\tau$, $T=50\tau$}\\

$\tau$ & Mean & RMSE & ARB & Size & Power
      & Mean & RMSE & ARB & Size & Power
      & Mean & RMSE & ARB & Size & Power
      & Mean & RMSE & ARB & Size & Power \\
\cmidrule(lr){2-6} \cmidrule(lr){7-11} \cmidrule(lr){12-16} \cmidrule(lr){17-21}

1 & 0.408 & 0.028 & 1.995 & 0.375 & 0.830 & 0.251 & 0.034 & 0.411 & 0.235 & 0.490 & 3.001 & 0.090 & 0.021 & 0.215 & 0.185 & 1.241 & 0.257 & 24.100 & 0.845 & 0.175 \\
2 & 0.412 & 0.020 & 2.932 & 0.530 & 0.995 & 0.248 & 0.019 & 1.006 & 0.185 & 0.855 & 3.000 & 0.057 & 0.012 & 0.255 & 0.190 & 1.251 & 0.258 & 25.080 & 0.990 & 0.130 \\
4 & 0.412 & 0.017 & 3.096 & 0.700 & 1.000 & 0.250 & 0.012 & 0.134 & 0.215 & 0.960 & 3.005 & 0.040 & 0.162 & 0.265 & 0.460 & 1.244 & 0.249 & 24.390 & 1.000 & 0.180 \\

\midrule
\multicolumn{21}{c}{\small \textbf{Panel B ($\pi_{u}=1/4$)}}\\

& \multicolumn{5}{c}{\small \textbf{$\rho=0.4$}}
& \multicolumn{5}{c}{\small \textbf{$\psi=0.25$}}
& \multicolumn{5}{c}{\small \textbf{$\beta_1=3$}}
& \multicolumn{5}{c}{\small \textbf{$\beta_2=1$}} \\
\cmidrule(lr){2-6} \cmidrule(lr){7-11} \cmidrule(lr){12-16} \cmidrule(lr){17-21}

\multicolumn{21}{c}{\small \textbf{Case I:} $N=25\tau$, $T=100\tau$}\\

$\tau$ & Mean & RMSE & ARB & Size & Power
      & Mean & RMSE & ARB & Size & Power
      & Mean & RMSE & ARB & Size & Power
      & Mean & RMSE & ARB & Size & Power \\
\cmidrule(lr){2-6} \cmidrule(lr){7-11} \cmidrule(lr){12-16} \cmidrule(lr){17-21}

1 & 0.411 & 0.037 & 2.751 & 0.505 & 0.485 & 0.249 & 0.048 & 0.290 & 0.260 & 0.250 & 2.992 & 0.156 & 0.279 & 0.375 & 0.060 & 1.318 & 0.343 & 31.790 & 0.830 & 0.080 \\
2 & 0.411 & 0.027 & 2.742 & 0.495 & 0.650 & 0.252 & 0.033 & 0.645 & 0.305 & 0.415 & 2.998 & 0.111 & 0.066 & 0.390 & 0.040 & 1.316 & 0.333 & 31.570 & 0.940 & 0.125 \\
4 & 0.414 & 0.024 & 3.388 & 0.655 & 0.740 & 0.248 & 0.022 & 0.941 & 0.260 & 0.770 & 2.996 & 0.085 & 0.139 & 0.405 & 0.105 & 1.304 & 0.316 & 30.410 & 0.995 & 0.085 \\

\multicolumn{21}{c}{\small \textbf{Case II:} $N=100\tau$, $T=25\tau$}\\

$\tau$ & Mean & RMSE & ARB & Size & Power
      & Mean & RMSE & ARB & Size & Power
      & Mean & RMSE & ARB & Size & Power
      & Mean & RMSE & ARB & Size & Power \\
\cmidrule(lr){2-6} \cmidrule(lr){7-11} \cmidrule(lr){12-16} \cmidrule(lr){17-21}

1 & 0.410 & 0.034 & 2.428 & 0.495 & 0.660 & 0.247 & 0.030 & 1.194 & 0.225 & 0.520 & 2.990 & 0.139 & 0.345 & 0.490 & 0.065 & 1.305 & 0.322 & 30.540 & 0.960 & 0.115 \\
2 & 0.409 & 0.024 & 2.321 & 0.595 & 0.845 & 0.249 & 0.017 & 0.313 & 0.185 & 0.900 & 3.003 & 0.098 & 0.095 & 0.530 & 0.095 & 1.308 & 0.319 & 30.820 & 1.000 & 0.115 \\
4 & 0.411 & 0.020 & 2.639 & 0.755 & 0.910 & 0.249 & 0.012 & 0.286 & 0.315 & 0.945 & 2.989 & 0.092 & 0.365 & 0.605 & 0.080 & 1.310 & 0.316 & 30.980 & 1.000 & 0.120 \\

\multicolumn{21}{c}{\small \textbf{Case III:} $N=50\tau$, $T=50\tau$}\\

$\tau$ & Mean & RMSE & ARB & Size & Power
      & Mean & RMSE & ARB & Size & Power
      & Mean & RMSE & ARB & Size & Power
      & Mean & RMSE & ARB & Size & Power \\
\cmidrule(lr){2-6} \cmidrule(lr){7-11} \cmidrule(lr){12-16} \cmidrule(lr){17-21}

1 & 0.408 & 0.034 & 1.940 & 0.455 & 0.605 & 0.250 & 0.032 & 0.090 & 0.240 & 0.585 & 3.001 & 0.131 & 0.035 & 0.385 & 0.070 & 1.302 & 0.321 & 30.200 & 0.870 & 0.070 \\
2 & 0.412 & 0.026 & 2.893 & 0.605 & 0.875 & 0.248 & 0.023 & 0.845 & 0.250 & 0.605 & 3.015 & 0.105 & 0.506 & 0.450 & 0.110 & 1.310 & 0.323 & 30.950 & 0.990 & 0.110 \\
4 & 0.412 & 0.021 & 2.875 & 0.655 & 0.780 & 0.250 & 0.016 & 0.090 & 0.270 & 0.895 & 3.007 & 0.087 & 0.247 & 0.530 & 0.175 & 1.311 & 0.320 & 31.060 & 1.000 & 0.125 \\

\multicolumn{21}{c}{}\\
\bottomrule
\end{tabular}
\begin{tablenotes}[para,flushleft]
\footnotesize
\textbf{Notes:} CCEX-IV refers to the IV estimator of \citet{ChenEtAl2022}, designed for large spatial, static panels with common factors and heterogeneous slope coefficients.
\end{tablenotes}
\end{threeparttable}
\end{table}
\end{landscape}

\newpage
\section*{Appendix C: Robustness Results for Empirical Illustration}
\label{sec-robustness}

\renewcommand{\thetable}{C.\arabic{table}}
\setcounter{table}{0}

Table \ref{tab:Robustness} presents results from a series of robustness checks designed to test the stability of the estimated coefficients using different spatial weighting matrices and sample data.\footnote{Compared to Table \ref{tab:MAIN}, notice here that $r_1+r_2$ is not reported because all the results are from the same heterogeneous parameter model as Column [6] of Table \ref{tab:MAIN}, where the MGIV estimator does not involve the estimation of $y$ factors in a second stage.}

Column [1] replicates the results from Column [6] of Table \ref{tab:MAIN}, serving as the benchmark. All other columns in Table \ref{tab:Robustness} have the same model specification as Column [1] but differ in either the spatial weighting matrix or sample composition. Columns [2] and [3] use Eq. \eqref{W_empirical} to construct the weighting matrix but modify the distance decay parameter $\zeta$ to $0.015$ and $0.01$, respectively. Lowering $\zeta$ results in a slower exponential decay with distance, assigning higher weights to more distant regions. Consequently, the influence of remote regions is amplified, leading to a weaker spatial localization effect. This adjustment reflects a scenario in which regional interactions are less constrained by geographical proximity and are more broadly distributed across the spatial network.

In Column [4], the spatial weighting matrix is specified as $w_{ii} = 0$, $w_{ij} = \frac{d_{ij}^{-2}}{\sum_{j=1}^{N} d_{ij}^{-2}}$ for $j \neq i$, a specification also explored in Ertur and Koch (2007). This formulation imposes a quadratic decay in spatial interactions, whereby closer regions exert significantly larger weights relative to those farther away. Column [5] modifies the sample by excluding 10 island regions within the EU. These regions, due to their physical separation, may exhibit weaker integration into the mainland spatial network. This adjustment assesses whether the inclusion of such isolated regions disproportionately influences the estimated spatial dynamics. Finally, Column [6] treats $\ln\left(inv_{it}\right)$ as weakly exogenous and therefore excludes instruments based on investment from the instrument set.

Some key observations are as follows. First, despite the variations in the weighting matrix and sample composition, the spatial spillover effects demonstrate remarkable consistency. Specifically, the coefficients of $\sum_{j=1}^Nw_{ij}\Delta \ln \left(gdp_{jt}\right)$, $\sum_{j=1}^Nw_{ij}\ln \left(gdp_{j,t-1}\right)$, $\sum_{j=1}^N w_{ij}\ln\left(inv_{jt}\right)$ and $\sum_{j=1}^N w_{ij}\ln\left(educ_{jt}\right)$ are similar in magnitude and significance across most configurations. This stability affirms the robustness of the estimates to different choices of the weighting matrix and alternative assumptions regarding the exogeneity of investment, and reinforces the conclusion that spillovers from neighboring regions play a substantial and persistent role in shaping a region's economic growth trajectory.
In addition, the estimated rate of conditional convergence, captured by the sum of the coefficients of $\ln \left(gdp_{i,t-1}\right)$ and $\sum_{j=1}^{N} w_{ij} \ln \left(gdp_{j,t-1}\right)$, also remains consistent across specifications. This invariance suggests that regional income convergence is robust to changes in the spatial weighting structure and sample adjustments.

On the other hand, noticeable variations arise in the impact of a region's own investment rate and population growth: for $\ln\left(inv_{it}\right)$, the direct effect is diminished in some specifications, while its spatial counterpart increases in magnitude. This pattern highlights the dominant role of investment spillovers relative to localized investment impacts, particularly when larger weights are assigned to interactions with distant regions. The coefficient on $\ln\left(n_{it}\right)$ also exhibits greater variability across specifications, reflecting the complex and context-dependent nature of demographic dynamics in influencing regional growth.

The robustness checks confirm the reliability of our key findings, particularly the significance of spillover effects. The consistent convergence rate further validates the framework's ability to capture regional growth dynamics.

\begin{table}[htbp]\centering
\caption{{Robustness results for empirical growth model} \index{Robustness Results for Empirical Growth Model 5.3}}
\label{tab:Robustness}
\begin{tabular}{lcccccc}
\hline
                        & [1]          & [2]          & [3]          & [4]          & [5]          & [6]          \\
\hline
$\ln\left(gdp_{i,t-1}\right)$ & -0.473***    & -0.481***    & -0.497***    & -0.517***    & -0.477***    & -0.437***    \\
                         & (0.028)      & (0.028)      & (0.030)      & (0.029)      & (0.029)      & (0.030)      \\

$\ln\left(inv_{it}\right)$ & 0.027**      & 0.023*       & 0.009        & 0.004        & 0.028**      & 0.020*      \\
                         & (0.013)      & (0.012)      & (0.013)      & (0.012)      & (0.014)      & (0.015)      \\

$\ln\left(n_{it}\right)$   & -0.582**     & -0.379*      & -0.297       & -0.145       & -0.684**     & -0.394*     \\
                         & (0.281)      & (0.218)      & (0.197)      & (0.201)      & (0.338)      & (0.237)      \\

$\ln\left(educ_{it}\right)$ & 0.007        & -0.006       & -0.016       & -0.024       & 0.008        & 0.011        \\
                         & (0.026)      & (0.024)      & (0.028)      & (0.025)      & (0.027)      & (0.036)      \\

$\ln\left(sci \& tech_{it}\right)$ & -0.003       & -0.008       & -0.027       & -0.019       & -0.005       & -0.013       \\
                         & (0.016)      & (0.015)      & (0.020)      & (0.014)      & (0.016)      & (0.022)      \\
\hline
\multicolumn{7}{l}{ } \\
$\sum_{j=1}^Nw_{ij}\Delta \ln \left(gdp_{jt}\right)$  & 0.628***     & 0.644***     & 0.661***     & 0.636***     & 0.629***     & 0.639***     \\
                                & (0.033)      & (0.035)      & (0.040)      & (0.038)      & (0.033)      & (0.035)      \\

$\sum_{j=1}^Nw_{ij} \ln\left(gdp_{j,t-1}\right)$ & 0.192***     & 0.191***     & 0.188***     & 0.165***     & 0.203***     & 0.202***     \\
                                & (0.052)      & (0.053)      & (0.061)      & (0.063)      & (0.054)      & (0.060)      \\

$\sum_{j=1}^Nw_{ij}\ln\left(inv_{jt}\right)$        & 0.071***     & 0.083***     & 0.099***     & 0.131***     & 0.072***     & 0.106***     \\
                                & (0.021)      & (0.021)      & (0.030)      & (0.034)      & (0.021)      & (0.023)      \\

$\sum_{j=1}^Nw_{ij}\ln\left(n_{jt}\right)$          & 0.643        & -0.242       & -0.369       & -0.696       & 1.102        & 0.600        \\
                                & (0.549)      & (0.668)      & (0.549)      & (0.521)      & (0.744)      & (0.603)      \\

$\sum_{j=1}^Nw_{ij}\ln\left(educ_{jt}\right)$       & 0.150***     & 0.176***     & 0.197***     & 0.248***     & 0.160***     & 0.055     \\
                                & (0.035)      & (0.033)      & (0.040)      & (0.053)      & (0.034)      & (0.042)      \\

$\sum_{j=1}^Nw_{ij}\ln\left(sci \& tech_{jt}\right)$ & 0.022        & 0.042        & 0.115**      & 0.108        & 0.009        & 0.004        \\
                                & (0.050)      & (0.041)      & (0.050)      & (0.081)      & (0.050)      & (0.049)      \\
\hline
$N_g$                  & 266          & 266          & 266          & 266          & 256          & 266          \\
$r_1$                  & 1            & 1            & 1            & 1            & 1            & 1            \\
\hline
\end{tabular}
\begin{minipage}{\textwidth}
\small
\textit{Notes:} Standard errors are in parentheses. * denotes significance at the 10\% level, ** at the 5\% level, and *** at the 1\% level. All the columns in the table have the same model specification as Column [6] in Table \ref{tab:MAIN} but may use a different spatial weighting matrix or sample composition.
Column [1] is the same as Column [6] in Table \ref{tab:MAIN}, repeated here for benchmarking. Columns [2] and [3] use a weighting matrix constructed by \eqref{W_empirical} but with $\zeta$ set to $0.015$ and $0.01$, respectively. In Column [4] the weighting matrix is given by $w_{ij} = \frac{d_{ij}^{-2}}{\sum_{j=1}^{N} d_{ij}^{-2}}$. Column [5] uses data that excludes 10 regions in the sample that are islands. Finally, Column [6] treats $\ln\left(inv_{it}\right)$ as weakly exogenous and therefore excludes instruments based on investment from the instrument set.
\end{minipage}
\end{table}

\end{document}